\documentclass[preprint,3p,times,twocolumn]{elsarticle}
\usepackage{graphicx}
\usepackage{setspace}
\usepackage[latin1]{inputenc}
\usepackage{color}															 
\usepackage{colortbl}
\usepackage{enumerate}
\usepackage{framed}
\usepackage{amsmath} 
\usepackage{amsthm}
\usepackage{amssymb}
\usepackage{lineno}
\usepackage{epigraph}
\usepackage[percent]{overpic}
\usepackage[usenames,dvipsnames]{xcolor}
\usepackage{morefloats}
\usepackage{textcomp}
\usepackage{color,soul}
\usepackage{wasysym}
\usepackage{caption}
\usepackage{placeins}
\usepackage{dcolumn}% Align table columns on decimal point
\usepackage{bm}
\newcommand{\avg}[1]{\left< #1 \right>} 
	\usepackage{tikz}
\usetikzlibrary{calc}
\newcommand{\squaree}{$\, \,$ \draw (0.02,0.02) rectangle (0.22,0.22);$\,$ $ \,$}
\newcommand{\fsquaree}{$\, \, $  \draw[black,fill=black] (0.02,0.02) rectangle (0.22,0.22);$\,  \, \,   \,$ }
\newcommand{\bcircle}{$\,$ $\,$\tikz\draw (0,0) circle (.7ex);$\,$ $\,$}
\newcommand{\bfcircle}{$\,$ $\,$\tikz\draw[black,fill=black] (0,0) circle (.7ex);$\,$ $\,$}

\newcommand{\tikzunderbrace}[2]{$\underbrace{\textrm{#1}}_{#2}$}
\newcommand{\tikzoverbrace}[2]{$\overbrace{\textrm{#1}}^{#2}$}
\usepackage{bbold}
\journal{Journal of Theoretical Biology}

\begin{document}

\begin{frontmatter}

\title{Self-tolerance and autoimmunity in a minimal model of the idiotypic network}

\author{Stefan Landmann}
\author{Nicolas Preuss}
\author{Ulrich Behn}%
\address{Institut f\"ur Theoretische Physik, Universit\"at Leipzig, POB 100 920, D-04009 Leipzig, Germany}
\ead{ulrich.behn@itp.uni-leipzig.de}
\begin{abstract}
We consider self-tolerance and its failure --autoimmunity-- in a minimal mathematical model of the idiotypic network. A node in the network represents a clone of B-lymphocytes and its antibodies of the same idiotype which is encoded by a bitstring. The links between nodes represent possible interactions between clones of almost complementary idiotype. A clone survives only if the number of populated neighbored nodes is neither too small nor too large. The dynamics is driven by the influx of lymphocytes with randomly generated idiotype from the bone marrow. Previous work has revealed that the network evolves towards a highly organized modular architecture, characterized by groups of nodes which share statistical properties. The structural properties of the architecture can be described analytically, the statistical properties determined from simulations are confirmed by a modular mean-field theory. To model the presence of self we permanently occupy one or several nodes. These nodes influence their linked neighbors, the autoreactive clones, but are themselves not affected by idiotypic interactions. The architecture is very similar to the case without self, but organized such that the neighbors of  self are only weakly occupied, thus providing self-tolerance. This supports the perspective that self-reactive clones, which regularly occur in healthy organisms, are controlled by anti-idiotypic clones. We discuss how perturbations, like an infection with foreign antigen, a change in the influx of new idiotypes, or the random removal of idiotypes, may lead to autoreactivity and devise protocols which cause a reconstitution of the self-tolerant state. The results could be helpful to understand network and probabilistic aspects of autoimmune disorders.  
\end{abstract}

\begin{keyword}
B lymphocytes \sep Idiotypic regulation \sep Immunological self \sep Autoimmune condition \sep Mathematical model



\end{keyword}

\end{frontmatter}

\section{Introduction}
\label{Introduction}
Self-tolerance is the state of an organism in which autoantigens do not elicit an immune response \cite{Cruse10Atlas}. The property of being self-tolerant is crucial for the proper function of organisms and its failure may lead to autoimmune diseases. It is a well established fact that autoreactive lymphocytes are present in healthy individuals and may recognize corresponding autoantigens \cite{avrameas1983studies,schwartz1986anti,tomer1988significance}. This is a considerable challenge to Burnets clonal selection theory (CST) based on positive and negative selection \cite{burnet1959clonal}. Furthermore the CST does not answer the question in which way the spontaneous production of immunoglobulins in non-immunized mice is induced \cite{pereira1986autonomous}.
\\ 
The paradigm of the idiotypic network developed by Jerne \cite{jerne74network}, see also Refs.~\cite{jerne1984idiotypic,Jerne85}, is able to explain the autonomous dynamics of the immune system not exposed to alien antigen. It also offers a mechanism of immunological memory and the network is thought to be crucial for the control of potentially autoreactive lymphocyte clones. For reviews on the concept and mathematical modeling of the idiotypic network see Refs.~\cite{behn2007idiotypic,behn2011idiotype}. The monograph by Tauber \cite{Tauber94} also discusses philosophical aspects of the concept of immunological self and non-self.
\\
The idiotypic network is generated by mutual interaction of the B-cells and antibodies. B-lymphocytes express receptor molecules, so called antibodies, on their surface. Those molecules have distinct binding sites determining their \it{idiotype}; \normalfont each receptor of one B-lymphocyte has the same idiotype. B-cells of different idiotype are produced at random in the bone marrow. The diversity of the potential repertoire is assumed to be larger than $10^{10}$ \cite{berek1988dynamic} while the expressed repertoire size is estimated to be within the order of magnitude of $10^8$ \cite{perelson1997immunology}. It is generated by somatic recombination of gene segments and mutations \cite{tonegawa1983somatic}. If the receptors of a B-cell are crosslinked by complementary structures, it is stimulated to proliferate and differentiate into memory B-cells and plasma cells. The plasma cells secrete huge amounts of antibodies possessing the idiotype of the original B-cell. If the B-cell is not stimulated it dies. Theoretical and experimental work has shown, that the stimulation-response curve is log-bell-shaped \cite{vogelstein1982specific}. Therefore, the concentration of structures which have the ability to crosslink the receptors should neither be too high nor too low in order to stimulate the cells. Complementary structures which are able to bind to the receptors are found on alien antigens but also on antibodies of complementary kind, which are called anti-idiotypic antibodies. Therefore the B-cells are able to stimulate each other via the secretion of antibodies, forming the idiotypic network. 
\\
In~\cite{coutinho1989beyond,varela1991second,coutinho2003walk} Varela and Coutinho proposed the concept of so-called second-generation networks which combine both paradigms of the CST and idiotypic networks. The architecture of these networks consists of a densely linked core  which is responsible for the autonomous dynamics of the immune system, and a sparsely linked periphery which causes the specific response to antigen.
\\
Since there are idiotypic interactions of T-cells with B-cells and between T-lymphocytes \cite{sim1986t}, the idiotypic network is not autonomous but coupled to a manifold of other networks.  Nevertheless a hypothetical autonomous B-cell system already comprises the features of random innovation, evolution and selection. Therefore, the structure of the idiotypic network can be understood as the outcome of an evolution throughout the life of an organism.
\\
The idea that idiotypic interactions may play an important role in the regulation of autoreactive lymphocytes is popular in theoretical as well as experimental studies on autoimmune disorders. The idiotypic network can control autoantibodies where other regulatory mechanism failed to suppress them \cite{UWFC77}. It has been observed that perturbations in the regulation of autoreactive clones can be associated with autoimmune diseases \cite{Hampe12,Avrameas91,SG97,Pendergraftetal04,McGH05,TR10,RT10,Shoenfeld04}, as e.g. in case of the B-cell related autoimmune disease Myasthenia gravis \cite{DVK86}. 
\\
\newline 
A huge amount of work has been performed on models of the idiotypic network and it is beyond the scope of this work to give an extensive review of this field. For an exhaustive discussion of mathematical models for idiotypic B-cell networks see \cite{behn2007idiotypic,schmidtchen2012randomly}. Here, only a few models dealing with the control of self-reactive lymphocytes are reviewed. 
\\
In the model introduced in Ref.~\cite{SV89} Stewart and Varela assumed an \textit{ad hoc} structure of second generation idiotypic networks \cite{varela1991second,coutinho2003walk} which was motivated by experimental findings \cite{kearney1987non}. An architecture composed of 26 clones was chosen, consisting of a multi-affine group A, two mirror groups  B and C without intra-group affinity but mutual coupling and a group D which only couples weakly to A. The dynamics of the B-cells and the corresponding antibodies in the presence of self is modeled by non-linear ordinary differential equations (ODEs) which are based on the proposed architecture \cite{SVC89}. The non-linear terms of the ODEs model the proliferation and maturation of B-cells which are activated by idiotypic interactions. Numerical solutions showed that the connectivity to other clones strongly influences the response of nodes coupled to self antigen. Weakly connected clones grow without limit, while highly connected clones show a large degree of tolerance.
\\
An analytic theory for the dynamics of the clones in the groups $B$ and $C$ was developed in Ref.~\cite{sulzer1994central} using a mean-field approach. In the frame of this theory one observes transitions between different fixed points corresponding to tolerant, autoimmune and neutral states. The transitions can be induced by an infection with alien antigen.
\\
Le\'on et al.~\cite{leon1998natural} proposed a network of B- and T-lymphocytes having the structure of a second generation idiotypic network as well. Hereby, the population dynamics of the B- and T-cell clones are modeled by differential equations. The network is based on a model proposed in \cite{carneiro1996modela,carneiro1996modelb}, but it is assumed that the average idiotypic connectivity is an explicit function of time and that the stimulation of T-cells is described by a log-bell shaped dose-response curve. In this model natural tolerance for antigens which are present during the neonatal period can be kept for an indefinite time after they have been removed from the network. In contrast, if the tolerance is induced in adults it is lost shortly after removal of the antigen. 
\\
In the frame of so-called spin-glass models, confer e.g. \cite{Agliari2013}, in Ref.~\cite{agliari2015anergy}, mechanisms also including B- and T-cell interactions which cause anergy of autoreactive B-lymphocytes are investigated.
\\
In \cite{menshikov2015idiotypic} a mathematical model of the immune network containing autoreactive clones was studied. It consists of a set of discrete equations constituting an idiotypic network of specific architecture which can be defined arbitrarily. The model assumes symmetric idiotype-anti-idiotype interactions and that autoreactive lymphocytes are normal components of the idiotypic network. The results for a network consisting of six clones arranged in a closed loop were compared with experimental data and a good agreement was found, strongly supporting the thesis that autoreactive lymphocytes are regulated by the idiotypic network. 
\\
Contrary to these approaches, an architecture having the characteristics of second generation networks evolves by itself in a model introduced in \cite{brede2003} which will be studied here. The model allows to examine idiotypic networks consisting of a number of B-cell clones which is large compared to the number of clones used in other models. In previous work a network comprised of 4096 clones has been investigated mainly but also larger networks have been examined. An analytical description for the building principles of the architectures observed in the model was developed in \cite{schmidtchen2012randomly} and a mean-field theory for the description of their statistical properties was introduced in \cite{schmidtchen2012Meanfield}.
In Ref.~\cite{SWB14} the emergence of self-tolerant states is examined in the frame of this model. There, one or several nodes are occupied permanently, playing the role of the self. The self nodes have a strong influence on the evolution of the network leading to the emergence of a self-tolerant steady state. These findings strongly support the thesis that idiotypic interactions are important in the regulation of autoreactive clones and their investigation could lead to a better comprehension of the causes behind autoimmunity.
In the present paper, the emergence of self-tolerance is examined further. Trying to understand possible mechanisms behind autoimmune diseases, the failure of self-tolerance in the network is studied. Therapeutic strategies are devised which are able to reconstitute the self-tolerant state. 
\\
\newline 
Besides idiotype-antiidiotype interactions there are other mechanisms which rest on T cells, confer e.g. \cite{Saeki20154,Blyuss201513}. We believe it is likely that the several mechanisms under consideration are not mutually exclusive, but are cooperating or are --in a sense-- redundant, working in parallel. For a most recent survey on theories and mathematical models of autoimmunity we refer to a thematic issue of Journal of Theoretical Biology, see especially the editorial by Root-Bernstein \cite{RootBernstein20151}.
\\
\newline
The plan of this work is as follows. For making the paper self-contained, in Secs.~\ref{Basics of the model}-\ref{Sec:Mean-Field theory} the model is introduced, the building principles of the observed patterns and a tool for their identification, the center of mass vector, are explained; furthermore, a mean-field approach for the description of the system is given. Section \ref{Self-tolerance} reviews the results of previous simulations and mean-field theory which examine the influence of self on the network \cite{SWB14}. These simulations are extended in Sec.~\ref{Self-tolerance and autoimmunity}, where it is shown that even if the self nodes are chosen at random the system evolves towards a self-tolerant state. The choice is performed under the immunological reasonable restriction that the idiotypes of the self are not allowed to differ arbitrarily much from each other.
In Sec. \ref{Sec:Transitions 12-group} the different types of possible transitions from self-tolerant states to autoimmune states are characterized and described mathematically. This is followed by Secs. \ref{Sec:Induced autoimmunity by variation of influx}-\ref{Sec:Removal of idiotypes}, where it is discussed how perturbations of the self-tolerant system may lead to such transitions using three biologically motivated examples, variations of the influx of new idiotypes produced in the bone marrow, the introduction of an infection to the system, and the random removal of occupied nodes. The corresponding average autoreactivities are determined using a mean-field approach. Section \ref{Remission of Autoimmunity and therapeutic protocols} presents two protocols, a variation of the influx and an infection, which can cause a reconstitution of the self-tolerant structure if the system has been in an autoimmune state before. 
\section{Basics of the model}
\subsection{The model}
\label{Basics of the model}
In this work a minimalistic model of the idiotypic network introduced in Ref.~\cite{brede2003} is considered. It is a coarse simplification of the original biological system but already provides many important features and shows a surprisingly complex behavior. Moreover the model has a minimal number of parameters which prevents overfitting and allows for an analytic understanding of many of its properties.
\\
Each node of the network stands for a clone of B-lymphocytes and antibodies of a given idiotype. Hereby each idiotype is encoded by a bitstring of length $d$ with entries 0 and 1. This gives a potential repertoire size of $2^d$. It is important to elucidate that a bitstring is not meant to represent the genetic code but is a caricature of the phenotype of the corresponding idiotype. Using this representation allows for an easy notion of complementarity. 
\\
Possible idiotypic interactions between the B-cell populations are represented by links between nodes of almost complementary idiotype. It is reasonable to allow for small variations since perfect complementarity of the receptor structures appears to be unrealistic. Thus, two nodes $v$ and $w$ are linked if their bitstrings are complementary up to $m$ mismatches. In other words, two nodes $v$ and $w$ are neighbors if their Hamming distance obeys $d_H(v,w)\ge d-m$. Then each node has $\kappa=\sum_{i=0}^{m}\binom{d}{i}$ neighbors. 
% We denote the undirected graph consisting of $2^d$ nodes labeled by bitstrings of length $d$ and having links between complementary nodes with up to $m$ mismatches as base graph $G_d^{(m)}$. 
\\
In the model, a node $v$ may either be occupied $n(v)=1$ or unoccupied $n(v)=0$, which corresponds to an idiotypic clone being present or absent, respectively. The expressed idiotypic network is comprised of the occupied nodes and thus does not involve the complete potential network.
\\
The dynamics on the graph is described in discrete time, whereby one time step is roughly the period in which an unstimulated B-cell dies and a stimulated B-cell proliferates.
The influx of new idiotypes produced in the bone marrow is simulated by occupying empty nodes at random with some probability $p$. Mimicking the stimulation-response curve, occupied nodes survive the update only if the number of their occupied neighbors is within a permitted window. This gives the following update rules:
 \begin{enumerate}[(i)]
 \itemsep-0.1em 
\item Influx: Occupy empty nodes with probability $p$.
\item Window rule: Count the number of occupied neighbors $n(\partial v)$ of all occupied nodes $v$. If $n(\partial v)$ is outside the window $[t_L,t_U]$, set the node $v$ empty: $n(v)=0$. This step is performed in parallel. 
\item Iterate.
\label{Update}
\end{enumerate}
This model can be classified as a probabilistic cellular automaton (as a relative of Conway's Game of Life) and as a Boolean network, confer Ref.~\cite{schmidtchen2012randomly} for an extensive discussion. It only has a minimum number of parameters: the length of the bitstrings $d$, the influx probability $p$, the permitted number of mismatches $m$ and the lower and upper threshold of the window rule $t_L$ and $t_U$. 
\\
Throughout this paper the following parameters are used: $d=12$, $m=2$, $t_L=1$, and $t_U=10$. Then, the network has $2^{12}=4096$ nodes and each node has $\kappa=79$ neighbors, making the linking neither too dense nor too sparse. The lower threshold is set to its minimum non-trivial value $t_L=1$, which means that a clone needs stimulation by at least one neighbored node in order to survive. Choosing the upper threshold $t_U=10$ excludes very regular patterns which are not of interest here. 
\\
Using this parameters, only the influx $p$ remains as control parameter. The architecture which is of most interest for this work emerges for an influx $p$ from $0.026 \dots 0.078$. We mostly choose an influx $p$ near $0.078$, where it is easiest to induce reorganization of patterns, but also broader ranges of $p$ were studied. 
\subsection{Architecture of patterns }
\label{Architecture of patterns}
\begin{figure}[h]
   \centering
   \includegraphics[width=0.8\linewidth ]{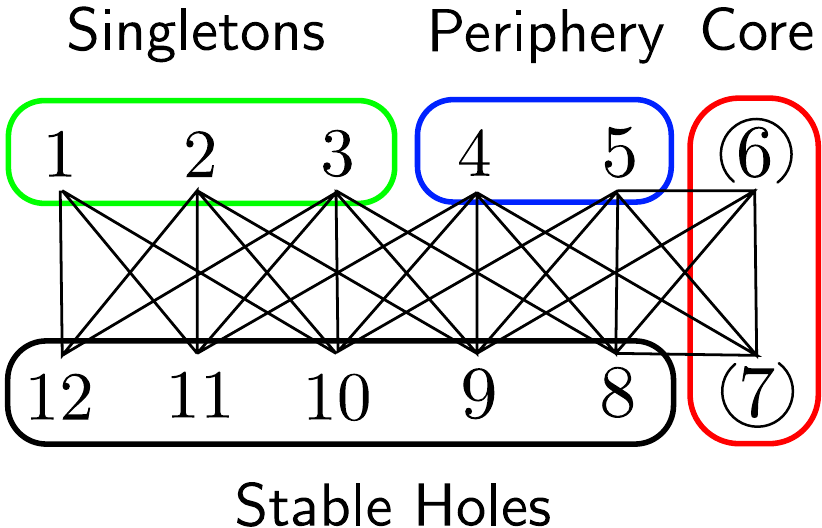}
   \caption{Schematic of the 12-group architecture. Links between the groups, i.e. possible idiotypic interactions between the corresponding nodes, are indicated by solid lines. According to their statistical properties, the groups can be sorted into larger phenomenological groups, as marked by the colored framing. Adapted from Fig.~3 in \cite{SWB14}.}
   \label{Fig:GroupStructure}
   \end{figure}

In this subsection we sketch the building principles of the patterns observed for typical parameters. A detailed derivation and discussion can be found in Ref.~\cite{schmidtchen2012randomly}.
\\
Simulations revealed that, starting from an empty base graph and applying the proposed iteration rules, a network with complex architecture emerges by itself \cite{brede2003}. In this architecture the nodes can be classified into groups which share statistical properties such as mean occupation $\avg{n(v)}$ and mean number of occupied neighbors $\avg{n(\partial v)}$. Homeostasis is implied by the fact, that the mean occupations of the nodes, the groups, and the complete base graph are stationary for typical observation times.
\\
Stated more precisely, the system is ergodic and the symmetry-breaking patterns are only quasistationary. The ergodicity can be seen easily considering the following scenario. For any influx $p>0$ there is a non-zero probability that all nodes are occupied after the influx. In this case the application of the window rule leads to an empty base graph since every node has too many occupied neighbors. Starting from the empty base graph every allowed pattern will be realized during an infinite observation time. However, in simulations we never observed such an 'extinction catastrophe' \cite{schmidtchen2012randomly} and the quasistationary patterns exist for very long periods of time.
\\
Varying the influx $p$ can cause transitions between different architectures. For small $p$ only static patterns are found, while one finds stationary dynamic patterns for intermediate influx. This architecture is the most interesting one. It includes a periphery, a densely connected core and isolated nodes (singletons) \cite{schmidtchen2012randomly,brede2003,schmidtchen2012Meanfield}, resembling the partition into a central and a peripheral immune system as proposed for second generation immune network models \cite{coutinho1989beyond,varela1991second,coutinho2003walk}. Figure \ref{Fig:GroupStructure} shows a schematic of this structure. The singleton groups are highly occupied and only linked to the permanently empty stable holes such that their occupation is determined by the random influx solely. The periphery groups are also highly occupied and linked to the core and additionally to the hole groups. The core groups are linked very densely also among their own nodes and have a low average occupation.

\begin{figure}[]
      \centering
      \includegraphics[width=1\linewidth ]{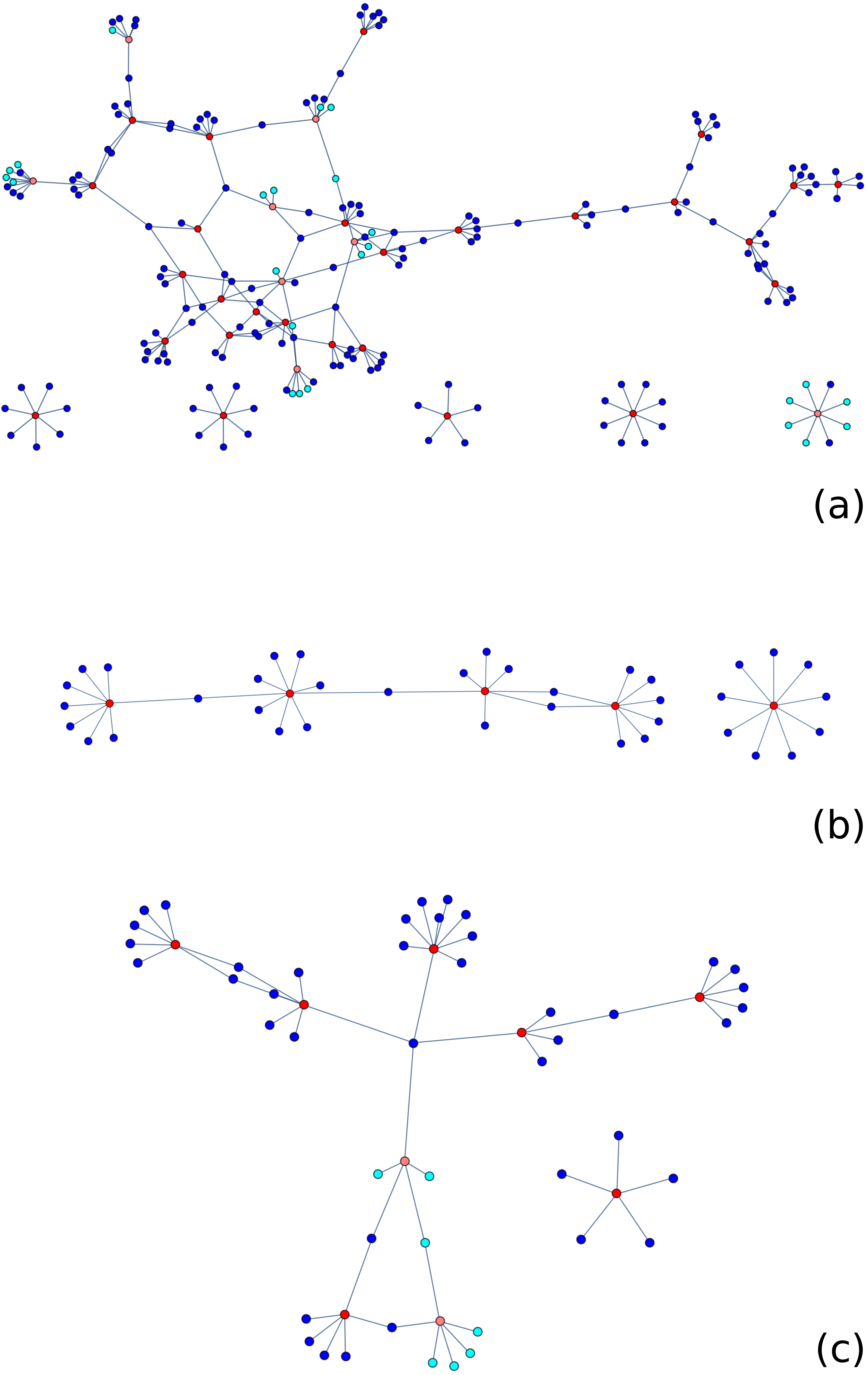}
      \caption{Snapshots of the connected parts of the network in a 12-group architecture for different values of the influx probability  $p=0.04$ (a), $p=0.05$ (b) and $p=0.074$ (c). Red and pink nodes belong to the core groups $S_6$ and $S_7$, the dark blue and light blue nodes to the periphery groups $S_5$ and $S_4$, respectively. One observes a strong linkage of the core-nodes, whereas the periphery-nodes have only very few neighbors. Occupied nodes without  occupied neighbors are not displayed. The simulations started from tabula rasa, the equilibration time was 2000 iterations.}
      \label{Network}
\end{figure}

Figure \ref{Network} shows snapshots of the connected parts of the network for different values of the influx $p$ chosen such that a 12-group architecture emerges.
One can clearly distinguish the strongly linked core nodes from the periphery nodes. The singletons are not displayed, since they have no occupied neighbors.   
\\
For larger influx $p$ the patterns become irregular and eventually chaotic. 
\\
For the classification of patterns the concept of determinant bit positions is useful \cite{schmidtchen2012randomly,schmidtchen2006randomly}. For an established architecture the number $d_M$ of the determinant bit positions is characteristic. The entries in the determinant bits determine the group membership of the nodes. Nodes in the group $S_1$ coincide in all determinant bit positions, while the entries in the non-determinant positions realize all $2^{d-d_M}$ possibilities. Nodes in group $S_2$ differ in one determinant bit position from nodes of $S_1$, while nodes from group $S_3$ differ in two determinant positions, and so on. Therefore the system consists of $d_M+1$ groups of size 
\begin{linenomath}
\begin{equation}
|S_g|=2^{d-d_M}\binom{d_M}{g-1},
\end{equation}
\end{linenomath}
with the number of links $L_{gl}$ of a node $v_g \in S_g$ to nodes in $S_l$, 
\begin{linenomath}
\begin{align}
L_{gl}=&\sum\limits_{k=0}^{m}\sum\limits_{r=0}^{k}\binom{g-1}{r}\binom{d_M-g+1}{l-1-r} \nonumber \\
	   &\times \binom{d-d_M}{k+l-1-2r-(d_M-g+1)}.
	   \label{Linkmatrix}
\end{align}
\end{linenomath}
The entries of the link matrix for the $d_M=11$ architecture are shown in Fig.~\ref{Fig:Linkmatrix}.
\begin{figure}[h!]
\centering 
\renewcommand{\arraystretch}{1.1}
 \setlength{\tabcolsep}{2pt}
\begin{tabular}{|c|ccc|cc|cc|ccccc|}
\hline
&  \multicolumn{3}{l|}{Singletons}   &  \multicolumn{2}{l|}{Periph.}   &  \multicolumn{2}{c|}{Core}  &    \multicolumn{5}{c|}{Stable holes }     \\ 
 & $S_1$ & $S_2$  & $S_3$ & $S_4$  & $S_5$ & $S_6$  & $S_7$ & $S_8$  & $S_9$ & $S_{10}$ & $S_{11}$ & $S_{12}$ \\ 
 \hline
$v_1$ &  &  &  &  &  &  &  &  &  & \multicolumn{1}{r}{55} & \multicolumn{1}{r}{22} & \multicolumn{1}{r|}{2} \\ 
$v_2$ &  &  &  &  &  &  &  &  & \multicolumn{1}{r}{45} & \multicolumn{1}{r}{20} & \multicolumn{1}{r}{12} & \multicolumn{1}{r|}{2} \\ 
$v_3$ &  &  &  &  &  &  &  & \multicolumn{1}{r}{36} & \multicolumn{1}{r}{18} & \multicolumn{1}{r}{20} &  \multicolumn{1}{r}{4} & \multicolumn{1}{r|}{1} 
\\
\hline
$v_4$ &  &  &  &  &  &  & \multicolumn{1}{r|}{28} & \multicolumn{1}{r}{16} & \multicolumn{1}{r}{26} & \multicolumn{1}{r}{6} & \multicolumn{1}{r}{3} & \multicolumn{1}{r|}{} \\ 
$v_5$ &  &  &  &  &  & \multicolumn{1}{r}{21} & \multicolumn{1}{r|}{14} & \multicolumn{1}{r}{30} & \multicolumn{1}{r}{8} & \multicolumn{1}{r}{6} &  &  \\ 
\hline
$v_6$ &  &  &  &  & \multicolumn{1}{r|}{15} & \multicolumn{1}{r}{12} & \multicolumn{1}{r|}{32} & \multicolumn{1}{r}{10} & \multicolumn{1}{r}{10} &  &  &  \\ 
$v_7$ &  &  &  & \multicolumn{1}{r}{10} & \multicolumn{1}{r|}{10} & \multicolumn{1}{r}{32} & \multicolumn{1}{r|}{12} & \multicolumn{1}{r}{15} &  &  &  &  \\ 
\hline
$v_8$ &  &  & \multicolumn{1}{r|}{6} & \multicolumn{1}{r}{8} & \multicolumn{1}{r|}{30} & \multicolumn{1}{r}{14} & \multicolumn{1}{r|}{21} &  &  &  &  &  \\ 
$v_9$ &  & \multicolumn{1}{r}{3} & \multicolumn{1}{r|}{6} & \multicolumn{1}{r}{26} & \multicolumn{1}{r|}{16} & \multicolumn{1}{r}{28} &  &  &  &  &  &  \\ 
$v_{10}$ & \multicolumn{1}{r}{1} & \multicolumn{1}{r}{4} & \multicolumn{1}{r|}{20} & \multicolumn{1}{r}{18} & \multicolumn{1}{r|}{36} &  &  &  &  &  &  &  \\ 
$v_{11}$ & \multicolumn{1}{r}{2} & \multicolumn{1}{r}{12} & \multicolumn{1}{r|}{20} & \multicolumn{1}{r}{45} &  &  &  &  &  &  &  &  \\ 
$v_{12}$ & \multicolumn{1}{r}{2} & \multicolumn{1}{r}{22} & \multicolumn{1}{r|}{55} &  &  &  &  &  &  &  &  &  \\ 
\hline
\end{tabular}
\renewcommand{\arraystretch}{1.0}
\caption{Entries of the link matrix $L_{gl}$ from Eq.~(\ref{Linkmatrix}) for the 12-group structure. Every entry shows how many neighbors a node $v_g$ has in the group $S_l$. Note that the core groups are the only groups which are linked to themselves. Missing entries are zero.}
\label{Fig:Linkmatrix}
\end{figure}

\subsection{Real time pattern identification }
\label{real time pattern identification}
\begin{figure}[h!]
      \centering
      \includegraphics[width=\linewidth ]{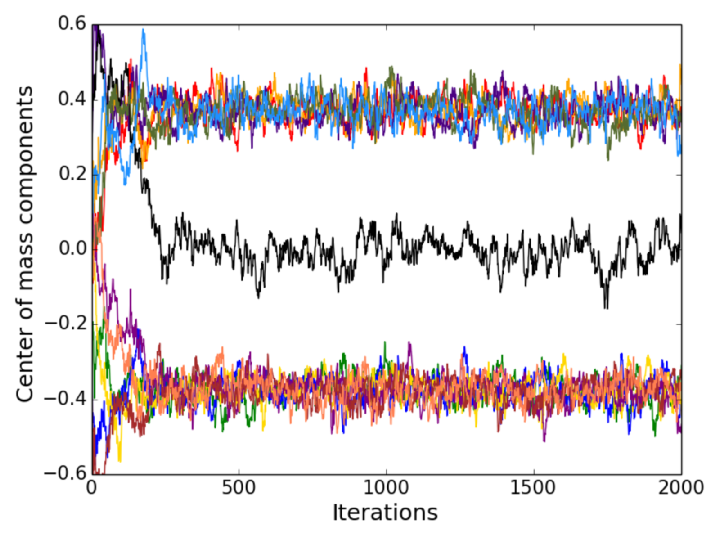}
      \caption{
 Pattern identification using the COM vector. The graph shows a time series of the COM components for an emerging 12-group structure. The simulation starts from tabula rasa; after an equilibration time of approximately 300 iterations the system reaches a quasistationary steady state. One observes that the black component $R_7$  fluctuates around zero, while the six components $R_1,R_3,R_4,R_9,R_{10},R_{11}$ fluctuate around $-0.39$ and the five components $R_2,R_5,R_6,R_8,R_{12}$ fluctuate around $0.39$. All components which fluctuate around a nonzero value correspond to determinant bit positions. Thus one can read off eleven determinant and one non-determinant bit positions and can conclude that the system has evolved towards a 12-group structure. The influx is $p=0.072$. }
      \label{COM}
      \end{figure}

The full information of the system is encoded in the time series of the $2^d$ nodes and allows to calculate quantities of interest, as e.g. the mean occupation of the graph. It is tedious to analyze this huge amount of data for the identification of patterns. By introducing the center of mass (COM) vector $\mathbf{R}$ of dimension $d$ a logarithmic reduction of information is reached, allowing for real time pattern identification and detection of pattern changes \cite{schmidtchen2012randomly}. It is defined as
\begin{linenomath}
\begin{equation}
\mathbf{R}=\frac{1}{n(G)}\sum\limits_{v}^{}n(v)\mathbf{r}(v),
\end{equation}
\end{linenomath}
with the position vector $\mathbf{r}(v)$ of a node $v$ represented by the bitstring $b_1b_2 \dots b_d$ having the components $r_i=2b_i-1$, $i=1, \dots ,d$. $n(G)$ denotes the total number of occupied nodes of the system. If the graph is randomly occupied without applying the update rule, the probability for an entry of a bitstring of an arbitrary node being zero is the same as for being one. This implies that $\mathbf{R}=0$ for such a configuration. Therefore it is easy to identify symmetry breaking patterns using the COM vector.
 \\
 Figure \ref{COM} shows the time series of the COM components for an evolving 12-group structure. 
 When the system is in a quasistationary state the COM components $R_i$ only show small fluctuations around its temporal average values $\overline{R_i}$. For any selection of determinant bits, the non-determinant bits realize all combinations of ones and zeros. Thus, assuming that every node of one group is occupied with the same probability, the average contribution of each group to the non-determinant components of $\mathbf{R}$ vanishes. Therefore, non-determinant bit positions are characterized by $\overline{R_i}\approx 0$. Observing that in Fig.~\ref{COM} only the black component $R_7$ fluctuates around zero, we conclude that a $d_M=11$ pattern has evolved. Furthermore we identify the entries of the determinant bits of the group $S_1$, which can be used to classify other nodes by comparison. The six components  $R_1,R_3,R_4,R_9,R_{10},R_{11}$ fluctuate around $-0.39$ corresponding to determinant bits with zeros as entries while the five components $R_2,R_5,R_6,R_8,R_{12}$ fluctuate around $0.39$ corresponding to determinant bits with ones as entries. This implies that nodes of the group $S_1$ have bitstrings of the form $\mathbf{0 \ 1 \ 0 \ 0 \ 1 \  1 \ \cdot \ 1 \ 0 \ 0 \ 0 \ 1}$, where the dot indicates the non-determinant bit position. Nodes of the group $S_2$ differ in the entry in one determinant bit position from this bitstring, nodes from the group $S_3$ differ in two determinant bit positions, and so on. For further details confer \cite{schmidtchen2012randomly}. 
\subsection{Mean-field theory}
\label{Sec:Mean-Field theory}
  When a certain architecture of the system, characterized by the number of groups, their size and linking, is established, it stays stationary for long periods of time and also over some interval of the influx $p$. Yet the statistical properties of nodes from the same group, as e.g. the mean occupation, depend on the actual value of $p$. Using a concept from statistical physics, the so called mean-field approach, one can determine these properties analytically in dependence of $p$ \cite{schmidtchen2012Meanfield}. \\
  In mean-field theory it is assumed that all nodes $v_g$ of a group $S_g$ are occupied independently with the same probability $P\left(n(v_g)=1\right)$. Considering an architecture with $d_M$ determinant bits there are $d_M+1$ groups. In the mean-field description the state of the system is defined by the set of average occupations $\mathbf{n}=(n_1,\dots,n_{d_M+1})^T$ of the groups instead of the set of occupations of all $2^d$ nodes. Applying the update rules (i-iii) to the state $\mathbf{n}$ gives the new state
  \begin{linenomath}
  \begin{equation}
  \mathbf{n}'=\mathbf{f(n)},
  \label{Update function}
  \end{equation}
  \end{linenomath}
  with the non-linear update function $\mathbf{f}$ depending on the described architecture and on the used update rules. Assuming an average occupation of $n_l$ in the group $S_l$, the average occupation after application of the influx with probability $p$ is given by $\tilde{n}_l=n_l+p(1-n_l)$. Furthermore, the probability for $k_l$ neighbored nodes in the group $S_l$ being occupied is given by the binomial distribution
  \begin{linenomath}
  \begin{equation}
  \binom{L_{gl}}{k_l}\tilde{n}_l^{k_l}(1-\tilde{n}_l)^{L_{gl}-k_l}.
  \label{factors}
  \end{equation}
  \end{linenomath}
 Assuming that the groups are independent, the probability for $\sum_{l=1}^{d_M+1}k_l$ neighbors being occupied in a micro-configuration with fixed neighborhood $k_l$, $l \in [1,\dots,d_m+1]$ is given by the product of the binomial distribution (\ref{factors}) for each group. Therefore, the probability that the number of occupied neighbors of a node $v_g$ in $S_g$ lies in the interval $[t_L,t_U]$ is determined by summing over all micro-configurations which are compatible with the window rule:
 \begin{linenomath}
 \begin{align}
 &\left[\sum\limits_{k_l=0}^{L_{gl}}\right]_{l=1}^{d_M+1}\mathbb{1}\left(t_L\leq\sum\limits_{l=1}^{d_M+1}k_l \leq t_U\right) \nonumber \\ \times &\prod\limits_{l=1}^{d_M+1}\binom{L_{gl}}{k_l}\widetilde{n_l}^{k_l}(1-\widetilde{n_l})^{L_{gl}-k_l}.
  \label{WindoProbability}
  \end{align}
  \end{linenomath}
  Here, the indicator function $\mathbb{1}(\cdot)$ ensures that only configurations contribute which fulfill the window rule. Finally, this expression has to be multiplied with the average occupation $\tilde{n}_g$ after the influx, giving the map
  \begin{linenomath}
  \begin{align}
  n'_g=&\tilde{n}_g\left[\sum\limits_{k_l=0}^{L_{gl}}\right]_{l=1}^{d_M+1}\mathbb{1}\left(t_L\leq\sum\limits_{l=1}^{d_M+1}k_l \leq t_U\right) \nonumber\\
  & \times \prod\limits_{l=1}^{d_M+1}\binom{L_{gl}}{k_l}\widetilde{n_l}^{k_l}(1-\widetilde{n_l})^{L_{gl}-k_l}.
  \label{MeanFieldIteration}
  \end{align}
  \end{linenomath}
 If Eq.~(\ref{MeanFieldIteration}) is iterated for all groups $g=1,\dots,d_M+1$ the state vector $\mathbf{n'}$ converges to a fixed point $\mathbf{n}^*$. Due to the non-linearity of $\mathbf{f(n)}$ there may exist several fixed points. In general, if the initial values are chosen close to the average values of stationary architectures, the fixed points correspond to results obtained from simulations. Other fixed points may also exist, but have not been found in simulations. For further details see Ref.~\cite{schmidtchen2012Meanfield}.
  \subsection{Self-tolerance}
  \label{Self-tolerance}
 
\begin{figure}[h!]
  \centering 
  \includegraphics[width=0.75\linewidth]{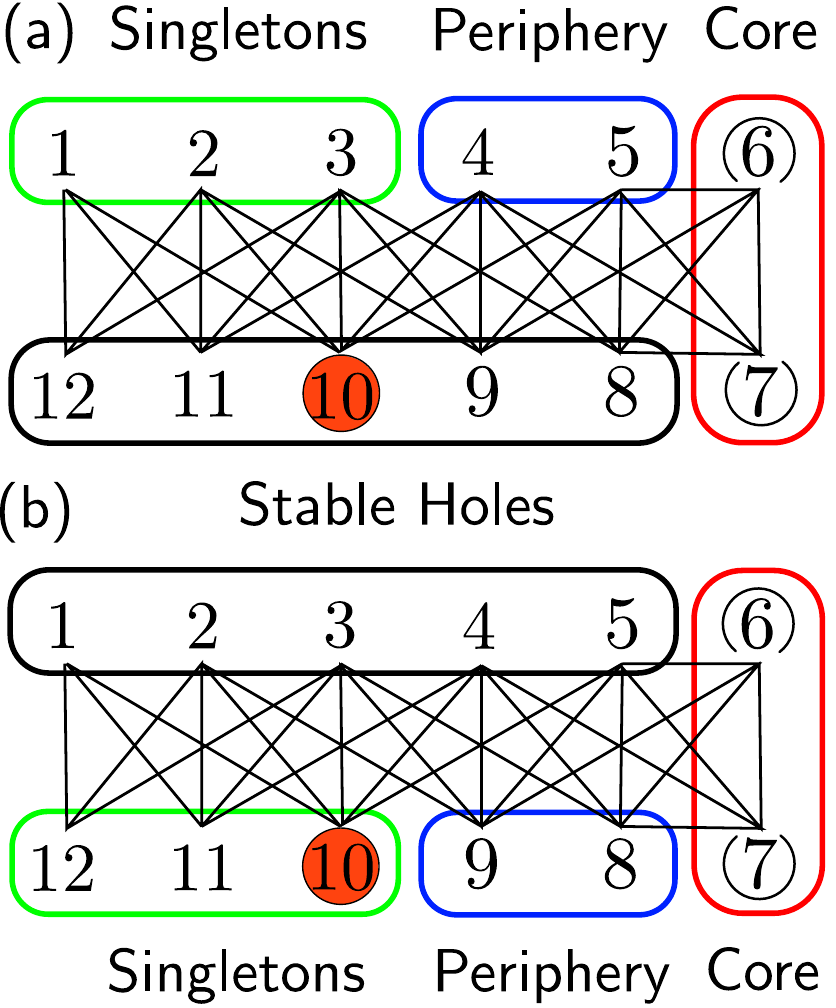}
  \caption{Rearrangement of the 12-group structure towards a self-tolerant state after permanently occupying the complete hole group $S_{10}$ with self (red circle). Starting from an established 12-group architecture, as already depicted in Fig.~\ref{Fig:GroupStructure}, the hole group $S_{10}$ is permanently occupied, mimicking self (a). This configuration is unstable and a new 12-group architecture evolves in which the self nodes belong to the singletons (b). This is immunologically favorable since the singletons are only linked to permanently empty hole groups such that the self can not be seen by other nodes. Adapted from Fig.~5 in \cite{SWB14}.
  }
  \label{caterpillar1}
  \end{figure}

  In Ref.~\cite{SWB14} several ways of introducing self into the network are discussed and realized in simulations and in mean-field theory. Here only the most instructive cases are reviewed.
  \\
  Self is mimicked by permanent occupation of one or several nodes. These nodes are not affected by idiotypic interactions but contribute to the number of occupied neighbors counted in the window rule. Two types of protocols were used: One, in which the self nodes are implemented in a fully developed 12-group architecture and another one in which the self nodes are inserted into an empty base graph and influence the evolution of the network from the beginning. 
  \\
  Inserting one self node in the hole group $S_{10}$ of an established 12-group architecture gives an immunologically unfavorable situation. A self node staying there would cause an autoimmune response, since the hole groups have many occupied neighbors. For high influx $p$ this node destabilizes the architecture and causes a rearrangement of the group structure. After this rearrangement a new 12-group structure evolves, in which the self node is found in a group which only has weakly occupied neighbors, as the singleton or periphery groups.\\
  If more than one node is occupied one observes a similar behavior. The reorganization takes place faster and the nodes are found in the singleton and periphery groups. This is also the case if all nodes of $S_{10}$ are permanently occupied as self nodes. In the steady state after the reorganization, all self nodes belong to singleton or periphery groups. Prevalently it occurs that in the new structure, the self nodes completely fill the singleton group $S_3$ which is of the same size as $S_{10}$. This situation is illustrated in Fig.~\ref{caterpillar1}. The complete occupation of group $S_{10}$ with self nodes (Fig.~\ref{caterpillar1}a) leads to a reorganization of the structure in which the former singleton and periphery groups become the new hole groups (Fig.~\ref{caterpillar1}b). The hole groups $S_8$ and $S_9$ become the new periphery, while the former hole groups $S_{10} \dots S_{12}$ form the new singleton groups. All groups of the same size have exchanged their phenomenological classification, such that the structure in Fig.~\ref{caterpillar1}b is a 'mirrored' version of the original architecture. This behavior is also observed in mean-field theory. \\
  Starting from a fully evolved 12-group architecture and permanently occupying one singleton or periphery group, this structure will persist for very long periods of time. 
  
  Simulations were also performed starting from an empty base graph while occupying several nodes permanently. Here the architecture emerges from the very beginning in a way such that the self nodes only have weakly occupied neighbors and therefore are tolerated. This corresponds to the suggestion in Ref.~\cite{leon1998natural} that autoreactive clones govern the growth process of the entire network during early ontogenesis and therefore determine the lymphocyte repertoire of the fully evolved system.
  \\
  In \cite{SWB14} it has been shown that the network evolves towards a self-tolerant state where the self-reactive clones are controlled by the network. These results support the perspective that idiotypic interactions may be essential in the control of autoreactive B-lymphocytes, cf. also the comment \cite{thomas2014complex} and references therein.
\section{Self-tolerance and spontaneous autoimmunity}
\label{Self-tolerance and autoimmunity}
\begin{figure}[h]
\includegraphics[width=1.1\linewidth]{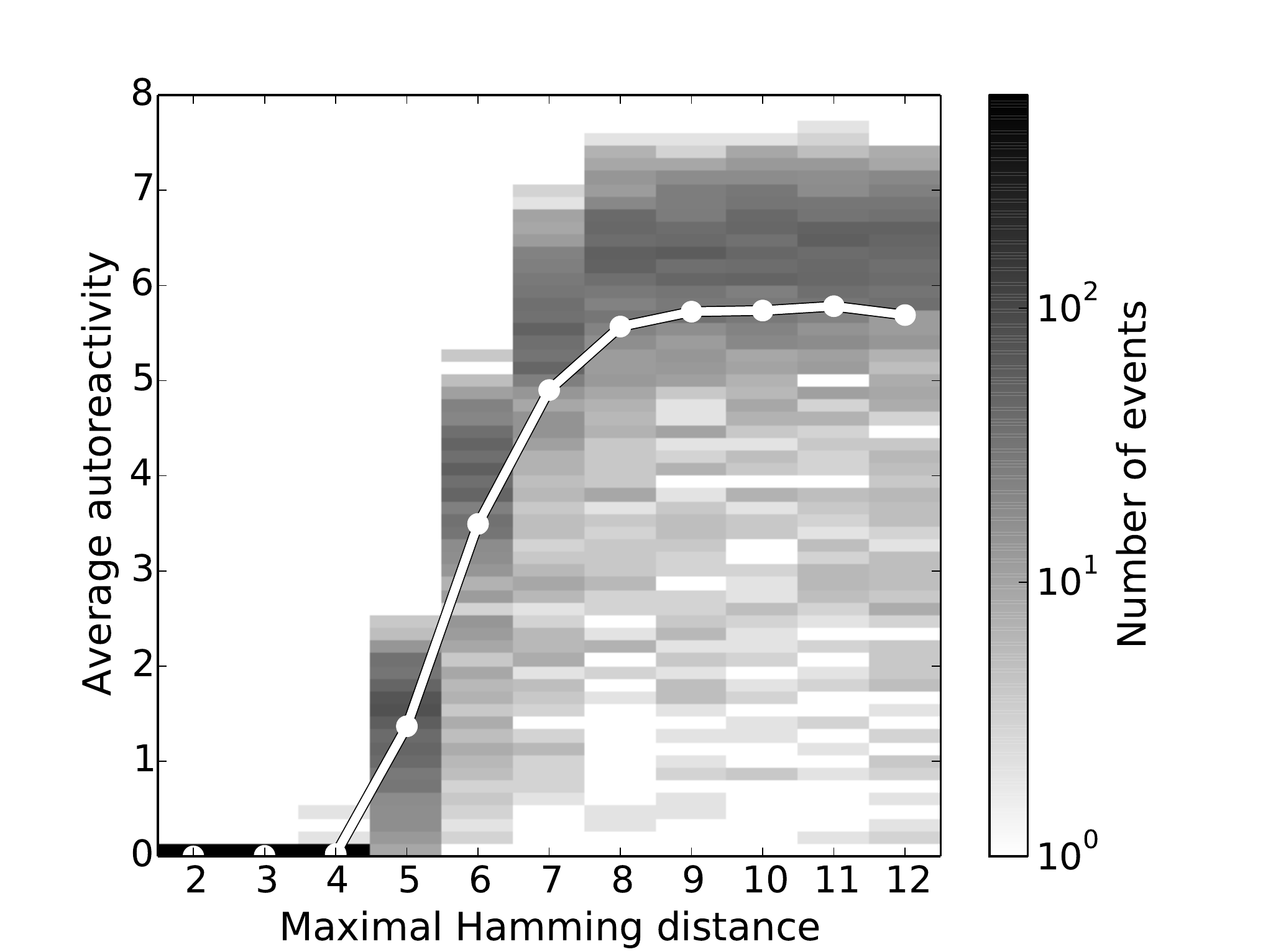}
\caption{Average autoreactivity for 50 randomly chosen self nodes with maximal Hamming distance $d_H^{\text{max}}$ to an arbitrary reference node. The self nodes are permanently occupied from the beginning of the simulations and strongly influence the emerging architecture. For small $d_H^{\text{max}}\leq 4$ the self nodes are only found in singleton and periphery groups and the average autoreactivity is negligible. For larger $d_H^{\text{max}}$ the average autoreactivity increases dramatically. The figure shows the height profile of histograms (gray scale) sitting on bins with fixed $d_H^{\text{max}}$ and of width $\Delta R_A=0.133$. For every $d_H^{\text{max}}$ 600 simulations were performed and the temporal average $\overline{R_A}$ over 30000 iterations (after an equilibration time of 2500) was taken. The dots are averages over all $\overline{R_A}$ for given $d_H^{\text{max}}$, the white line guides the eye.
}
\label{Autoreact_Histo}
\end{figure}

If several nodes of the network are occupied from the beginning of the simulations, playing the role of self, a self-tolerant architecture emerges, as it was shown in Ref.~\cite{SWB14}. Hereby the nodes were always chosen in a way that they fitted in an anticipated singleton group $S_3$. In this section we study whether self-tolerant structures also emerge when the self nodes are chosen at random and one does not anticipate an architecture. 
\\
In order to do so we first introduce a measure of autoreactivity. The autoreactivity per self node $R_A$ is defined as the average of the number of occupied neighbors over all self nodes
\begin{linenomath}
\begin{equation}
R_A=\frac{1}{|S_{\text{Self}}|}\sum\limits_{v_\text{Self}}^{}n(\partial v_{\text{Self}}).
\end{equation}
\end{linenomath}
If no self node has occupied neighbors, the self is not 'seen' by the network and therefore the autoreactivity $R_A$ is zero. On the contrary, if the self nodes have a lot of occupied neighbors, as it is e.g. the case for the hole group $S_{10}$, the self is 'seen' by the network and the autoreactivity $R_A$ is high. Since one node has $\kappa$ neighbors it holds that $R_A \in [0,\kappa]$. Defining $R_A$ as an intensive quantity has the advantage that its actual value does not depend on the rather arbitrary choice of the number of self nodes.
\\
As mentioned above, the system evolves towards a self-tolerant structure if several nodes of the anticipated group $S_{3}$ or the complete group itself are occupied from the beginning. An important question is, whether the network is able to build up a self-tolerant state if no final structure is anticipated and the self nodes are chosen in a random fashion. A reasonable restriction to this randomness is to assume that the self nodes have a similar idiotype, mimicking the resemblance of endogenous material. 
 \\
To shed light on this question, simulations were performed using the following protocol: In the beginning a random node of the network is chosen as the 'self-seed'. Then, fifty nodes are selected randomly, whereby they are not allowed to exceed a maximal Hamming distance $d_H^{\text{max}}$ to the self-seed. These nodes constitute the self and are occupied from the beginning of the simulations. We use different values of the maximal Hamming distance $d_H^\text{max}$ to examine varying similarity of the self nodes.
\\ 
Figure \ref{Autoreact_Histo} depicts histograms of the temporal average $\overline{R_A}$ of the autoreactivity found in stationary states for different values of $d_H^\text{max}$. 
One observes that for small allowed maximal Hamming distance $d_H^{\text{max}} \le 4$, i.e. high similarity of the self nodes, the autoreactivity is negligibly small. When large  Hamming distances $d_H^{\text{max}}>4$ are permitted, the autoreactivity increases dramatically.

In the first case, all self nodes are found in periphery and/or singleton groups, which are only linked to weakly occupied nodes and therefore cause weak autoreactivity. For $d_H^{\text{max}}=2$ it has even been observed that in approximately 15$\%$ of the simulations all self nodes end up in singleton groups only, corresponding to an autoreactivity of zero.
\\
In the second case the Hamming distance between the self nodes is too large for the network to build up an architecture in which they are distributed over the singleton and the periphery groups only. Thus, some of the self nodes are found in the core or even the hole groups. Nodes in these groups have a lot of occupied neighbors, causing high autoreactivity. The threshold of $d_H^{\text{max}}$ above which the self nodes do not fit into the singleton and periphery groups anymore is derived analytically in ~\ref{appendix_max_hamming_distance}.
\\
In summary, we have extended the results of Ref.~\cite{SWB14} for a more general setting: Usually the network evolves towards a self-tolerant architecture. This is, for example, the case when all self nodes are chosen to be in one hole or singleton group or if their Hamming distance is not allowed to exceed an appropriately small maximal value. Only if the idiotypes of the self differ too much, the system is not able to establish a structure in which all self nodes can be found in immunologically favorable groups. 
\\
\newline
In addition to the autoreactivity $R_A$ the Shannon entropy $H$ \cite{shannon2001mathematical} proved useful for examining perturbations of the network. Here, we use a form which neglects correlations between the nodes
\begin{linenomath}
\begin{equation}
H=-\sum\limits_{i=0}^{2^d}\left[n_i\cdot \text{ln}(n_i)+(1-n_i)\cdot\text{ln}(1-n_i) \right],
\end{equation}
\end{linenomath}
with $n_i$ being the average occupation of the node $v_i$. The terms of the sum are symmetrical around and maximal at $n_i=1/2$. If the nodes show a rather constant occupation the entropy is small, indicating static patterns. If the nodes show a rather variable occupation $H$ is large, indicating dynamical though stationary patterns.
\\
Extensive simulations revealed that a transition from one stationary state to another stationary state generically involves a lowering of $H$. Examples are transitions from one pattern to another one, which are caused by variations of the influx $p$, or a reordering of the group structure induced by permanently occupying a whole group $S_i$ with self, as discussed above in more detail. 
\\
The lowering of entropy is easy to understand in the situation depicted in Fig.~\ref{caterpillar1}. Here, the hole group $S_{10}$ is permanently empty in the original structure and its nodes give no contribution to $H$. Since permanently occupied nodes give no contribution to $H$, occupying $S_{10}$ with self would not change $H$ if no reordering occurred. Due to the rearrangement, all groups are interchanged with other groups of the same size (the structure is 'mirrored'), which would not alter $H$ if no self was present. For example, in the new structure, $S_{11}'$ gives the same contribution to $H$ as $S_2$ in the original one. This does not work for $S_3$ since $S_{10}'$ is permanently occupied. Therefore, $H$ is lowered by the amount of entropy constituted by $S_3$ in the original structure. 
\section{Self-tolerance and induced autoimmunity}
\label{Sec:Self-tolerance and induced autoimmunity}
Having examined the emergence of self-tolerant architectures, we proceed by studying reasons for its possible failure. Here, we want to investigate transitions from a healthy self-tolerant state to an autoimmune state. To induce such a transition, we tested several ways of perturbing the self-tolerant system.
\\
At first we will discuss transitions between different 12-group architectures in general. Then three biologically motivated types of perturbations and the corresponding transitions are presented and discussed: strong variations of the influx parameter $p$, the implementation of an antigen population, and the random removal of idiotypes.
 \subsection{Transitions in the 12-group architecture}
 \label{Sec:Transitions 12-group}
Starting from an established 12-group architecture a perturbation of the system can cause a transition to a new 12-group architecture. We denote the groups of the new structure by a prime: $S_i'$. It occurs that nodes of the group $S_i$ can be found in $S_j'$ with $i \ne j$ after a perturbation. This may lead to dramatic changes of the statistical properties of these clones. Suppose e.g. that the nodes have been in the weakly occupied core group $S_6$. Then the system is perturbed and they are found e.g. in the periphery group $S_{5}'$ afterwards. Since the periphery groups are densely occupied, the average occupation of the considered nodes increases dramatically.
 \\
 As it was explained in Sec.~I B, it is possible to determine the group membership of a node by comparing its bitstring with the entries of the determinant bit positions of nodes from $S_1$. If they differ in $n$ positions, the node is in the group $S_{n+1}$. Therefore, the architecture can be completely classified by the entries in the determinant bit positions of $S_1$. The determinant bits of nodes from $S_1$ can be read off from the center of mass vector.  
 \\
 To facilitate further considerations it proved useful to introduce a symbolic way of illustrating bitstrings according to the group membership of their corresponding nodes. Supposing a $d_M=11$ pattern we visualize the entries of determinant bit positions of a node in $S_1$ by empty circles and the non-determinant bit position by a square. Therefore, the bitstring of a node $v$ in $S_1$ looks like
 \vspace{2mm}
\begin{flushleft}
	\begin{tikzpicture}
  	\text{$\hphantom{ALA} \hphantom{LLL}$    }   \bcircle \bcircle \bcircle \bcircle \bcircle \bcircle \bcircle \bcircle \bcircle \bcircle \bcircle \squaree $\, \, \, \, .$
  	\end{tikzpicture}
\end{flushleft}
Since there are no distinguished determinant bit positions we can always order them as above.
The entries of the bit positions of another node which are complementary to those of $S_1$ are illustrated as filled circles. Therefore, the bitstring of a node $v \in S_{12}$  has the form
\vspace{2mm}
\begin{flushleft}
	\begin{tikzpicture}
  	\text{$\hphantom{ALA} \hphantom{LLL}$    }   \bfcircle \bfcircle \bfcircle \bfcircle \bfcircle \bfcircle \bfcircle \bfcircle \bfcircle \bfcircle \bfcircle \squaree $\, \, \, \, ,$
  	\end{tikzpicture}
\end{flushleft}
since it is complementary in all its entries of the determinant bits compared to bitstrings of nodes in $S_1$.
\\
If a $d_M=11$ pattern is perturbed and evolves towards another $d_M=11$ pattern one can classify the transition by the number $a$ of entries in determinant bit positions of nodes in $S_1$ which have changed their value. One also has to consider if the non-determinant bit position has changed or not. If it has changed, the actual position of the non-determinant bit in the new structure is not of importance for the reordering. 
\\
For the perturbations studied in this section, only transitions with a change of the non-determinant bit position occurred. Nevertheless, for certain settings one can also observe transitions where the non-determinant bit position does not change, e.g. for high influx $p \gtrsim 0.078$ or in case of the perfect mirroring depicted in Fig.~\ref{caterpillar1}. 
\\
Using the notation introduced above one can illustrate a transition by writing down the bitstrings of nodes from the group $S_1'$ in the new structure compared to the bitstrings of nodes from $S_1$ in the original structure. A reordering where the non-determinant bit position and the entries in $a$ determinant positions change is denoted as a $T_a$ transition. Below, an example for a $T_0$ transition is shown:
\vspace{2mm}
\begin{flushleft}
	\begin{tikzpicture}
  	\text{$\hphantom{A}v \in S_1': \hphantom{L}$    } \bcircle \bcircle \bcircle \bcircle \bcircle \bcircle \bcircle  \bcircle \squaree$\, \, $ $'$ \bcircle \bcircle \bcircle $.$
  	\end{tikzpicture}
\end{flushleft}
Here, the non-determinant bit (marked by a primed square) of nodes in the group $S_1'$ is in another position than for nodes in the original group $S_1$. Since the entries of the determinant bit positions do not change for a $T_0$ transition,  these bits are depicted by empty circles.
We now have a look at a $T_1$ transition, where also one entry of a determinant bit position changes:
\vspace{2mm}
\begin{flushleft}
	\begin{tikzpicture}
  	\text{$\hphantom{A}w \in S_1': \hphantom{L}$    }   \bcircle \bcircle \bcircle \bfcircle \bcircle \bcircle \bcircle \bcircle \bcircle \squaree $\, \, $ $'$  \bcircle \bcircle  $ .$
  	\end{tikzpicture}
\end{flushleft}
In this example the non-determinant bit position has been shifted and the entry in the fourth determinant bit position has changed its value, such that it is complementary to the corresponding entry of nodes in $S_1$ in the original structure.
\\
For a better understanding of how a perturbation may induce autoimmunity it is important to know how many nodes of a group $S_i$ can be found in a group $S_j'$ after a transition. For a $T_a$ transition this is given by the elements of the transition matrix $\mathbf{T}(a)$ as
\begin{linenomath}
\begin{equation}
	T_{ij}(a)\negmedspace  =\negthickspace    \negmedspace \sum\limits_{y,y'=0}^{1}  \negmedspace \binom{a}{\frac{a+i-j+y-y'}{2}} \binom{d_M-1-a}{i-1-y'-\frac{a+i-j+y-y'}{2}},
	\label{Transitionmatrix}
\end{equation} 
\end{linenomath}
where the binomial coefficients are zero if one of their entries is not a non-negative integer.
For a detailed derivation of the elements of $\mathbf{T}(a)$ confer ~\ref{Derivation of the transitionmatrix}.
\\
As an example we consider a $T_0$ transition. Then the number of nodes which e.g. are in $S_4$ in the original and can be found in $S_5'$ in the new structure is $T_{4,5}(0)=\binom{10}{3}=120$. 
\\
Figure \ref{TransitionMatrices} shows the transition matrices for a $T_0$ and a $T_1$ transition, respectively.
\begin{figure*}[t]
        \centering                           
   \begin{minipage}[ht]{\columnwidth}
   \centering
                               	$\mathbf{T}(0)$
                               	\\
\centering \hspace{-0.3cm}	\scalebox{0.95}{
	 
                            	\renewcommand{\arraystretch}{1.2}
                            \setlength{\tabcolsep}{1.5pt}
                           		 \begin{tabular}{|l|rrr|rr|rr|rrrrr|}
                           		
                                    		                        	         		\hline 
                                    		                        	         		& \multicolumn{1}{l}{$S_1$} & \multicolumn{1}{l}{$S_2$} & \multicolumn{1}{l|}{$S_3$} & \multicolumn{1}{l}{$S_4$} & \multicolumn{1}{l|}{$S_5$} & \multicolumn{1}{l}{$S_6$} & \multicolumn{1}{l|}{$S_7$} & \multicolumn{1}{l}{$S_8$} & \multicolumn{1}{l}{$S_9$} & \multicolumn{1}{l}{$S_{10}$} & \multicolumn{1}{l}{$S_{11}$} & \multicolumn{1}{l|}{$S_{12}$} \\ \hline
                                    		                        	         		$S_1'$ & 1 & 1 & &  &  & &  &  &  &  &  &  \\ 
                                    		                        	         		$S_2'$ & 1 & 11 & \textcolor{red}{10} &  &  &  &  &  &  &  &  &  \\ 
                                    		                        	         		$S_3'$ &  & 10 & \textcolor{red}{55} & 45 &  &  &  &  &  &  &  &  \\ \hline
                                    		                        	         		$S_4'$ &  &  & \textcolor{red}{45} & 165 & 120 &  &  &  &  &  &  &  \\ 
                                    		                        	         		$S_5'$ &  &  &  & 120 & 330 & 210 & &  &  &  &  &  \\ \hline
                                    		                        	         		$S_6'$ &  &  &  &  & 210 & 462 & 252 &  &  &  &  &  \\ 
                                    		                        	         		$S_7'$ &  &  &  &  &  & 252 & 462 & 210 &  &  &  &  \\ \hline
                                    		                        	         		$S_8'$ &  &  &  &  &  &  & 210 & 330 & 120 &  &  &  \\ 
                                    		                        	         		$S_9'$ &  &  &  &  &  &  &  & 120 & 165 & 45 &  &  \\ 
                                    		                        	         		$S_{10}'$ &  &  &  &  &  &  &  &  & 45 & 55 & 10 &  \\ 
                                    		                        	         		$S_{11}'$ &  &  &  &  &  &  &  &  &  & 10 & 11 & 1 \\ 
                                    		                        	         		$S_{12}'$ &  &  &  &  &  &  &  &  &  &  & 1 & 1 \\ \hline 
                                    		                        	         		   
                                    		                        	         	\end{tabular}
                                    		                        	         	}
                                    		                        	         	\hspace{0.5cm}
                            	\end{minipage}
                            	\hspace{0.65cm}
   \begin{minipage}[ht]{\columnwidth}
    \centering
     	$\mathbf{T}(1)$
                                      	\\	                        	         	
                           \centering           	\scalebox{0.95}{
                                      	
                                      	\renewcommand{\arraystretch}{1.2}
                                      \setlength{\tabcolsep}{1.5pt}
                                     	\begin{tabular}{|l|rrr|rr|rr|rrrrr|}
                                      		                      	      		\hline 
                                      		                        	      		&\multicolumn{1}{l}{$S_1$} & \multicolumn{1}{l}{$S_2$} & \multicolumn{1}{l|}{$S_3$} & \multicolumn{1}{l}{$S_4$} & \multicolumn{1}{l|}{$S_5$} & \multicolumn{1}{l}{$S_6$} & \multicolumn{1}{l|}{$S_7$} & \multicolumn{1}{l}{$S_8$} & \multicolumn{1}{l}{$S_9$} & \multicolumn{1}{l}{$S_{10}$} & \multicolumn{1}{l}{$S_{11}$} & \multicolumn{1}{l|}{$S_{12}$} \\ \hline
                                      		                        	      		$S_1'$ &  & 1 & \textcolor{red}{1} &  &  &  &  &  &  &  &  &  \\ 
                                      		                        	      		$S_2'$ & 1 & 2 & \textcolor{red}{10} & 9 &  & &  &  &  &  &  &  \\ 
                                      		                        	      		$S_3'$ & 1 & 10 & \textcolor{red}{18} & 45 & 36 &  &  &  &  &  &  & \\ \hline
                                      		                        	      		$S_4'$ &  & 9 & \textcolor{red}{45} & 72 & 120 & 84 &  &  &  &  &  &  \\ 
                                      		                        	      		$S_5'$ &  &  & \textcolor{red}{36} & 120 & 168 & 210 & 126 &  &  &  &  &  \\ \hline 
                                      		                        	      		$S_6'$ &  &  &  & 84 & 210 & 252 & 252 & 126 &  &  &  &  \\ 
                                      		                        	      		$S_7'$ &  &  &  &  & 126 & 252 & 252 & 210 & 84 &  &  &  \\ \hline
                                      		                        	      		$S_8'$ &  & &  & &  & 126 & 210 & 168 & 120 & 36 &  &  \\ 
                                      		                        	      		$S_9'$ &  & &  &  &  &  & 84 & 120 & 72 & 45 & 9 &  \\ 
                                      		                        	      		$S_{10}'$ &  &  &  &  &  &  &  & 36 & 45 & 18 & 10 & 1 \\ 
                                      		                        	      		$S_{11}'$ &  &  &  &  &  &  &  &  & 9 & 10 & 2 & 1 \\ 
                                      		                        	      		$S_{12}'$ &  &  &  &  &  &  &  &  &  & 1 & 1 &  \\  \hline 
                       	\end{tabular}
                  		}
   \end{minipage}

	\caption{Transition matrices for the $T_0$ and the $T_1$ transition. The groups $S_i$ represent  the original structure while the groups $S_i'$ stand for the structure after rearrangement. An entry at position $(i,j)$ indicates the number of nodes which are shifted from $S_i$ to $S_j'$. Nodes of the group $S_3$ represent the self and are colored red. Note that some of the self nodes are found in the periphery groups. Missing entries are zero.}
	\label{TransitionMatrices}
\end{figure*}       
They represent the two smallest possible transitions with a change of the non-determinant bit position, causing the nodes from a certain group in the original structure being the least scattered over other groups in the new structure. They are of most interest here. 
If the self fills up the complete singleton group $S_3$ these two transitions do not lead to self nodes ending up in core groups. For the perturbations which are discussed here, we only observed such transitions, which supports the thesis that structures with low autoreactivity are preferred. Nevertheless, the original state is completely self-tolerant, while the new state shows autoreactive behavior. This low autoreactivity is caused by the self nodes which are found in periphery groups after the perturbation.   	       
\subsection{Induced autoimmunity by variation of influx}
\label{Sec:Induced autoimmunity by variation of influx}
\begin{figure*}[ht]
                                                          	\begin{minipage}[ht]{\columnwidth}
                                                          		\centering
                                                          		\includegraphics[width=\textwidth]{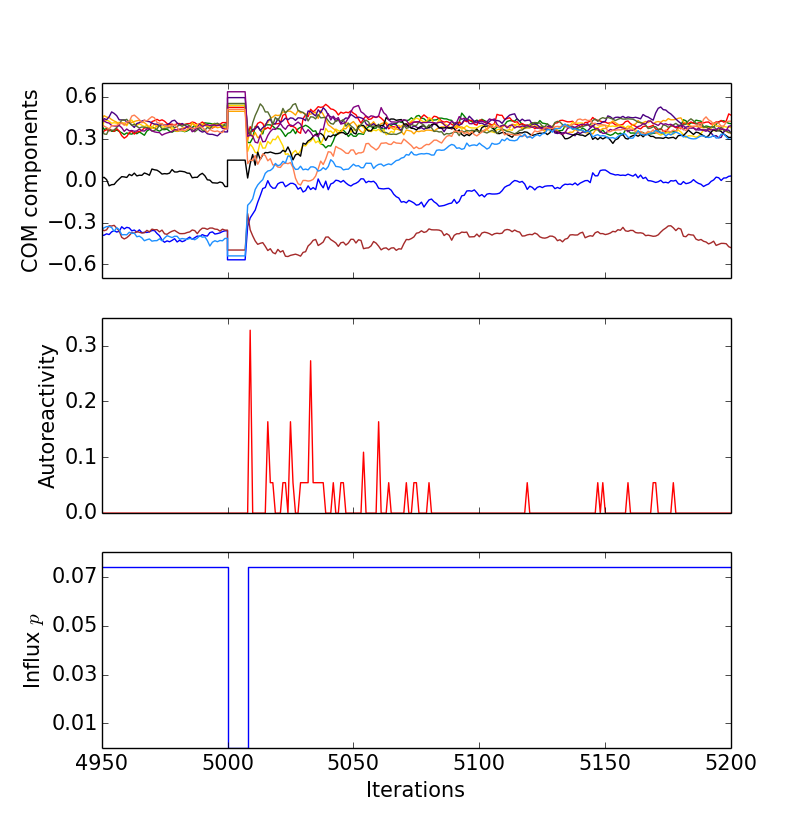}
                                                          			\vspace{-1.1cm}
																	\caption*{\hspace{6cm}\large(a)}
                                                          		\label{Fig:TurnOff_abrupt}
                                                          	\end{minipage}
                                                          	\hspace{1.5cm}
                                                          	\begin{minipage}[ht]{0.91\columnwidth}
                                                          		\centering
                                                          		\vspace{0.05cm}
                                                          		\includegraphics[width=\textwidth]{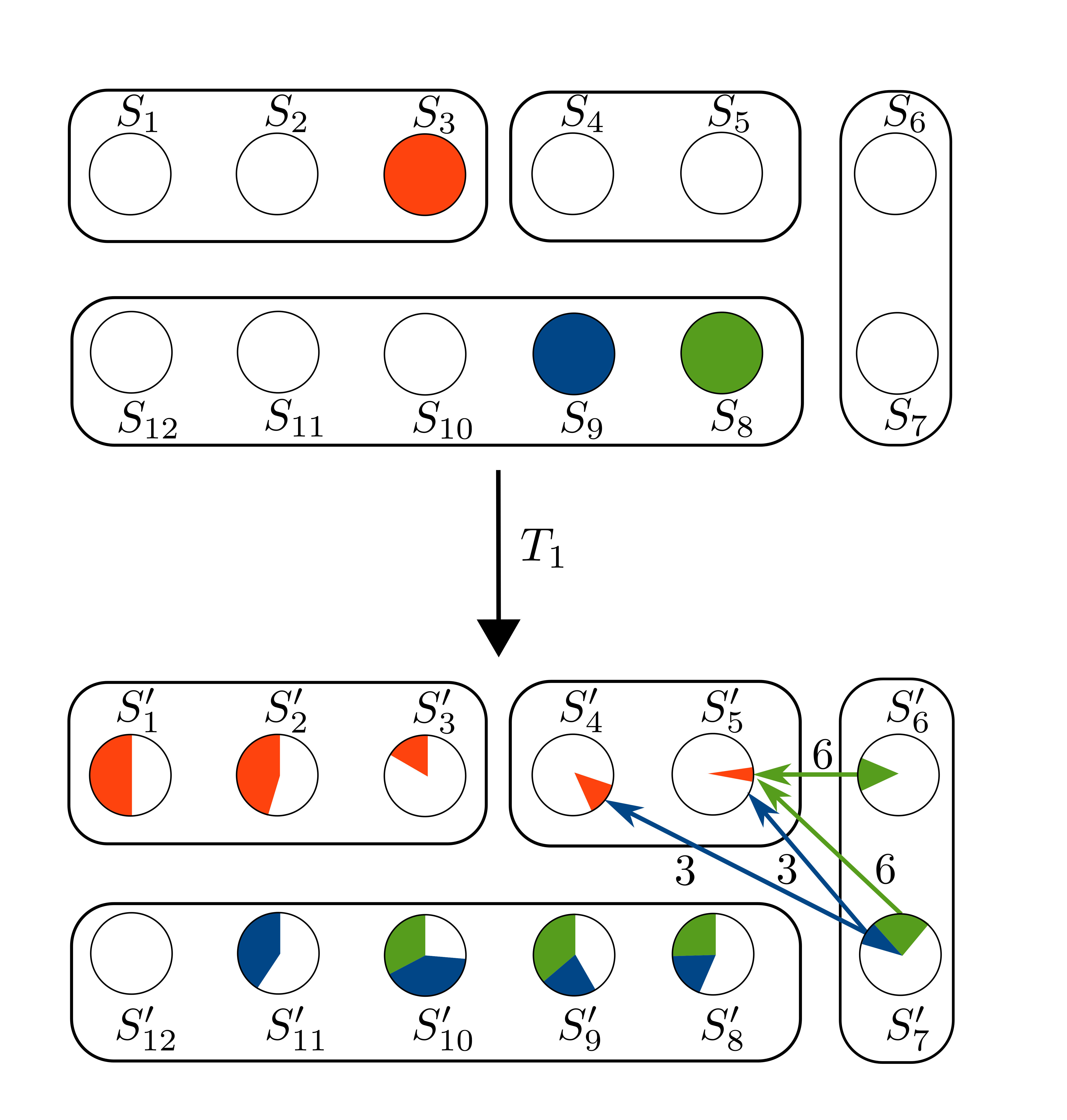}
     																	\vspace{-0.75cm}    				\caption*{\hspace{5cm}\large (b)}
     															\end{minipage}

     	\caption{Induction of autoimmunity due to strong variation of influx $p$. (a) Simulation results for a sudden stop of the influx $p$. At the 5000th iteration the influx $p$ is set to zero for 8 iterations, afterwards it is reset to the original value $p=0.074$. The COM vector indicates a change of the group structure corresponding to a $T_1$ transition. This transition leads to a quasistationary state in which self nodes have neighbors with non-vanishing average occupation such that autoreactive bursts occur as depicted in the time series of the autoreactivity. The self was placed in the complete group $S_3$.
     	(b)
     	Distribution of self (red) over the groups caused by a $T_1$ transition and linkage of potentially autoreactive clones to the self (arrows) in the new structure. After the transition parts of the hole groups $S_8$ (green) and $S_9$ (blue), which are linked to self in the periphery groups $S_4'$ and $S_5'$, are found in the new core groups $S_6'$ and $S_7'$. These are populated and thus cause autoreactivity.
 				}
     \label{Fig:T1_noantigen}
     \end{figure*}

\begin{figure}[t]
	     	\centering 
	     	\includegraphics[width=1\linewidth ]{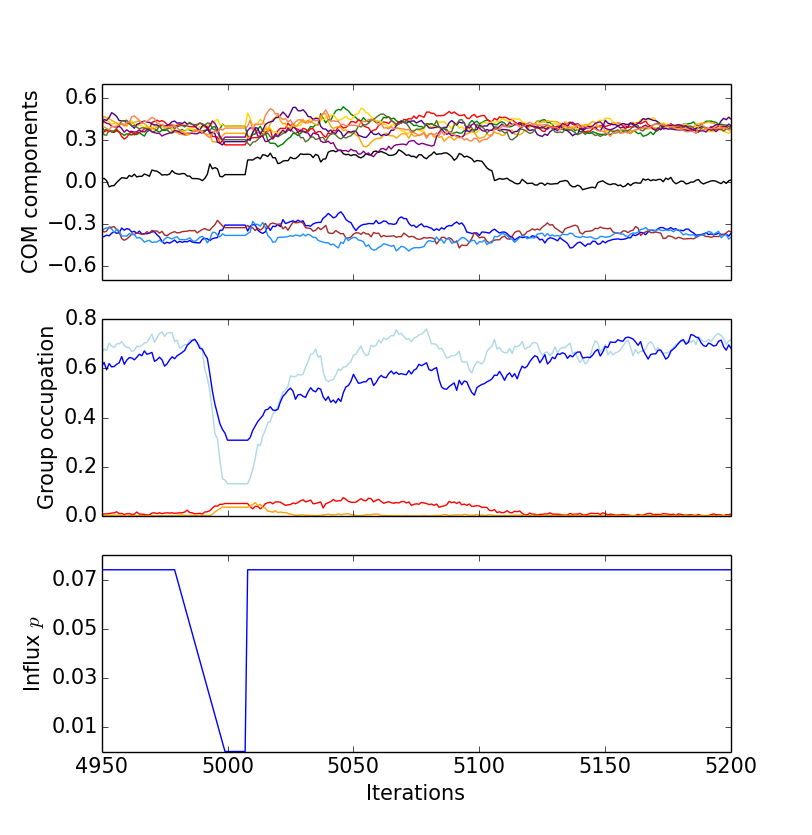}
	     	\caption{Maintenance of the self-tolerant state during a smooth decrease and a sudden increase of the influx (protocol II). The upper diagram shows the COM components, the diagram in the middle the occupation of the core groups $S_6$ (red) and $S_7$ (orange) and the periphery groups $S_4$ (light blue) and $S_5$ (dark blue), while the lower one shows the protocol of the influx $p$. 
	     	Starting from the 4980th iteration the influx $p$ is set to zero in 20 steps and it is waited for 8 iterations. Afterwards $p$ is reset to the original value $p=0.074$.  In contrast to Fig.~\ref{Fig:T1_noantigen}a we do not observe a change of the group structure and therefore no autoreactivity, $R_A=0$. A possible mechanism explaining this observation is the strong increase in the number of core nodes during the decrease of $p$, which stabilizes the original structure. The same seed as in Fig.~\ref{Fig:T1_noantigen}a was used.
	 		}
	     	\label{Fig:TurnOff_smooth}
	     \end{figure}

The influx of randomly generated idiotypes from the bone marrow, modeled by the parameter $p$, is not necessarily constant over the lifespan of an individual. Experimental studies showed, that the B lymphopoiesis may decrease with increasing age \cite{linton2004age}.
\\
For the examination of age-induced effects, we considered a decreasing influx rate $p$ and investigated if transitions from self-tolerant states to autoimmune states occur. Starting from an established 12-group structure with $p=0.074$ the influx was reduced by $10 \dots 30 \%$ in a timespan of 3000 iterations. Since the time represented by one iteration lies in the order of magnitude of 10 days this timespan roughly corresponds to the lifetime of a human. Throughout all simulations the system showed no autoreactive behavior.
\\
Now we follow a more radical approach and model a perturbation by simply stopping the influx $p$. This could model a radiation therapy in which the bone marrow of an individual is destroyed and therefore the influx of B-lymphocytes is paused \cite{henderson1969treatment}. The question is whether the idiotypic network attains its original, self-tolerant structure, when the influx is set back to its original value.
In \cite{SWB14} it already has been stated, that only the connected parts of the network survive if the influx is set to zero. The singletons  depopulate immediately, as they have no occupied neighbors and therefore are only sustained by the influx. The temporal evolution stops and the connected parts 'freeze in'. 
\\
Figure~\ref{Fig:T1_noantigen}a shows the center of mass components, the autoreactivity $R_A$ and the influx $p$ for such a protocol with the self completely filling $S_3$. Until the 5000th iteration, the influx is kept constant at $p=0.074$. One observes a 12-group structure, since only one component of the COM vector fluctuates around zero. The system has evolved to a self-tolerant state and the autoreactivity $R_A$ is zero. Then the influx is suddenly stopped for 8 iterations. Afterwards it is set back to its original value. Due to this perturbation, the group structure of the system changes, as one can conclude from the COM components. The black COM component, which fluctuated around zero prior to the perturbation is interchanged with the dark blue component, indicating a change of the non-determinant bit position of the bitstrings of nodes in $S_1$. The light blue component which fluctuated around a negative average value before the transition fluctuates around a positive average value afterwards which implies that one entry in a determinant bit position of the bitstrings of nodes in $S_1$ has changed. Therefore the system has performed a $T_1$ transition.       
     \\
This change of the group structure leads to a new steady state in which some of the self nodes have neighbors which are not permanently empty. Figure~\ref{Fig:T1_noantigen}b illustrates the rearrangement of the group structure induced by this transition. Prior to the perturbation the self (red) fills the periphery group $S_3$ completely. The perturbation causes a $T_1$ transition after which the self nodes are found in the groups $S_1' \dots S_5'$ as one concludes from the transition matrix $\mathbf{T}(1)$ (see Fig.~\ref{TransitionMatrices}). The distribution of the self over these groups is depicted in the lower part of Fig.~\ref{Fig:T1_noantigen}b.
\\
The self nodes in the periphery groups $S_4'$ and $S_5'$ are linked to nodes of the populated core groups $S_6'$ and $S_7'$  (see Fig.~\ref{Fig:GroupStructure}) and therefore cause autoreactive bursts, as observed in Fig.~\ref{Fig:T1_noantigen}a. One can see that the autoreactivity only attains quantized values.
   \\      
Before discussing this quantization a new notation is introduced: If a node $v$ is in the group $S_i$ prior to a transition and found in the group $S_j'$ afterwards, we write $v \in S_i \cap S_j'$. The self nodes which have been in $S_3$ and are found in $S_4'$ after the perturbation therefore make up the subgroup $S_3 \cap S_4'$. This subgroup is illustrated as the red part of $S_4'$ in Fig.~\ref{Fig:T1_noantigen}b. The nodes which have the potential to 'see' self constitute the subgroups $S_8 \cap S_7'$,  $S_8 \cap S_6'$, and  $S_9 \cap S_7'$. Every node from $S_8$ and $S_9$ has six links to nodes from $S_3$ as can be concluded from the entries of the link matrix in Fig.~(\ref{Fig:Linkmatrix}). Furthermore, the neighbors of a node remain neighbors. Therefore, the nodes in $S_8 \cap S_7'$,  $S_8 \cap S_6'$, and  $S_9 \cap S_7'$  are still linked to six self nodes each. 
\\
If a node of one of these subgroups is occupied, it counts as active neighbor for six self nodes of $S_3 \cap S_4'$ or $S_3 \cap S_5'$. This corresponds to an autoreactivity of $R_A=\frac{1}{|S_{\text{Self}}|}\sum_{v_\text{Self}}n(\partial v_{\text{Self}})=\frac{1}{110}\cdot 6 \approx 0.055$, which is the value of quantization observed in Fig.~\ref{Fig:T1_noantigen}a. Since the core groups $S_6'$ and $S_7'$ are occupied very weakly, it is unlikely that several potentially autoreactive nodes are occupied at the same time, such that one does not observe multiples of this value of quantization in Fig.~\ref{Fig:T1_noantigen}a in the new steady state. A similar discussion explains the quantization in case of a $T_0$ transition which will occur in the next subsection.
\\
We also investigated if transitions occur as well when other protocols are applied for setting the influx down to zero:
\begin{center}

\begin{enumerate}[(I)]
\itemsep-0.1em 
\item A smooth decrease and a smooth increase of $p$.
\item A smooth decrease and an abrupt increase of $p$.
\item An abrupt decrease and a smooth increase of $p$.
\end{enumerate}
\end{center}
Surprisingly, the first two protocols did not cause a change of the group structure. Figure \ref{Fig:TurnOff_smooth} exemplarily shows simulation results for the second protocol.   
Here the influx is decreased in 20 steps, which causes a growth of the core groups. Therefore larger and densely linked parts of the network survive. This stabilizes the original structure when the influx is reset to its original value. One can conclude that smoothly reducing the influx helps to recover the original self-tolerant state when $p$ is set back to its initial value.
\\
In the next subsection it is discussed how perturbations can be modeled by implementation of an antigen population.        
           
\subsection{Antigen induced autoimmunity}
\label{Antigen induced autoimmunity}
It is well known that certain autoimmune diseases may be triggered by infections \cite{Bach2002Infections,Rose1998rolfeofinfections,Bach2005Infectionsandautoimmdiseases,beyerlein2016infections}. Famous examples are rheumatic fever, the Guillain-Barr\'e syndrome, type 1 diabetes and multiple sclerosis \cite{davidson2001autoimmune,Wucherpfennig2001autoimmunitybyinfection,cole1993triggering,Rees1995GuillainBarre}. In this subsection it will be studied in the frame of our model whether an infection can lead to a transition from a self-tolerant to an autoimmune state.
\subsubsection{Modeling infections}
\begin{figure}[!h]
     	\centering 
     	\includegraphics[width=\linewidth ]{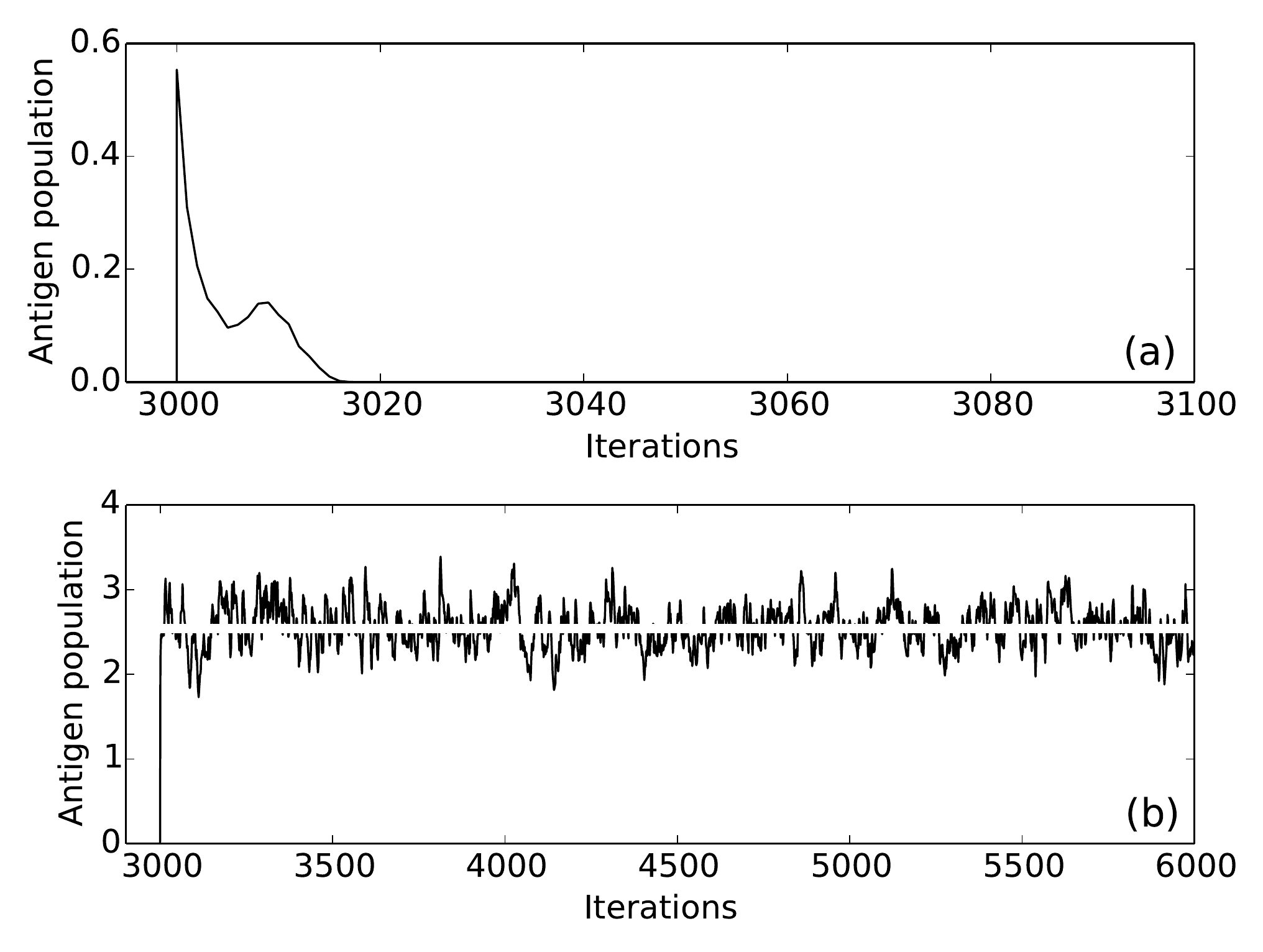}
     	\caption{
     Time course of an infection with antigen inserted at iteration 3000 into the hole group $S_{12}$ for different values of the efficacy parameter $l=0.085$ (a) and $l=0.05$ (b), $A_{\text{start}}=0.55$, $A_{\text{max}}=4$, $p=0.074$. (a) shows an acute infection which is defeated after less than 20 iterations. (b) depicts the case of a chronic infection where the antigen fluctuates around the value of the nontrivial stable fixed point of Eq.~(\ref{Iteration}) given by $A^*=A_{\text{max}}-l \avg{N}$ (indicated by the white line) where $\avg{N}=49.2$ is the average number of neighbors of the antigen determined from simulation. The same seed was used in both cases, no self was inserted.
     }
     	\label{Fig:Cases}
     \end{figure}

For the implementation of the infection by a foreign antigen one node is chosen to represent its idiotype. The antigen population $A$ can proliferate and attains continuous values from zero to $A_{\text{max}}$. If a node is linked to the antigen idiotype, the antigen population $A$ counts as additional neighbors for the window rule. 
\\
For modeling the population dynamics of the antigen we start with a version of the logistic map with an additional suppressing term
\begin{linenomath}
\begin{equation}
A_{k+1}=A_k+A_k\cdot(A_{\text{max}}-A_k)-lN_kA_k.
\label{naive}
\end{equation}
\end{linenomath}
Here, $A_k$ is the antigen population at the $k$th iteration and $N_k$ the number of occupied neighbors which the antigen idiotype possesses before application of the window rule. The parameter $l$ describes the efficacy of the immune response: the larger $l$, the stronger the suppression of the antigen population due to neighbored occupied nodes.  
\\
Equation (\ref{naive}) can be understood as a naive discretization of the logistic differential equation. It shows oscillating and chaotic behavior not found in analytical solutions of the logistic differential equation. Therefore we use instead a non-local discretization where the quadratic term $A_k^2$ is replaced by $A_k\cdot A_{k+1}$ \cite{mickens1994nonstandard}. This gives the  difference scheme
\begin{linenomath}
\begin{equation}
A_{k+1}=\frac{1+A_{\text{max}}-lN_k}{1+A_k}\cdot A_k.
\label{Iteration}
\end{equation}
\end{linenomath}
In case of an autonomous antigen population, i.e. ${N_k=0}$, and a positive starting value $A_0>0$ the population approaches its maximum value $A_\text{max}$ when iterating Eq.~(\ref{Iteration}). The presence of occupied neighbors effectively reduces $A_\text{max}$ to $A_\text{max}^\text{eff}=A_\text{max}-lN$.
When assuming that the number of occupied neighbors attains an approximately constant value of $N_k \approx \avg{N}$ after an initial time, Eq.~(\ref{Iteration}) has two fixed points. A trivial one which is $A^*=0$ and a nontrivial one given by
\begin{linenomath}
\begin{equation}
 A^*=A_{\text{max}}-l\avg{N}.
 \label{FixedPoint}
\end{equation}
\end{linenomath}
The most interesting behavior is found if the antigen population is placed in the hole groups. Here it has a lot of occupied neighbors and a strong interaction between the network and the antigen takes place. Simulations showed that inserting antigen into the groups $S_{11}$ and $S_{12}$ generically does not cause rearrangements while inserting into the groups $S_{7} \dots S_{10}$ does. 
\\
In general, one observes three courses of infection in dependence of $A_{\text{max}}$, $l$, and the average number of neighbors of the antigen $\avg{N}_0$ for vanishing antigen population. Which course typically occurs for a certain choice of parameters can be deduced from the numerator in Eq.~(\ref{Iteration}).
For $1+A_{\text{max}}  \ll l\avg{N}_0$ a sub-clinical infection usually appears which is defeated after one or two iterations. An acute infection occurs if $1+A_{\text{max}} \approx l\avg{N}_0$ and a chronic infection for $1+A_{\text{max}} \gg l\avg{N}_0$. \\
Figure \ref{Fig:Cases} shows two typical courses of infection for the insertion of antigen in the group $S_{12}$ without self. In Fig.~\ref{Fig:Cases}a one observes an acute infection which only lasts for approximately 15 iterations, while Fig.~\ref{Fig:Cases}b shows a chronic infection where $A$ fluctuates around the non-trivial fixed point given by Eq.~(\ref{FixedPoint}) with ${\avg{N} \approx 49.2}$.

\begin{figure*}[t]
       	\centering
       	\includegraphics[width=\linewidth]{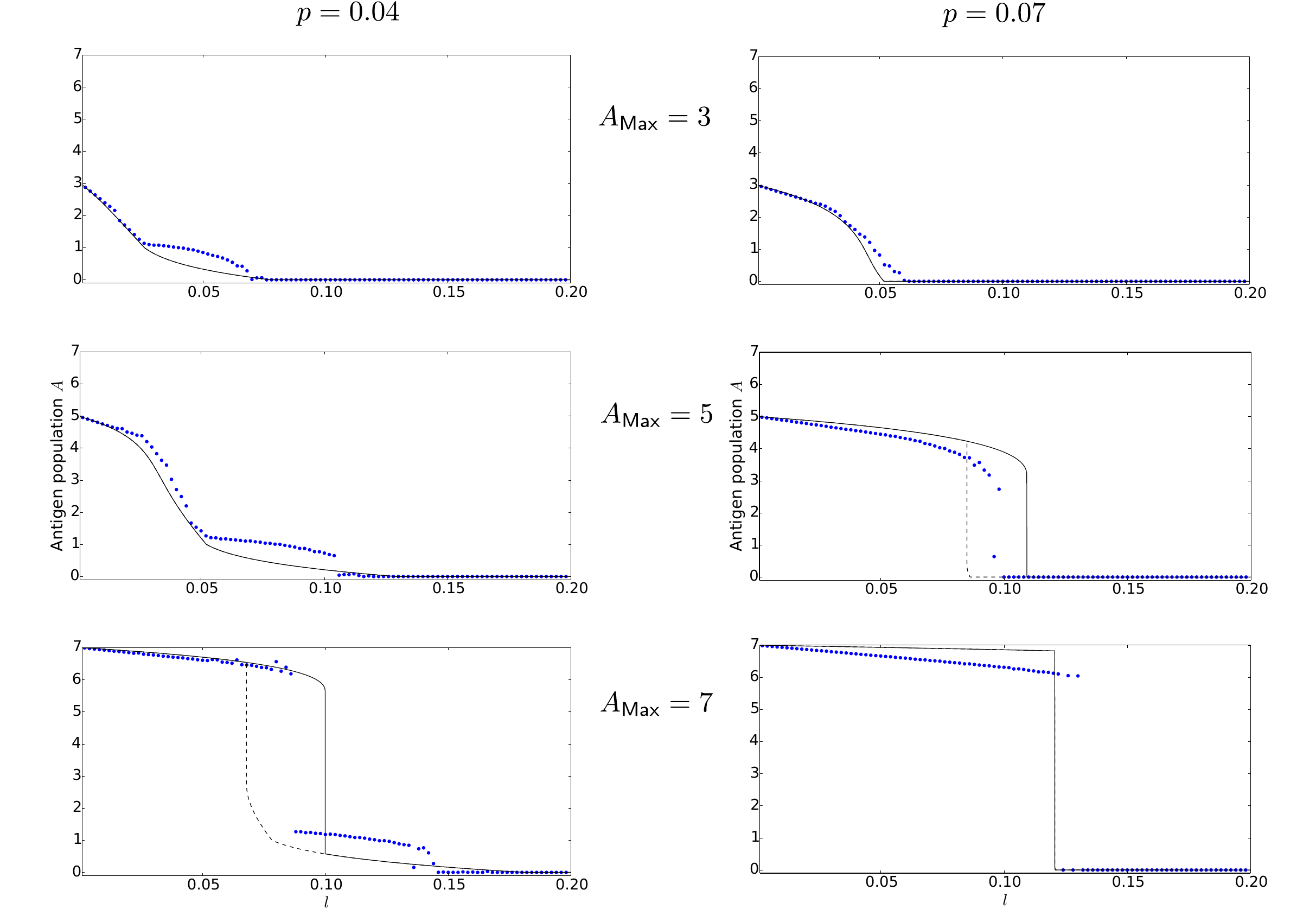}	
        \caption{  
        		Steady state antigen population in the hole group $S_{12}$ as function of the efficacy $l$ of the immune response for different values of $p$ and $A_{\text{max}}$ without self. The dots are the averages of the antigen population obtained from 5000 iterations after an equilibration period of 1110 iterations and therefore only show steady state phenomena (chronic infection or healthy state). In the first simulation $A_\text{start}=1$ was chosen, then the result of the previous simulation served as initial value. A different seed was used for every simulation and it was always checked that a 12-group structure had emerged. The solid (dashed) lines depict the mean-field results obtained from Eq.~(\ref{Iteration}) and Eqs.~(\ref{MFA_Singleton},\ref{FinalIteration}) for increasing (decreasing) values of $l$, where the result of the previous $l$ was used as initial value for the next choice of $l$. Only in case of hysteresis the dashed line is visible, i.e. for $(p,A_\text{max})=(0.7,5)$ and $(0.04,7)$. In simulations one finds bistability in the regions where the mean-field results show hysteresis. Since one state is always attained much more preferentially than the other the bistability can not be seen in the average values. For the insertion of antigen into the hole group $S_{12}$ we did not observe rearrangements of the group structure.
         		}
        	\label{Antigen_MFT_1}
        \end{figure*}

Before starting the discussion of transitions caused by antigen in $S_7 \dots S_{10}$ it seems promising to get a better understanding of the behavior of antigen populations in $S_{12}$ without self. For doing so, we calculate the average antigen population depending on $l$, $A_{\text{max}}$ and $p$ in a mean-field approach.
\\
An idiotype in the group $S_{12}$ only has neighbors in the singleton groups $S_1 \dots S_{3}$ as can be concluded from the link matrix Eq.~(\ref{Linkmatrix}), see also Fig.~\ref{Fig:GroupStructure}. The singletons do not possess any occupied neighbors after application of the window rule. Therefore, the probability of a singleton to survive the window rule is determined by the influx alone as
\begin{linenomath}
\begin{align}
P_{\text{Sing}}^\mathbb{W}=\sum\limits_{n=t_L}^{t_U}\binom{\kappa}{n}p^n(1-p)^{\kappa-n}.
\label{Singleton_Prob1}
\end{align}
\end{linenomath}
If an antigen population is neighbor of the singletons the situation changes. Then the probability of a singleton to survive is also influenced by the antigen population $A$, changing the limits of the sum in Eq.~(\ref{Singleton_Prob1}). Since $A$ is a continuous variable we rewrite the cumulative binomial distribution $B_C$ with help of the regularized incomplete Beta function $I_x(a,b)=B(x;a,b)/B(a,b)$ where $B(x;a,b)$ is the incomplete Beta function and $B(a,b)$ the Beta function \cite{olver2010nist} as
\begin{linenomath} 
\begin{align}
	B_C(\kappa,p,t_U)=& \sum\limits_{n=0}^{ t_U }\binom{\kappa}{n}p^n(1-p)^{\kappa-n}\nonumber
	 \\ =&\text{I}_{1-p}(\kappa-t_U,t_U+1).
	\label{Kumulative Binomialverteilung}
\end{align}
\end{linenomath}
The sum in Eq.~(\ref{Singleton_Prob1}) can be rewritten as $B_C(\kappa,p,t_U)-B_C(\kappa,p,t_L-1)$. Furthermore, the presence of an antigen population effectively shifts the limits of the sum to ${t_U-A}$ and $t_L-A$.
Thus, the probability for the survival of a singleton neighbored to the antigen for $A<t_L$ is
\begin{linenomath}
\begin{align}
	  P_{\text{Sing}}^\mathbb{W}(A) \nonumber =& \text{Prob}(t_L-A \leq n(\partial v_{\text{Sing}}) \leq t_U-A)\nonumber \\
	=&\text{I}_{1-p}(\kappa-t_U+A,t_U+1-A) \nonumber  \\ & -\text{I}_{1-p}(\kappa-t_L+1+A,t_L-A),
\end{align}
\end{linenomath}
and for $A \ge t_L$
\begin{linenomath}
\begin{align}
	  P_{\text{Sing}}^\mathbb{W}(A)  =\text{I}_{1-p}(\kappa-t_U+A,t_U+1-A),
\end{align}
\end{linenomath}
which is continuous in $A$.\\
One obtains the average occupation of the singletons by multiplying the average occupation after the influx $n^\text{Sing}_{k-1}+p(1-n^\text{Sing}_{k-1})$ with the probability to survive the window rule $P_{\text{Sing}}^\mathbb{W}(A_{k})$ which gives
\begin{linenomath}
\begin{align}
n^\text{Sing}_{k}(A_k)=\left[n^\text{Sing}_{k-1}+p(1-n^\text{Sing}_{k-1})\right]P_{\text{Sing}}^\mathbb{W}(A_{k}).
\label{MFA_Singleton}
\end{align}
\end{linenomath}
One gets the iteration rule for $A$ by replacing the number of occupied neighbors in (\ref{Iteration}) with
\begin{linenomath}
\begin{equation}
N_{k}=\kappa \cdot n^\text{Sing}_{k}(A_k).
	\label{FinalIteration}
\end{equation}
\end{linenomath}
Figure~\ref{Antigen_MFT_1} compares the average steady state antigen population obtained in simulations with solutions of the mean-field equation in dependence of the efficacy $l$ for different values of $A_\text{max}$ and $p$. The average is taken over 5000 iterations after an equilibration time of 1100 iterations. Doing so only steady state phenomena (chronic infections or healthy states) are described, whereas transient phenomena (acute infections) are not visible. For the first simulation a starting value $A_\text{start}=1$ was chosen, afterwards the result of the previous simulation was used as starting value. The solid (dashed) lines depict the mean-field results obtained for increasing (decreasing) $l$. For low and high values of $l$ one observes a good agreement between simulation and mean-field results. For intermediate ranges the simulation results are only qualitatively reproduced by the mean-field solutions. 
\\
In all cases it can be seen that for low $l$ the antigen population attains a high average value, while it vanishes completely for high $l$. For a higher $A_\text{max}$ the network needs a higher efficacy $l$ to overcome the antigen. Furthermore, Fig.~\ref{Antigen_MFT_1} reveals that the transition from the case of a chronic infection with non-vanishing antigen population to a healthy steady state becomes sharper with increasing $A_{\text{max}}$. Interestingly, the mean-field solutions show hysteresis for certain choices of the parameters. In the regions where the mean-field approximation shows hysteresis also the simulations reveal bistable behavior. However, this bistability has no influence on the average results for the antigen population since one of the states is always established much more preferentially than the other and a switching towards the less preferred state occurs only very seldom.
\\
For some parameter settings, e.g. $(p,A)=(0.04,5)$, the average value of $A$ obtained from simulations falls to zero abruptly above a certain value of $l$. This is due to the fluctuations of $A$ which are already large enough to reach zero then. Since the mean-field solutions do not fluctuate they do not show such a behavior.
        
        \FloatBarrier
\subsubsection{Antigen induced transitions}
\begin{figure}[h!]
                          	\begin{minipage}[ht]{\linewidth}
                          		\centering
                          		\includegraphics[width=\textwidth]{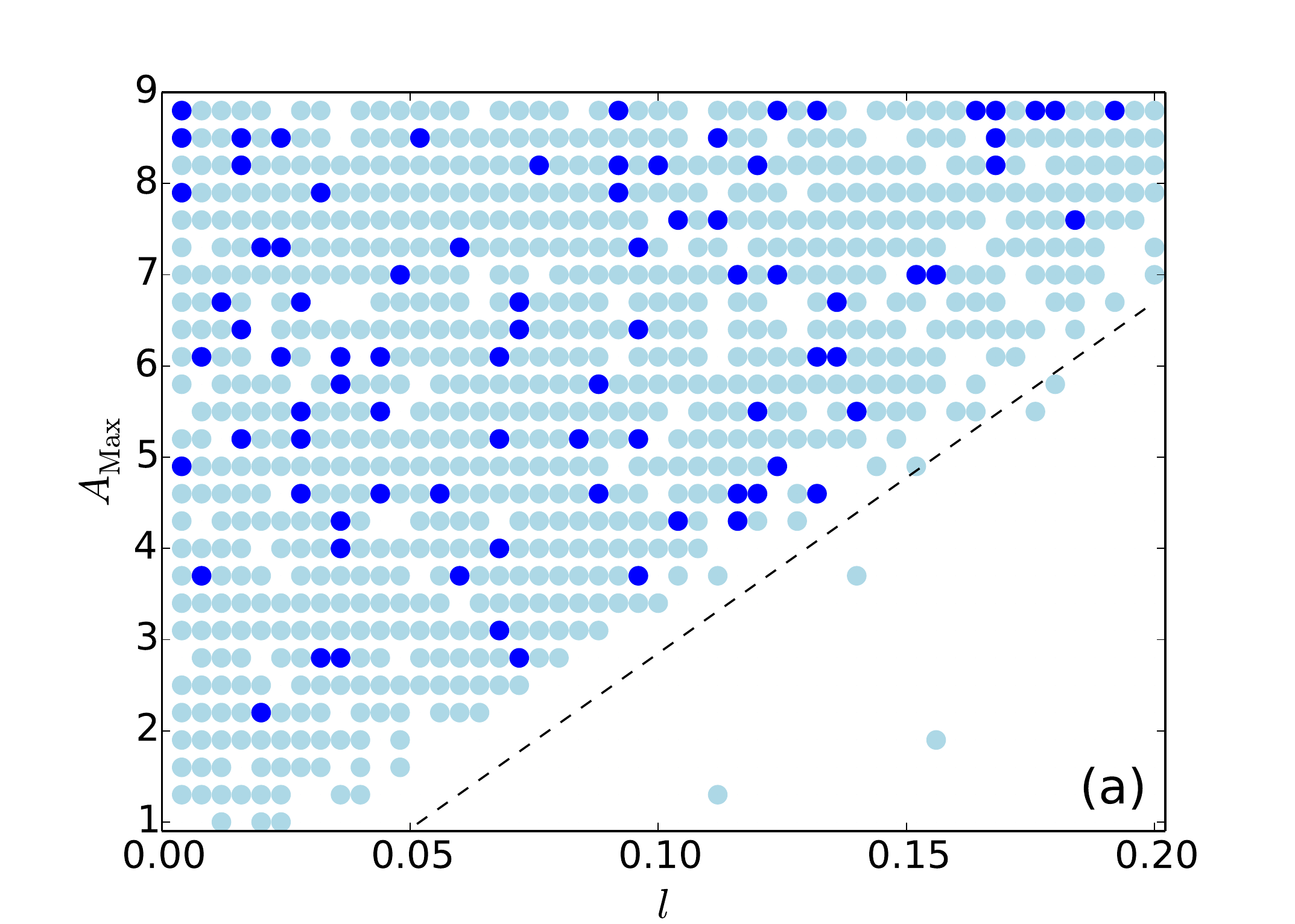}
                          		\label{S10}
                          	\end{minipage}
                          	\begin{minipage}[ht]{\linewidth}
                          		\centering
        	                  		\includegraphics[width=\textwidth]{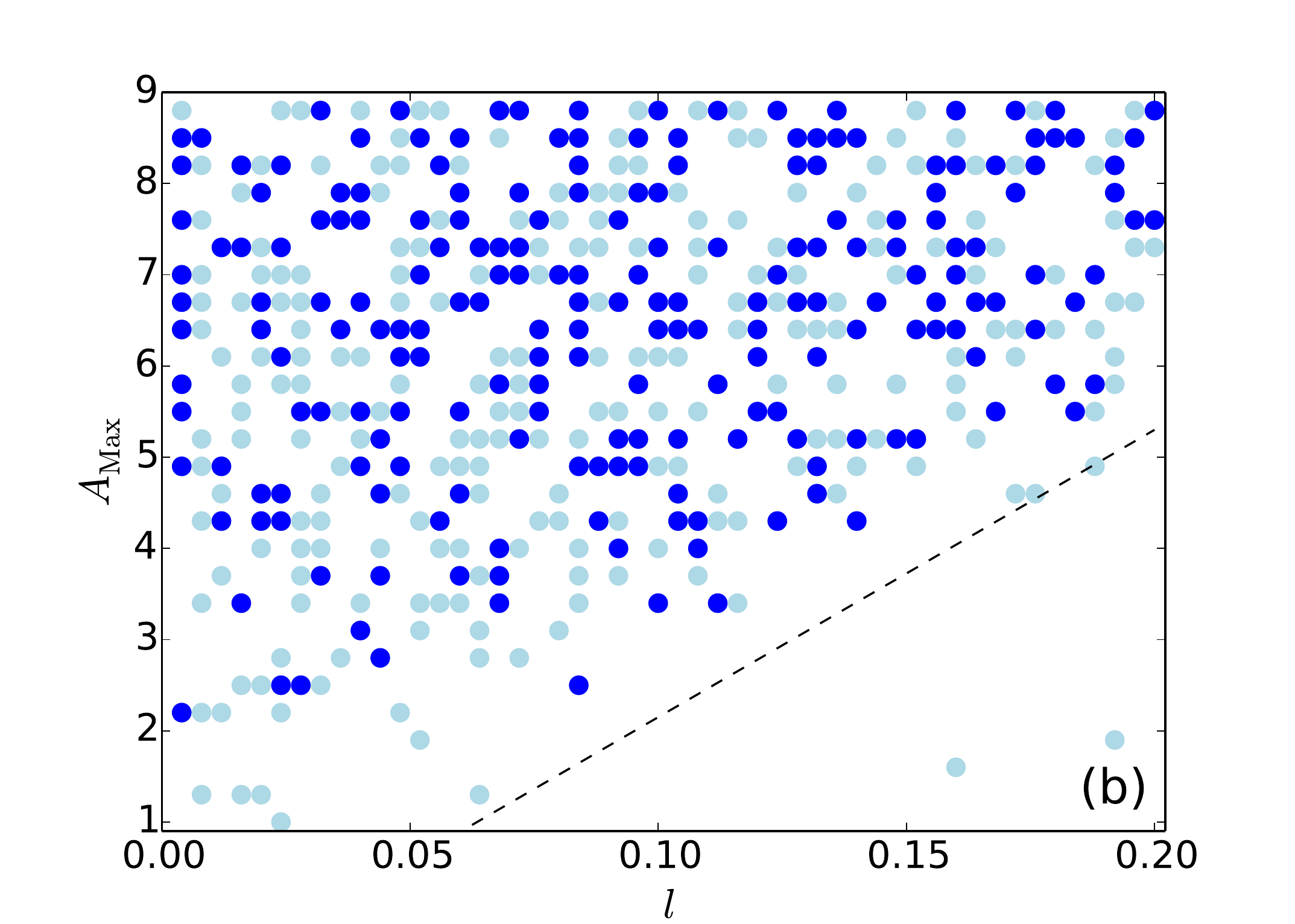}
                          		\label{S9}
              	\end{minipage}
              	\caption{
              	Antigen induced transitions of the 12-group architecture with self placed in the singleton group $S_3$. The figures show the occurrence and the type of transitions with the antigen inserted into the hole groups $S_{10}$ (a) and $S_9$ (b) for different values of $l$ and $A_{\text{max}}$. Every dot indicates an observed transition in one simulation of 30000 iterations, light blue dots encode a $T_0$ transition, blue dots a $T_1$ transition (see text), missing dots mean that no transition was observed. For every simulation a different seed was used. The dashed lines depict $A_\text{max}=l \avg{N}_0-1$, where $\avg{N}_0$ is the average number of occupied nodes which are neighbors of the antigen idiotype for vanishing antigen population. This lines divide the parameter plane in a region where almost no transitions occur and a region where transitions are frequent.
              	}
               	\label{Transitions}
                \end{figure}

Having discussed the behavior of antigen populations in the hole group $S_{12}$ without self, we now examine the behavior in the presence of self. We place self into the singleton group $S_3$ and insert antigen into the hole groups $S_9$ or $S_{10}$, what can induce autoimmunity. At first it is studied which type of transitions occur, thereafter a detailed discussion of how they cause autoimmunity is given. 
\\
Figure \ref{Transitions} shows for which choices of parameters $A_{\text{max}}$ and $l$ transitions occur and which kind of transitions the systems performs. Here, every dot shows the result of one simulation. Light-blue dots indicate that a $T_0$ transition and blue dots that a $T_1$ transition has taken place, while no transitions occurred in white regions. As already mentioned in Subsec.~\ref{Sec:Transitions 12-group}, one only observes $T_0$ and $T_1$ transitions. Rearrangements which distribute the self farther would result in self nodes ending up in the core groups, which would have very negative consequences in the immunological context.
\\
Figure~\ref{Transitions} reveals that the parameter plane spanned by $l$ and $A_\text{max}$ is divided into a region where transitions occur often and a region where nearly no transitions occur. The border separating this regions is given by the root of the numerator in Eq.~(\ref{Iteration}) $A_{\mathrm{max}}= l\avg{N}_0-1$ (depicted by the dashed lines) with $\avg{N}_0$ being the average number of occupied neighbors of the antigen idiotype for vanishing antigen population.
 In case of insertion of antigen into $S_{10}$ (Fig.~\ref{Transitions}a) the number of $T_0$ transitions (light-blue dots) is much higher than the number of $T_1$ transitions (blue dots). In case of insertion into $S_{9}$ (Fig.~\ref{Transitions}b) the total number of transition is lower than for insertion into $S_{10}$, while $T_1$ transitions (blue dots) occur more frequently.
 
\begin{figure*}[t]
                                                      	\begin{minipage}[ht]{\columnwidth}
                                                      		\centering
                                                      		\includegraphics[width=\textwidth]{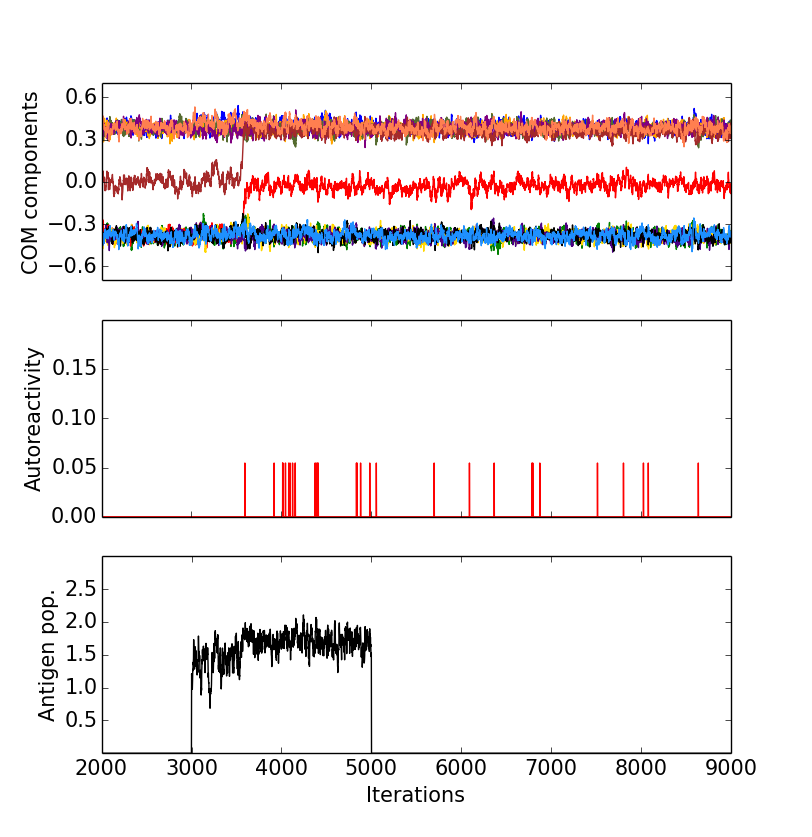}
                                                      					\vspace{-1.1cm}
         							\caption*{\hspace{6cm}\large(a)}
                                                      		\label{Antigen_in_S10}
                                                      	\end{minipage}
                                                      	\hspace{1.5cm}
                                                      	\begin{minipage}[ht]{0.81\columnwidth}
                                                      		\centering
                                                      			\vspace{0cm}
           	\includegraphics[width=\textwidth]{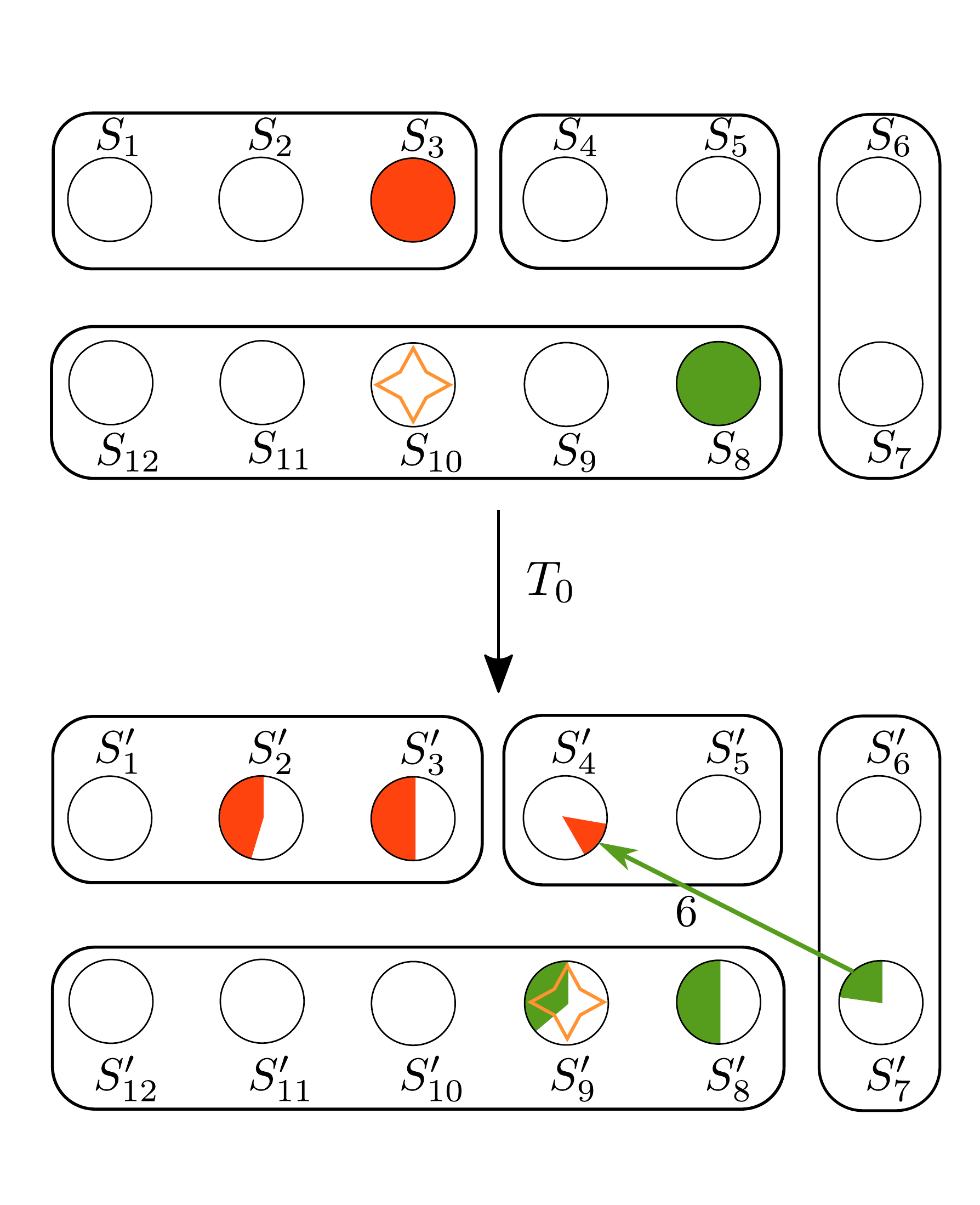}
        	\vspace{-0.9cm}    	
         	\caption*{\hspace{5cm}\large (b)}
                                              	\end{minipage}
                                                \caption{Antigen induced autoimmunity. Starting from a self-tolerant state antigen is inserted into the hole group $S_{10}$ causing a transition to an autoimmune state. Even when the antigen population is set to zero, modeling a successful treatment of the infection, the autoreactivity persists. (a) shows simulation results for an antigen induced $T_0$ transition and (b) a schematic of the corresponding change of the group structure. Until the 3000th iteration the system is in a self-tolerant state and no autoreactivity is detectable. Then antigen is inserted into $S_{10}$ which causes a rearrangement of the system as can be seen by the COM components. Afterwards one can observe autoreactive bursts which persist when the antigen population is set to zero. (b) illustrates how the self (red) is distributed over the groups by a $T_0$ transition (see Fig.~\ref{TransitionMatrices} and text) and how the potentially autoreactive clones are linked to the self (arrows). The transition is caused by antigen (orange star) which is inserted into $S_{10}$ and found in $S_9'$ after the rearrangement. After the transition a part of the hole group $S_8$ (green), which is linked to self, is found in the new core group $S_7'$ which is populated and thus causes autoreactivity.
                                                 Used parameters: $p=0.074$, $A_\mathrm{start}=2.5$, $A_\mathrm{max}=2.8$, $l=0.04$.
                                                  }
                                              \label{Pie1}
                                                      \end{figure*}

We now have a closer look on how the antigen population induces autoreactivity. Figure \ref{Pie1}a shows time series of the center of mass components, the autoreactivity $R_A$, and the antigen population $A$  inserted into $S_{10}$. The system evolves towards a self-tolerant architecture, such that $R_A$ is equal to zero in the beginning. At the 3000th iteration the antigen population is inserted. The parameters are chosen such that the antigen is not defeated and causes a rearrangement of the group structure after approximately 600 iterations. This can be concluded from the COM components, where the non-determinant bit position changes, indicating a $T_0$ transition. Immediately after the rearrangement autoreactive bursts are observable. 
\\
If the antigen population is set to zero, modeling a successful treatment of the infection, the bursts continue with lower frequency.

\begin{figure*}[]
                                              	\begin{minipage}[ht]{1\columnwidth}
                                              		\centering
                                              		\includegraphics[width=\textwidth]{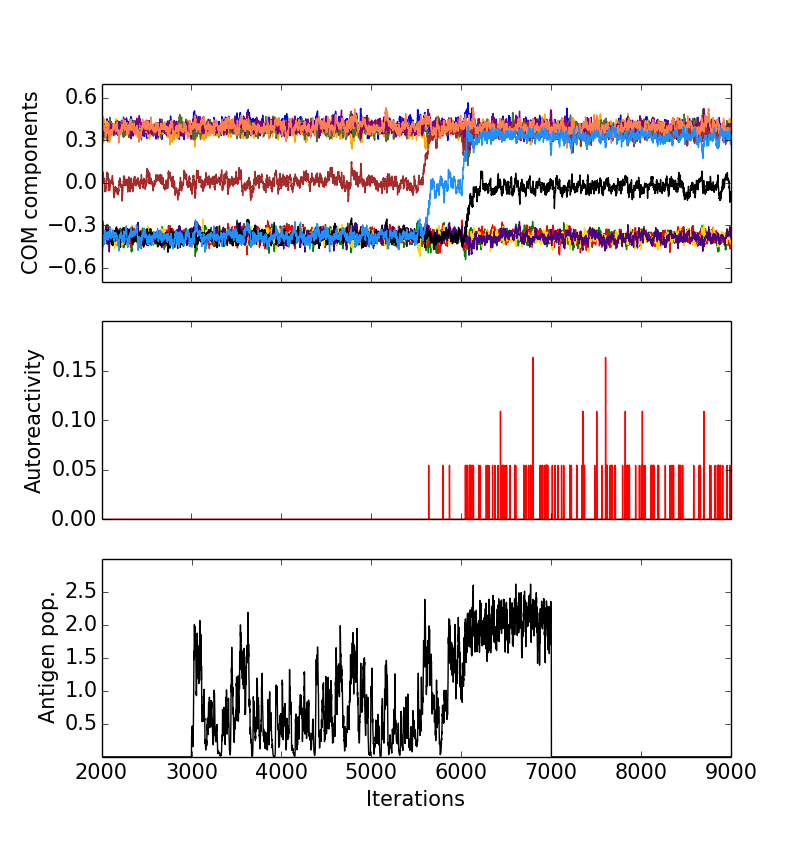}
                                              			\vspace{-1.1cm}
                                              		  	\caption*{\hspace{6cm}\large(a)}
                                              		\label{}
                                              	\end{minipage}
                                              	\hspace{1.5cm}
                                              	\begin{minipage}[ht]{0.87\columnwidth}
                                              		\centering
                                              	\vspace{0.2cm}	\includegraphics[width=\textwidth]{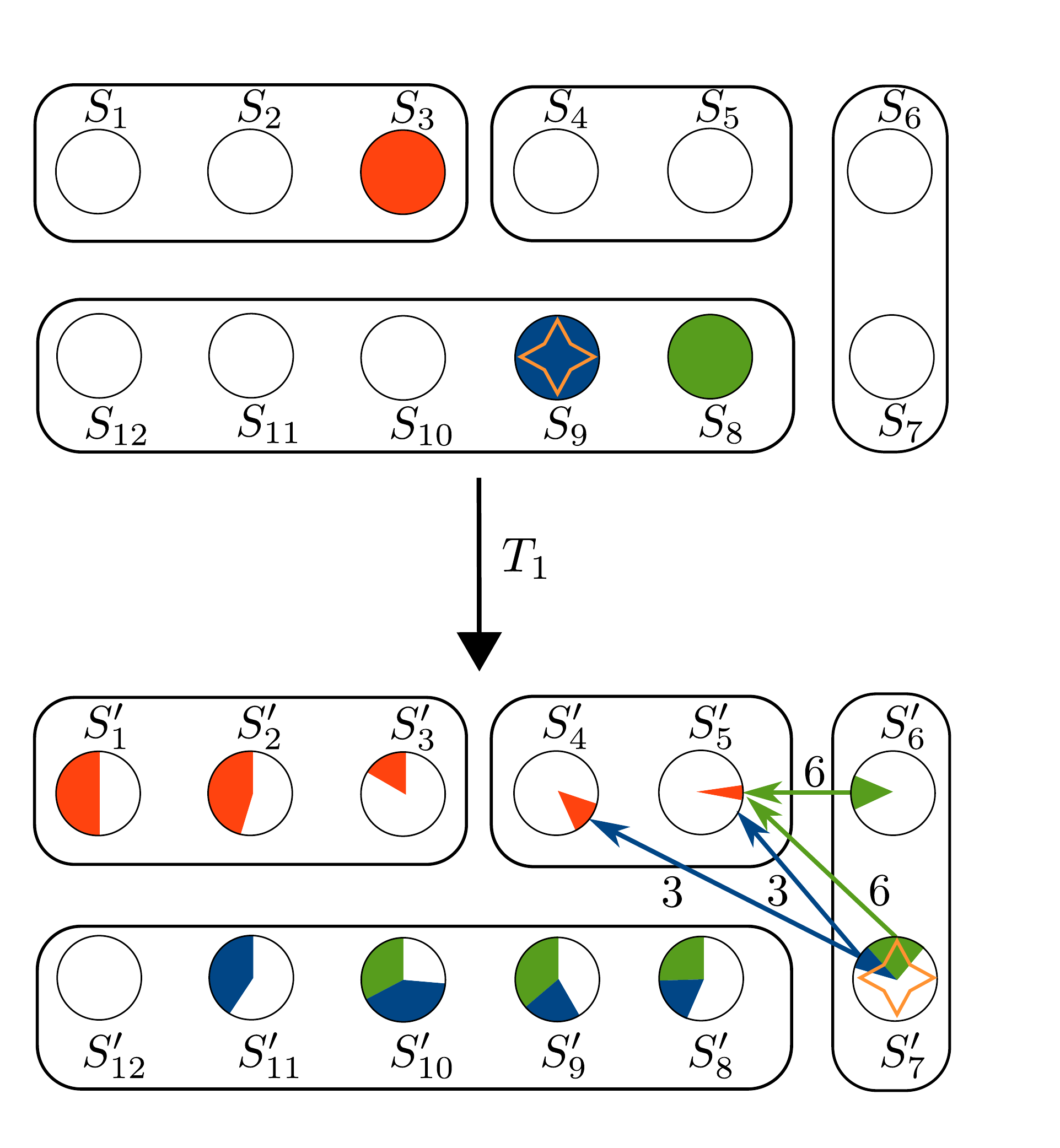}
        										\vspace{-0.7cm}    	
         										\caption*{\hspace{4.6cm}\large (b)}                                              		
                                              	\end{minipage}
                                              	\caption{
                                              	Antigen induced autoimmunity. The same protocol and parameters are used as in Fig.~\ref{Pie1} apart from antigen being inserted into the hole group $S_9$ instead of $S_{10}$ and using a different seed. Here, the antigen population induces a $T_1$ transition (see Fig.~\ref{TransitionMatrices} and text) which spreads the self nodes farther than in case of the $T_0$ transition. After the rearrangement a part of the hole group $S_8$ (green) which is linked to self is found in the core groups $S_7'$ and even in $S_6'$ which are both populated. Some nodes of $S_9$ (blue) are also found in $S_7'$. In this structure, the neighbors of the self nodes have a higher occupation than in case of a $T_0$ transition, causing autoreactive bursts to occur more frequently than in Fig.~\ref{Pie1}a. The antigen population is located in group $S_7'$ in the new steady state.                                    
                                              	}\label{Pie2}
                                              	\end{figure*}

\begin{figure}[h]
             	\centering 
             	\includegraphics[width=0.785\linewidth ]{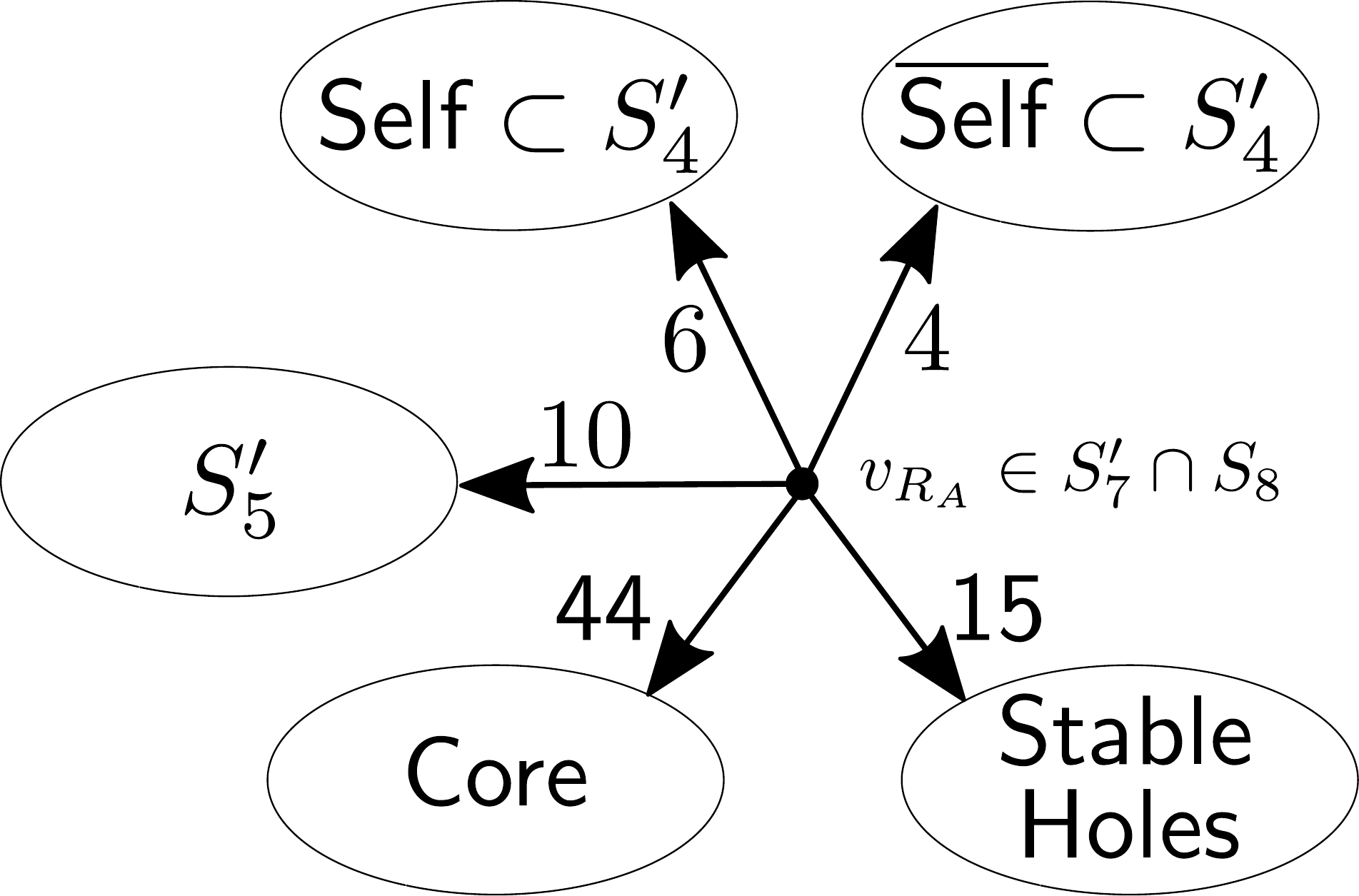}
             	\caption{Neighborhood of a potentially autoreactive clone $v_{R_A} \in S_7' \cap S_8$ after a $T_0$ transition. The arrows show the links, which $v_{R_A}$ has to nodes in other groups. This depicts the neighborhood of potentially autoreactive nodes in Fig.~\ref{Pie1} which constitute the green part of $S_7'$.}
             	\label{Autoreacivity_dm0'_schematic}
\end{figure}

\begin{figure}[h]
             	\centering 
             	\includegraphics[width=0.565\linewidth ]{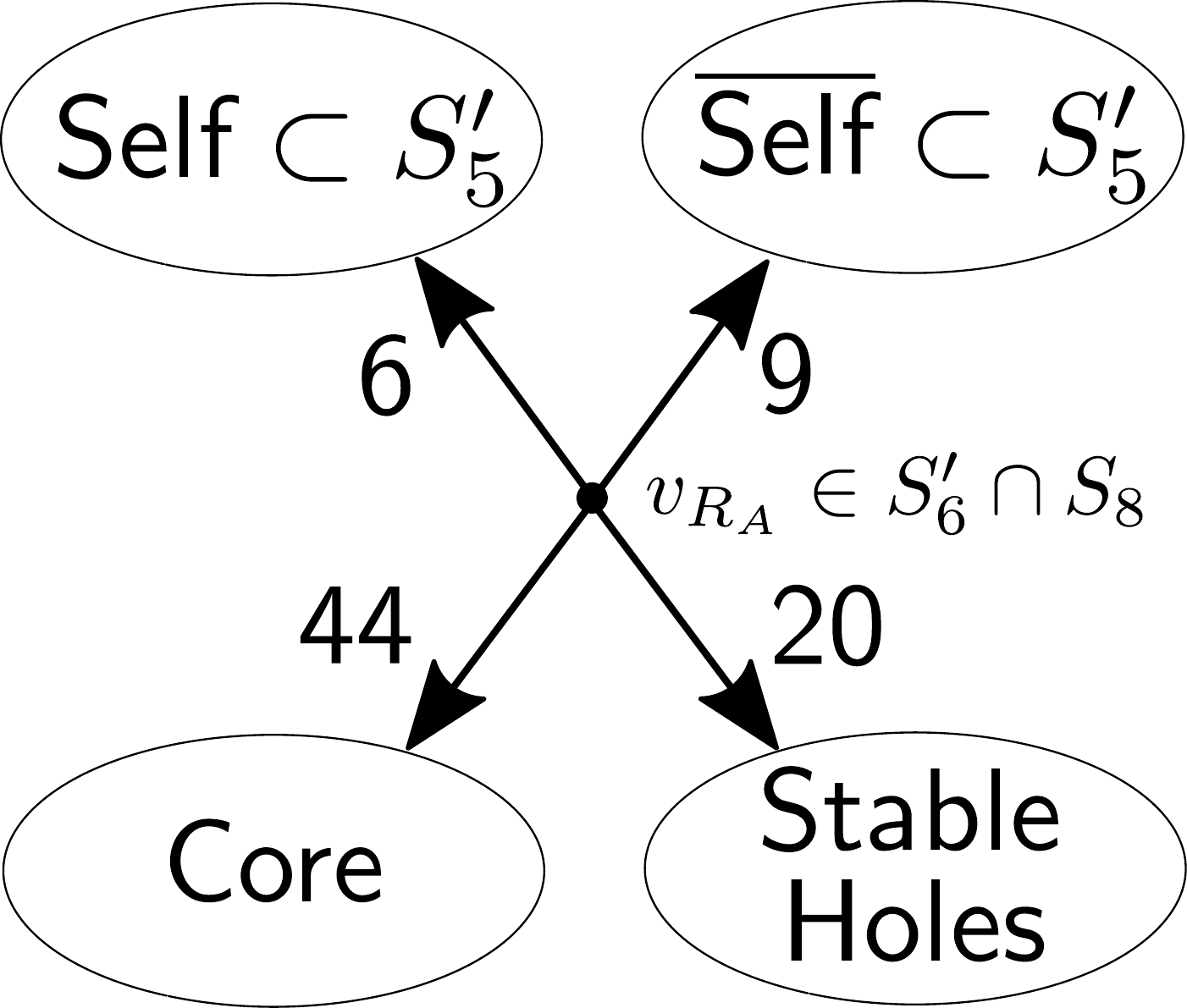}
             	\caption{Neighborhood of a potentially autoreactive clone $v_{R_A} \in S_6' \cap S_8$ after a $T_1$ transition. The arrows show the number of links which $v_{R_A}$ has to nodes in other groups. This  depicts the neighborhood of potentially autoreactive nodes in Fig.~\ref{Pie2} which constitute the green part of $S_6'$.}
             	\label{Autoreacivity_dm12'}
  \end{figure}

Figure \ref{Pie1}b shows a schematic of how the transition causes autoreactivity. In the upper part of the figure the system is depicted in the self-tolerant state. The self nodes (red) fill up $S_3$ completely and no autoreactivity can be detected. Then antigen is inserted into the group $S_{10}$ as marked by the orange star. This causes a rearrangement of the group structure and a $T_0$ transition can be observed. When the new steady state is reached, the antigen is in the group $S_9'$ and the self nodes are distributed over the groups $S_2'$, $S_3'$ and $S_4'$ as illustrated in the lower part of Fig.~\ref{Pie1}b. At the same time, the nodes of the group $S_8$ which are linked to the self in $S_3$ (see Fig.~\ref{Fig:GroupStructure}) are found in $S_7'$, $S_8'$ and $S_9'$, cf. the transition matrix in Fig.~\ref{TransitionMatrices}. Since the core nodes in $S_7'$ have a non vanishing average occupation it occurs that self nodes in $S_4'$ have occupied neighbors and therefore are seen by the network, which causes the autoreactive bursts. Since the nodes of the hole groups $S_8'$ and $S_9'$ are permanently unoccupied they do not induce autoreactivity.              
\\
\newline
Figure \ref{Pie2}a shows the center of mass components, autoreactivity, and antigen population for a $T_1$ transition caused by insertion of antigen into the group $S_9$. Again, the antigen is inserted at the 3000th iteration. Here, it takes approximately 3000 iterations until the transition is initiated. Since one determinant COM component (blue) changes its average value and the non-determinant COM component (brown) is exchanged, a $T_1$ transition takes place. Compared to Fig.~\ref{Pie1}a a much higher average autoreactivity is seen.
\\
This can be explained considering Fig.~\ref{Pie2}b:
In the original structure, the self nodes constitute the singleton group $S_3$ and the autoreactivity is equal to zero. This time the antigen is implemented in the hole group $S_9$ and triggers a $T_1$ transition, after which the antigen population is found in the core group $S_7'$. This transition causes more nodes to change their group and distributes them farther than the $T_0$ transition. This can be seen very clearly comparing the matrices in Fig.~\ref{TransitionMatrices}. In the new steady state we observe that self nodes can be found not only in $S_4'$ but also in $S_5'$ which are both periphery groups. Therefore, there are three subgroups which are linked to self having a non vanishing average occupation: $S_8 \cap S_7'$, $S_9 \cap S_7'$ and $S_8 \cap S_6'$. Especially nodes from $S_8 \cap S_6'$ cause frequent autoreactive bursts, since having a much higher average occupation than nodes in $S_7'$. 
\\
To get a better insight into the mechanism of autoimmunity in this model, we calculate the average autoreactivity found after the two types of rearrangement in mean-field approximation.
\\
At first we consider the $T_0$ transition as it is depicted in Fig.~\ref{Pie1}. It is necessary to characterize the neighborhood of the potentially autoreactive clones $v_{R_A}$ which are not permanently empty and linked to self. In the case of a $T_0$ transition these clones constitute the subgroup $S_8 \cap S_7'$ which is depicted as the green part of $S_7'$ in Fig.~\ref{Pie1}b. The neighborhood of these clones is illustrated in Fig.~\ref{Autoreacivity_dm0'_schematic}.
A node in $S_7'$ has 44 links to the weakly occupied core nodes, 15 to the permanently empty hole groups and 10 links to the strongly occupied periphery group $S_5'$, as can be seen in Fig.~\ref{Fig:Linkmatrix} which shows the entries of the link matrix for the $d_M=11$ architecture. Six of the ten links to the group $S_4'$  connect to self nodes from $S_3 \cap S_4'$, as it is also indicated in the lower part of Fig.~\ref{Pie1}b.
\\
For determining the probability $P_{v_{R_A}}^{\mathbb{W}}$ that a potentially autoreactive node $v_{R_A}$ survives the window rule one has to consider that $v_{R_A}$ already has six permanently occupied self nodes as neighbors. Therefore, $P_{v_{R_A}}^{\mathbb{W}}$ is given by the probability that the number of occupied neighbors excluding the self nodes is less or equal to $t_U-6=4$
\begin{linenomath}
\begin{align}
P_{v_{R_A}}^{\mathbb{W}}=\text{Prob}\left\{ n(\partial v_{R_A}\notin \text{Self}) \leq 4 \right\}.
\end{align}
\end{linenomath}
%Simplifying notation we introduce a shorthand for the cumulative binomial distribution
%\begin{equation}
%	B_C(n,p,N)=\sum\limits_{i=0}^{N}\binom{n}{i}p^i(1-p)^{n-i}.
%\end{equation}               
The probability of survival is simply given by the sum over the probabilities of all states in which the autoreactive node has less than $t_U$ occupied neighbors. We approximate the weakly occupied core nodes as permanently empty. Including the hole nodes this gives 59 permanently empty neighbors in total. The probability that $N$ of these nodes are occupied after the influx is given by the cumulative binomial distribution $B_C(59,p,N)$. Since the periphery groups are strongly occupied  it is a good approximation that the probability to have $N'$ occupied neighbors in, e.g., $S_5'$ is given by $B_C(10,n_5',N')$. Here, $n_5'$ is the average group occupation of $S_5'$ excluding the self nodes, which can be determined in mean-field approximation. This gives the final result                     
\begin{linenomath}                
          \begin{align}
          	 P_{v_{R_A}}^{\mathbb{W}}=&\sum\limits_{N+N'+N''\leq 4}^{}B_C(59,p,N) \nonumber  \\ &\times  B_C(10,n_5',N') \cdot B_C(4,n_4',N'').
          	\label{Survivalprobability_Autoreactive_in_S7'}
          \end{align}      
\end{linenomath}             
Now, the average autoreactivity can be determined using the fixed point  $n_{R_A}^*$ of the iteration rule
\begin{linenomath}
\begin{equation}
	n_{{R_A}}'=\left[n_{R_A}+(1-n_{R_A})p \right] P_{v_{R_A}}^{\mathbb{W}},
	\label{MFA autoreactivity}
\end{equation}
\end{linenomath}
and inserting it into the equation for the autoreactivity
\begin{linenomath}
\begin{align}
	\avg{R_A}_{\text{MFA}}=&\frac{1}{|S_3|}\cdot L_{8,3}\cdot T_{8,7}(0) \cdot n_{R_A}^* \nonumber \\ =&\frac{1}{110}\cdot 6\cdot 210 \cdot n_{R_A}^*,
	\label{Final expectation value autoreactivity dm0'}
\end{align}
\end{linenomath}
where the transition matrix element $T_{8,7}(0)$ can be read off Fig.~\ref{TransitionMatrices}.
With the values of the mean occupation in the new structure $n_4'=0.683$ and $n_5'=0.671$ for $p=0.074$ in mean-field approximation we get the result
\begin{linenomath}
\begin{equation}
	\avg{R_A}_{\text{MFA}}=1.11\cdot10^{-4}.
\end{equation}
\end{linenomath}
If we determine the occupations from simulations we get $n_4'=0.683$ and $n_5'=0.659$ which gives
\begin{linenomath}
\begin{equation}
	\avg{R_A}_{\text{MFA}}'=1.20\cdot10^{-4}
\end{equation}
\end{linenomath}
differing only slightly from the value above. Determining the average autoreactivity in simulations by performing an average over 100000 iterations after causing a $T_0$ transition, as depicted in Fig.~\ref{Pie1}, gives
\begin{linenomath}
\begin{equation}
	\avg{R_A}_{\text{Sim}}=1.59\cdot 10^{-4}.
\end{equation}
\end{linenomath}
 Considering the low frequency at which the autoreactive bursts occur, the calculated values give a good approximation of the average simulation results. One can expect better results for the $T_1$ transition, where the autoreactive bursts are much more frequent. 
\\
 For the $T_1$ transition, as it is depicted in Fig.~\ref{Pie2}, the same line of argument is followed. Yet, the discussion is more intricate, since there are three groups of nodes, which can cause autoreactivity. The first group is $S_8 \cap S_7'$ for which Eq.~(\ref{Survivalprobability_Autoreactive_in_S7'}) from above can be used. The characterization of the neighborhood of $S_8 \cap S_6'$ is simple. Since the only group containing self nodes to which $S_6'$ is linked is $S_5'$ every node of $S_8 \cap S_6'$ has six links to self nodes in $S_5'$. This gives the neighborhood as depicted in Fig.~\ref{Autoreacivity_dm12'}. 
 
 The neighborhood of $S_9 \cap S_7'$ is more complicated to characterize. To start with, the bitstrings of nodes in $S_1$, $S_9$ and $S_3$ are illustrated in Fig.~\ref{Fig:S_3,S_7}. Remember that a node can be classified by comparison of its entries in the determinant bit positions with those of nodes from $S_1$. For the 12-group architecture there is one non-determinant bit position whose entry has no influence on the group membership of a node. This non-determinant bit position is marked by a square.
       
                \begin{figure}[h]
                	\flushleft
                \begin{tikzpicture}
                \text{$\hphantom{A}v \in S_1: $    }   \bcircle \bcircle \bcircle \bcircle \bcircle \bcircle \bcircle \bcircle \bcircle \bcircle \bcircle \squaree 
                \end{tikzpicture}
                \linebreak 
                \begin{tikzpicture}
                \text{$\hphantom{A}v \in S_9: $}  \bcircle \bcircle \bcircle \bfcircle  \bfcircle \bfcircle \bfcircle \bfcircle \bfcircle \bfcircle \bfcircle \squaree 
                \end{tikzpicture}
                \linebreak
                \begin{tikzpicture}
                \text{$\hphantom{A}v \in S_3:$    }   \bcircle \bcircle \bcircle \bcircle \bcircle \bcircle \bcircle \bcircle \bcircle \bfcircle \bfcircle \squaree 
                \end{tikzpicture}
                \caption{Bitstrings of nodes from the groups $S_1, \, S_9$ and $S_3$ in the original structure. Empty circles represent entries of determinant bit positions which coincide with the respective entries of a node in $S_1$. Squares stand for the non-determinant bit position, which has no influence on the group membership of a node. Filled circles stand for entries in determinant bit positions which are complementary to the corresponding entries of the group $S_1$.}
                \label{Fig:S_3,S_7}
                \end{figure}
Assume that a $T_1$ transition leads to the situation depicted in Fig.~\ref{Fig:S_3',S_7'}. Here, the non-determinant bit has changed its position and one of the entries of the determinant bit positions has changed its value. In this new structure the nodes, represented in Fig.~\ref{Fig:S_3,S_7}, belong to different groups than in the original architecture: The node from $S_9$ is found in $S_7'$ and the node from $S_3$ in $S_5'$. The entries of the underlined bit positions can not be interchanged with entries of a different value without changing the group membership in the original or new structure.

  \begin{figure}[h]
             	\flushleft
             \begin{tikzpicture}
             \text{$\hphantom{Aaaa}v \in S_1': \hphantom{,}$    }   \bcircle \bcircle \bcircle \bcircle \bcircle \bcircle \bcircle  \squaree $\, \, $ $'$  \bfcircle   \bcircle \bcircle \bcircle
             \end{tikzpicture}
             \linebreak 
             \begin{tikzpicture}
             \text{$v \in S_9 \cap S_7': \hphantom{,}$    } \bcircle \bcircle \bcircle \bfcircle  \bfcircle \bfcircle  \bfcircle  \underbar{\fsquaree$ ' $} \underbar{\bfcircle} \bfcircle \bfcircle \underbar{\bcircle} 
             \end{tikzpicture}
             \linebreak
             \begin{tikzpicture}
             \text{$ v \in S_3 \cap S_5': \hphantom{,}$    }   \bcircle \bcircle \bcircle \bcircle \bcircle  \bcircle \bcircle \underbar{\squaree$\, \, ' $ } \underbar{\bcircle} \bfcircle \bfcircle \underbar{\bcircle} 
             \end{tikzpicture}
             \caption{Nodes of the groups $S_1'$, $S_9 \cap S_7'$ and $S_3 \cap S_5'$ in the new structure  caused by a $T_1$ transition. The entries of the bit positions which are underlined can not be interchanged with entries of a different value without changing the membership of the nodes in the original or new structure. One can observe that there are three nodes in $S_3 \cap S_5'$ which are linked to a node in $S_9 \cap S_7'$. Compare with Fig.~\ref{Pie2} where the nodes from $S_9 \cap S_7'$ make up the blue part of the pie chart of $S_7'$. }
             \label{Fig:S_3',S_7'}
             \end{figure}
\begin{figure}[h]
         	\centering 
         	\includegraphics[width=0.885\linewidth ]{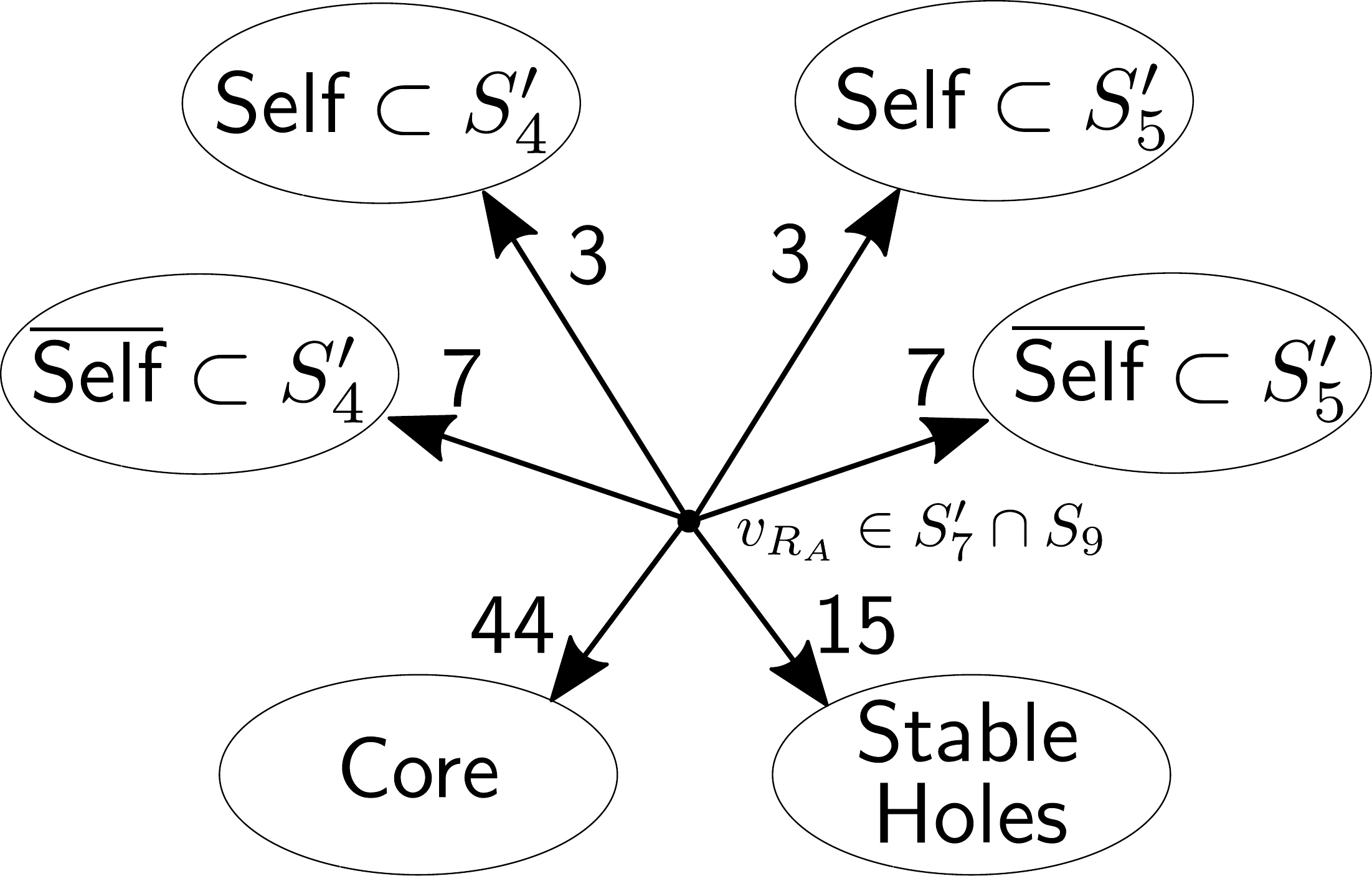}
         	\caption{Neighborhood of a node $v_{R_A} \in S_7' \cap S_9$ which causes autoreactivity after a $T_1$ transition. The arrows show the number of links which $v_{R_A}$ has to nodes in other groups. One can see that the self nodes to which $v_{R_A}$ is linked do not lie in one group, but are split up between $S_4'$ and $S_5'$. This figure shows the neighborhood of autoreactive nodes in Fig.~\ref{Pie2} which are in $S_7'$ in the new structure and have been members of $S_9$ in the original structure.}
         	\label{Autoreacivity_dm3'}
         \end{figure}

 
 Therefore, there is one fixed bit position in which the nodes coincide. Interchanging two of the first three entries of $ v \in S_3 \cap S_5'$ with the two complementary ones (filled circles), makes $ v \in S_3 \cap S_5'$ a neighbor of $v \in S_9 \cap S_7'$. As a reminder: Two nodes are neighbors if their bitstrings are complementary up to $m=2$ mismatches. There are $\binom{3}{2}=3$ possibilities for doing so, meaning that every node in $S_9 \cap S_7'$  has three self nodes in $S_5'$ as neighbors. This implies, that every node in $S_9 \cap S_7'$ also has three self nodes in $S_4'$ as neighbors, since this is the only other group containing self and the number of neighbored self nodes for a node from $S_9$ is constantly equal to six. Finally, this gives the neighborhood as shown in Fig.~\ref{Autoreacivity_dm3'}.
Considering this neighborhood and using similar arguments for $v_{R_A} \in S_8 \cap S_6'$ one gets the probabilities 
\begin{linenomath}
\begin{eqnarray}
\hspace{-0.6cm}P_{v_{R_A} \in S_8\cap S_7'}^{\mathbb{W}}=&\sum\limits_{N+N'+N''\leq 4}^{}B_C(59,p,N) \nonumber  \\ &\hspace{-0.6cm}\times  B_C(4,n_5',N') \cdot B_C(10,n_4',N''),
\label{Survivalprobability_Autoreactive_in_S7' cap S9}
\\
\hspace{-0.2cm}P_{v_{R_A}\in S_9 \cap S_7'}^{\mathbb{W}} =&\sum\limits_{N+N'+N''\leq 4}^{}B_C(59,p,N) \nonumber \\ & \hspace{-0.2cm}\times  B_C(7,n_5',N')\cdot B_C(7,n_4',N'')
\label{Survivalprobability_Autoreactive_in_S7' from S_9},
\\
\hspace{-0.2cm}P_{v_{R_A}\in S_8 \cap S_6'}^{\mathbb{W}}=&\hspace{-0.4cm}\sum\limits_{N+N'\leq 4}^{}B_C(64,p,N)\cdot B_C(9,n_5',N').
\label{Survivalprobability_Autoreactive_in_S6'}
 \end{eqnarray}
 \end{linenomath}
  For all these three sub-groups one needs to iterate the mean-field equation (\ref{MFA autoreactivity})
 to get the fixed points $n_{R_{A} \in S_i \cap S_j'}^*$.
 Finally the average autoreactivity can be calculated as
 \begin{linenomath}
 \begin{align}
 	\avg{R_A}_{\text{MFA}}&=\frac{6}{|S_3|} \cdot \biggl[ T_{9,7}(1)\cdot n_{R_{A}\in S_9 \cap S_7'}^* \nonumber  
 	\\
 	 &+ T_{8,7}(1)  \cdot  n_{R_{A}\in S_8 \cap S_7'}^*+T_{8,6}(1)\cdot n_{R_{A}\in S_8 \cap S_6'}^*  \biggr] \nonumber
 	\\
 	&= \frac{6}{110} \cdot \biggl[ 210  \cdot  n_{R_{A}\in S_9 \cap S_7'}^*\nonumber 
 	\\
 	&+84\cdot n_{R_{A}\in S_8 \cap S_7'}^*+ 126\cdot n_{R_{A}\in S_8 \cap S_6'}^*  \biggr].
 \end{align}
 \end{linenomath}
 Using the mean-field results $n_4'=0.683$ and $n_5'=0.667$ we get
 \begin{linenomath}
 \begin{equation}
 	\avg{R_A}_{\text{MFA}}=0.0027,
 \end{equation}
 \end{linenomath}
  When using the average occupations $n_4'=0.687$ and $n_5'=0.665$ from simulations one obtains
  \begin{linenomath}
  \begin{equation}
  	\avg{R_A}_{\text{MFA}}'=0.0029.
  \end{equation}
  \end{linenomath}
 Both results are close to the autoreactivity averaged over 100000 iterations during a simulation as shown in Fig.~\ref{Pie2}
 \begin{linenomath}
 \begin{equation}
 \avg{R_A}_{\text{Sim}}=0.0023.
 \end{equation}
 \end{linenomath}
 The relative deviation of the average autoreactivity obtained in mean-field approximation with respect to the simulation results is smaller for the $T_1$ transition as for the $T_0$ transition, which already has been suspected above.
 \subsection{Removal of idiotypes}
 \label{Sec:Removal of idiotypes}
\begin{figure}[h!]
   	\centering 
   	\includegraphics[width=\linewidth ]{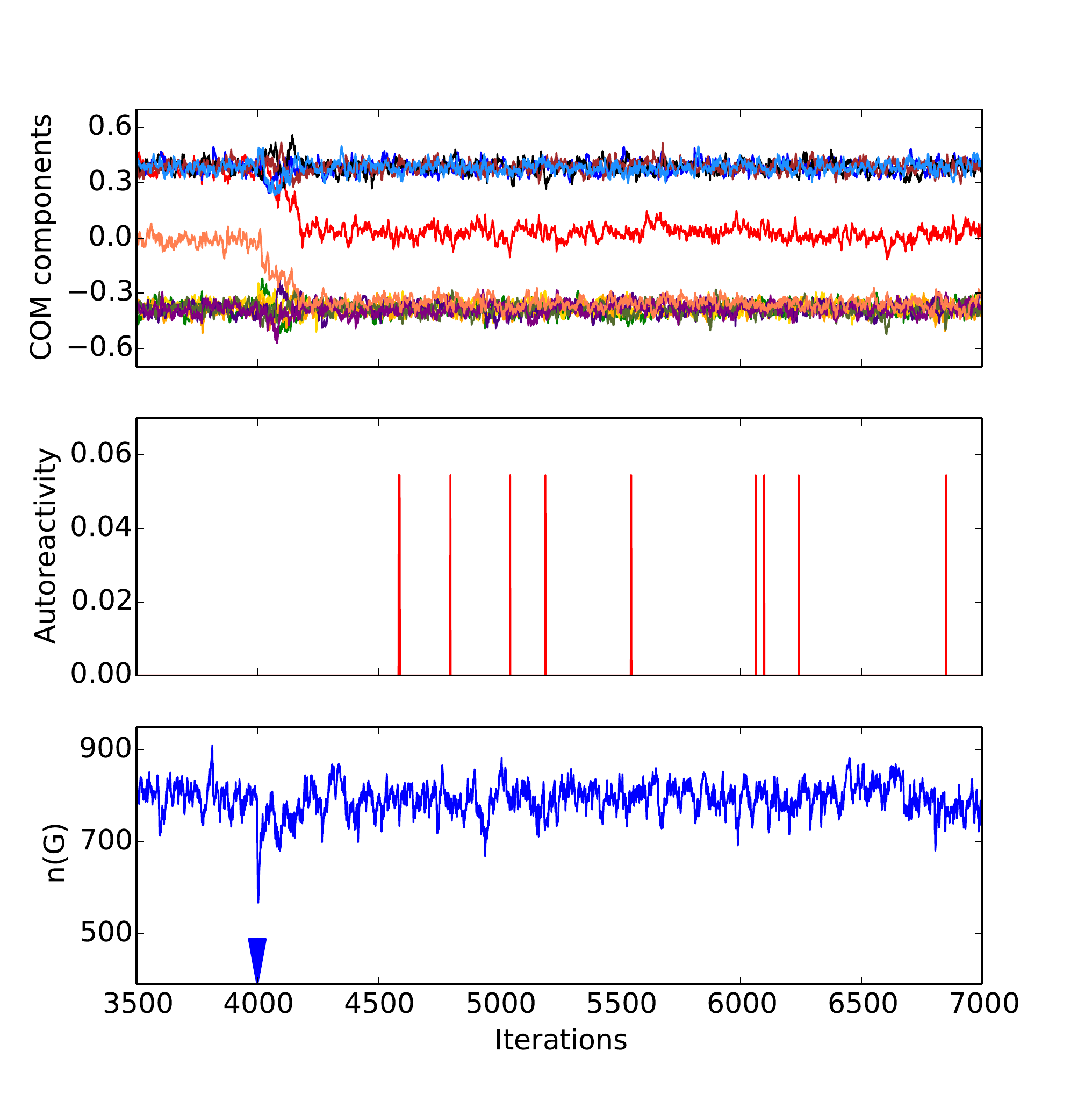}
   	\caption{
   				Induction of autoimmunity by repeated deletion of random nodes, modeling a series of blood extractions. Beginning from the 4000th iteration (as marked by the blue triangle) for five successive iterations 10 \% of all occupied nodes are chosen randomly and deleted. This leads to a large decrease of the total number of occupied nodes $n(G)$ and causes a $T_0$ transition, as can be seen by the center of mass components. Afterwards autoreactive bursts are observable. The influx for this simulation was $p=0.074$.
  			}
   	\label{fig:BlutTransf}
   \end{figure}

 In this subsection we investigate whether autoimmune states can be induced by random deletion of a certain fraction of all occupied nodes. The motivation behind this perturbation is to mimic the random removal of lymphocytes and antibodies from the organism as it occurs in large amounts during the course of \textit{autologous blood donations}. 
 \\
 It is common practice that patients donate blood for themselves prior to a non-emergency surgery. If loss of blood occurs during the operation this blood can be used for reinfusion. This so-called autologous blood donation has a lot of advantages compared to allogenic blood donations where blood is received from a foreign donor. 
 The use of autologous blood donations became common due to feared transmissions of infectious diseases like HIV. Even though the risk for the transmission of infections via allogenic transfusion decreased strongly in the last decades, the autologous transfusion is still favorable having a lower probability of causing complications like allergic and febrile reactions, alloimmunizations and hemolytic reactions \cite{Kumar2013}. Due to the repeated blood donations previous to the operation the erythropoiesis of the patient is enhanced, which helps to recover a normal level of erythrocytes after potential blood losses during surgery \cite{EndelePhD2005}. 
 \\
 Nevertheless, in extremely few cases very severe outcomes were also observed for autologous blood donations \cite{AutologousTransfusionPoposvsky}. The possible mechanisms for the observed complications could be caused by contamination of the stored blood, by storage effects but also by immunological processes. 
 \\
 Here we want to investigate if, in the frame of the model, autoreactive behavior can be observed during a modeled autologous blood transfusion. First we only examine the influence of the blood extraction on the system. For doing so, we start from an established 12-group structure and delete $10 \%$ of all occupied nodes. This is done for five successive iterations respectively. Every deletion models the extraction of $500 \, \text{ml}$ of blood i.e. approximately $10 \%$ of the total amount of blood. The timespan which is represented by one iteration lies in the order of magnitude of 10 days, which is also the typical amount of time between two blood extractions.
 \\
 Figure \ref{fig:BlutTransf} shows results for this protocol, which begins at the 4000th iteration. The decrease of the total number of occupied nodes $n(G)$ can be seen clearly in the lower graph. Looking at the center of mass components one finds that this perturbation causes a $T_0$ transition. After the transition autoreactive bursts occur as one observes in the graph showing the autoreactivity $R_A$.
\\
 Here we found that it is possible to induce autoreactive behavior by repeated random removal of occupied nodes.
 For this choice of parameters a transition was observed in approximately $5 \%$ of all simulations. Increasing the percentage of deleted nodes per iteration causes transitions to occur more often. 
 \\
Figure \ref{fig:BlutTransf} shows that the total occupation $n(G)$ has been reduced by approximately $32 \%$ due to the repeated deletion of $10 \%$ of all occupied nodes. Simulations revealed that if the same fraction of nodes is deleted in one iteration transitions occur approximately as often as in the case of repeated removal.
 \\
 In addition to these findings we examined whether autoreactive bursts occur more frequently or further transitions take place when we re-occupy the deleted nodes.
 Under the assumption that the collected blood is not leukoreduced \cite{blajchman2006clinical}, meaning that the leukocytes are not filtered out during blood extraction, the removed lymphocytes will be reinserted into the organism when the blood is reinfused. Therefore, to mimic the reinfusion of the collected blood we simply occupy the nodes which were emptied due to the blood extraction. Hereby, we varied the number of iterations between the last deletion of nodes and their re-insertion between 3 and 50 iterations. Nevertheless we did neither observe further transitions nor a higher frequency of occurrence of autoreactive bursts due to the re-occupation of nodes.
   
 \section{Remission of Autoimmunity and Therapeutic protocols}
 \label{Remission of Autoimmunity and therapeutic protocols}
 In the previous sections the emergence and loss of self-tolerance has been studied. It is of course of great interest to examine if there are protocols which reconstitute the self-tolerant state after a perturbation. In this section two protocols are presented: The first restores the self-tolerant state by an increase of the influx $p$, the second by insertion of antigen with suitable idiotype.
\subsection{Variation of influx}
\begin{figure}[h]
         \centering
          \includegraphics[width=\linewidth ]{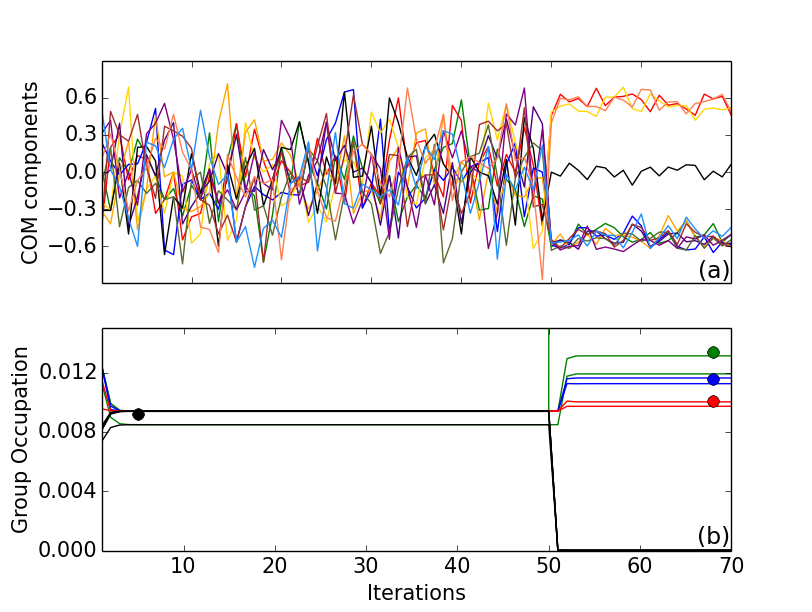}
           \caption{Formation of a 12-group architecture even for high influx $p=0.2$ due to the insertion of self into $S_3$ at the 50th iteration. (a) shows the COM components obtained by simulation, while (b) illustrates the average occupations in the different groups obtained in MFA using the same protocol. Here the singleton groups are represented in green, periphery groups in blue, core groups in red and hole groups in black. The colored points represent simulation results for the average occupation of singletons (excluding self), periphery, core and the one-group structure in the steady states respectively. The iteration starts from tabula rasa  without self.
          }   
            \label{Meanfield1}
        \end{figure}

The dynamics of the network without self becomes completely dominated by the influx if $p>p_c\approx0.13$ (cf. \cite{schmidtchen2012randomly}). In this case, there is no observable group structure and all nodes of the system share the same statistical properties. 
\\
Nevertheless, one can expect to observe a more interesting behavior when self is present in the network, e.g. by permanent occupation of $S_3$. We conducted the following protocol: Starting from tabula rasa without self and an influx $p=0.2$, which is far above the critical value $p_c$, we let the system evolve for 50 iterations. A certain 12-group structure is anticipated and all nodes of $S_3$ permanently occupied. Figure \ref{Meanfield1} shows simulation results and mean-field solutions for this protocol. Up to the 50th iteration all center of mass components  fluctuate around zero, which implies a symmetrically occupied graph. This means, that all nodes have the same statistical properties and there is no observable group structure. The same conclusion can be made from the mean-field results for the group occupation. Despite of the groups $S_1$ and $S_{12}$, which are too small to give good results in mean-field approximation, all groups have the same average occupation. This implies that, in fact, there is one group only. 
\\
The implementation of self in the anticipated group $S_3$ at the 50th iteration drastically changes this picture. The random fluctuations of the center of mass components weaken and the vector attains a form indicating a 12-group structure. This is in accordance with the mean-field results for the group occupations. Here one observes a strong decrease of the hole group occupation, followed by an increase and differentiation of the occupations for singletons (green), periphery (blue) and core (red).
\\
These observations can be explained as follows: Before the implementation of self, all nodes have the same average occupation. The average occupation is relatively low, since most of the nodes have too many occupied neighbors. Therefore, the permanently occupied self nodes have a suppressing effect on its neighbors. Since $S_3$ is only linked to the hole groups (see Fig.~\ref{Fig:GroupStructure}), its permanent occupation leads to a suppression and thus to a reconstitution of the stable holes. This reduces the number of neighbors of all other groups, leading to an increase of their average occupation. 

\begin{figure}[h]
      \centering
               \vspace{-0.31cm}
      \includegraphics[width=1.0\linewidth ]{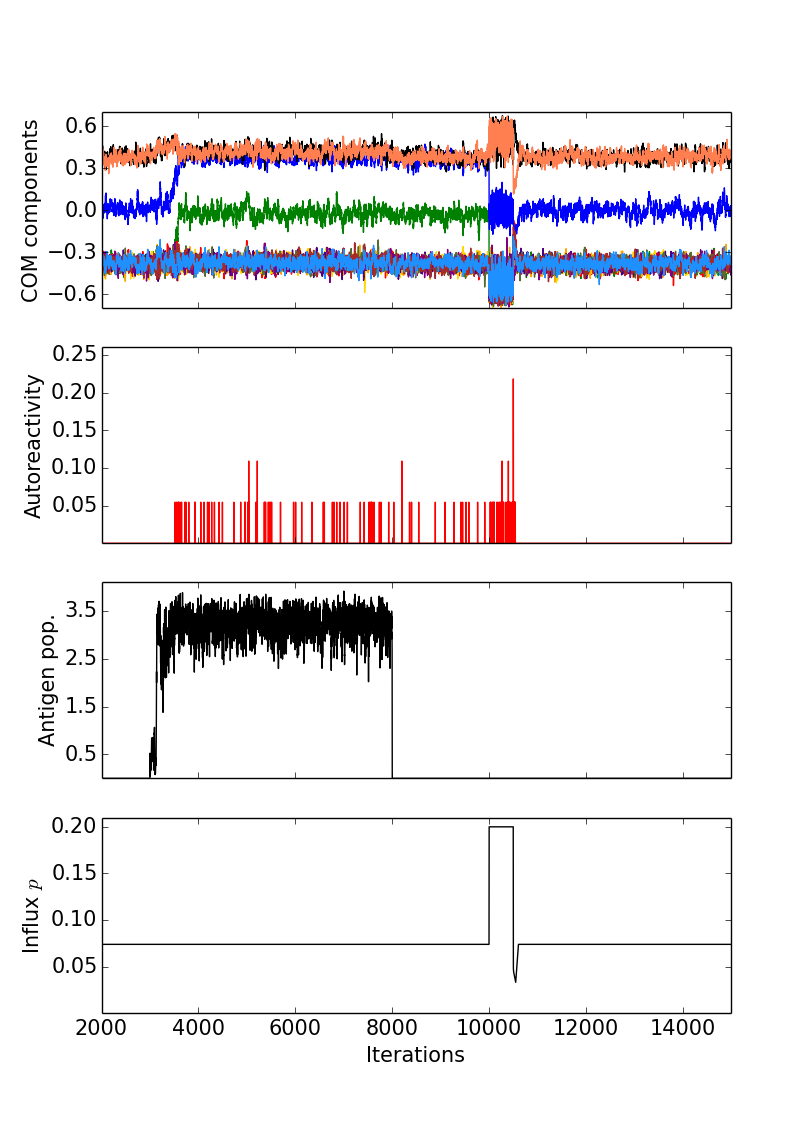}
      \caption{Remission of autoreactivity caused by variation of the influx $p$. The system evolves undisturbed with self placed in $S_3$ and an influx of $p=0.074$. At the 3000th iteration antigen is placed in the group $S_9$ which causes a $T_1$ transition, as can be seen by the COM components. This induces outbreaks of autoreactivity which also continue after the deletion of the antigen at the 8000th iteration. The original state is reconstituted by an increase of the influx up to $p=0.2$ followed by a decrease down to $p=0.035$ and a final increase up to $p=0.074$ in 50 steps. After execution of this protocol the COM vector has the same form as in the original state and the autoreactivity is zero, which shows that the self-tolerant state is reached again.}
       \label{Wiederherstellung}
       \end{figure}

Thus, by insertion of self into the network it is possible to observe group structures for values of the influx $p$ far above the critical value $p_c$. 
\\
This result was used to find a protocol for the reconstitution of the self-tolerant state after a transition. Figure \ref{Wiederherstellung} shows the center of mass components, the autoreactivity, the antigen population and the influx $p$ for such a protocol. As usual, the system evolves towards a self-tolerant architecture at first. Then antigen is inserted into $S_9$, causing a $T_1$ transition leading to an autoreactive state. Afterwards the antigen population is set to zero, modeling a successful treatment of the infection. This is followed by an increase of $p$ up to $p=0.2$. Considering the center of mass components one sees that the original structure is reconstituted, with the center of mass components showing large fluctuations. Then the influx is decreased to $p=0.035$ in 50 steps, followed by an increase up to the original value in 50 steps. Instead of setting $p$ back to its original value directly it proved more successful to use a protocol of this form. In the end, the original state is restored and no autoreactivity detectable. 
\subsection{Antigen induced tolerance}
In the previous sections it has been shown that an antigen population has the ability to induce an autoimmune state, in accordance with experimental findings. Paradoxically, it also has been observed that infections may protect its host from the development of autoimmune diseases \cite{Bach2002Infections,Bach2005Infectionsandautoimmdiseases}.
\\
We follow this approach and try to 'heal' an antigen induced autoimmune state by inserting a second antigen. For doing so, it is first necessary to work out how the first and the second antigen have to be related such that a reconstitution of the self-tolerant state is possible. 
\\
In principle, a transition to the self-tolerant state can be caused by inserting antigen independently of how the autoimmune state has been reached. In this more general case it is a larger effort to find the right idiotype for the antigen to cause a remission.
\\
Here, we consider a $T_1$ transition which has been induced by antigen in the group $S_9$. The idiotype $v_1$ of the antigen can be found in $S_7'$ after the rearrangement. In order to revoke the changes due to this reordering and to restore the self-tolerant state, the second antigen $v_2$ has to cause a $T_1$ transition which is 'inverse' to the first one. Knowing that antigen placed in $S_9'$ usually causes a $T_1$ transition one has to choose $v_2$ such that it is in $S_9'$ in the new structure and in $S_7$ in the original structure. Therefore one can conclude that $v_1 \in S_9 \cap S_7'$ and $v_2 \in S_9' \cap S_7$. Figure~\ref{Fig:2nd infection_1} depicts the bitstrings of these nodes  for an exemplary $T_1$ transition. 

\begin{figure}[h!]
 	\flushleft
 	\begin{tikzpicture}
 	\text{$\hphantom{Aaaa}v \in S_1: \hphantom{,}$    }   \bcircle \bcircle \bcircle \bcircle \bcircle \bcircle \bcircle \bcircle \bcircle \bcircle \bcircle \squaree 
 	\end{tikzpicture}
 	\linebreak 
 	\begin{tikzpicture}
 	\text{$\hphantom{AAA}w \in S_1': \hphantom{,}$    }   \bcircle \bcircle \bcircle \bcircle \bcircle \bcircle \bcircle \bcircle  \bcircle \bfcircle  \squaree$\, \, ' \hspace{0.03cm}$ \bcircle
 	\end{tikzpicture}
 	\linebreak 
 	\begin{tikzpicture}
 	\text{$ v_1  \in S_9 \negthinspace \cap \negthinspace S_7' \hspace{-0.03cm}: \hspace{0.18cm}$    }  \bcircle \bcircle \bcircle \bfcircle  \bfcircle \bfcircle \bfcircle \bfcircle  \bfcircle \underline{\bfcircle \fsquaree$ ' \hspace{0.035cm}$ \bcircle }
 	\end{tikzpicture}
 	\linebreak
 	\begin{tikzpicture}
 	\text{$ v_2 \in S_9'\negthinspace \cap \negthinspace S_7 \hspace{-0.03cm} : \hspace{0.18cm} $    }  \bcircle \bcircle \bcircle \bfcircle  \bfcircle \bfcircle \bfcircle \bfcircle  \bfcircle \underline{\bcircle \squaree$\, \, ' \hspace{0.035cm}$\bfcircle} 
 	\end{tikzpicture}
 	\caption{Bitstrings of the reference nodes $v \in S_1$ and $w \in S_1'$ for an exemplary $T_1$ transition and bitstrings of the nodes $v_1 \in S_7 \cap S_7'$ and $v_2 \in S_7' \cap S_9$ which represent the idiotypes of the first and the second antigen. The underlined entries of the bit positions can not be changed without changing the group membership of $v_1$ or $v_2$ in the original or in the new group structure. One can determine the maximal and the minimal Hamming distance between the nodes $v_1$ and $v_2$ by permutation of the entries of the bit positions which are not underlined, see text.}
 	\label{Fig:2nd infection_1}
 \end{figure}
\begin{figure}[t]
 	\centering 
 	
 	\includegraphics[width=\linewidth ]{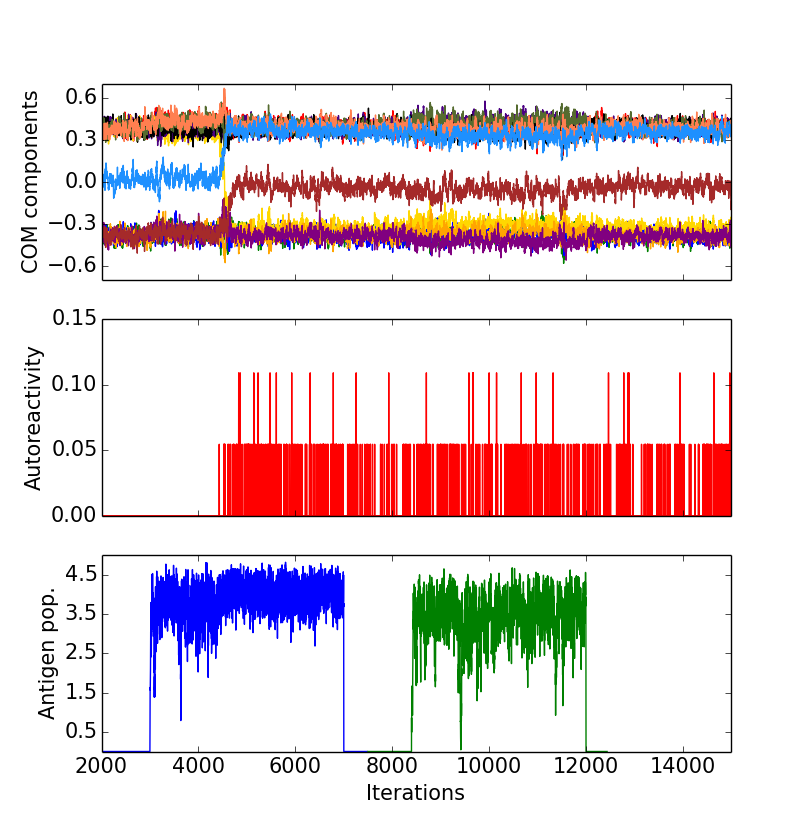}
 	
 	\caption{
 Unsuccessful attempt to reconstitute the self-tolerant state by infection with a second antigen. The first antigen population is depicted in blue, the second in green. In the presence of the second antigen the yellow and the blue COM components exhibit larger fluctuations towards zero indicating an attempt to restore the original structure. However, no transitions occur and the original state is not reconstituted.
    Parameters: $l=0.15$, $A_{\text{Start}}=2.5$, $A_{\text{max}}=5$, $p=0.075$.}
 	\label{Restoring1}
 \end{figure}
\begin{figure}[h!]
 	\centering 
 		\vspace{-0.31cm}
 	\includegraphics[width=\linewidth ]{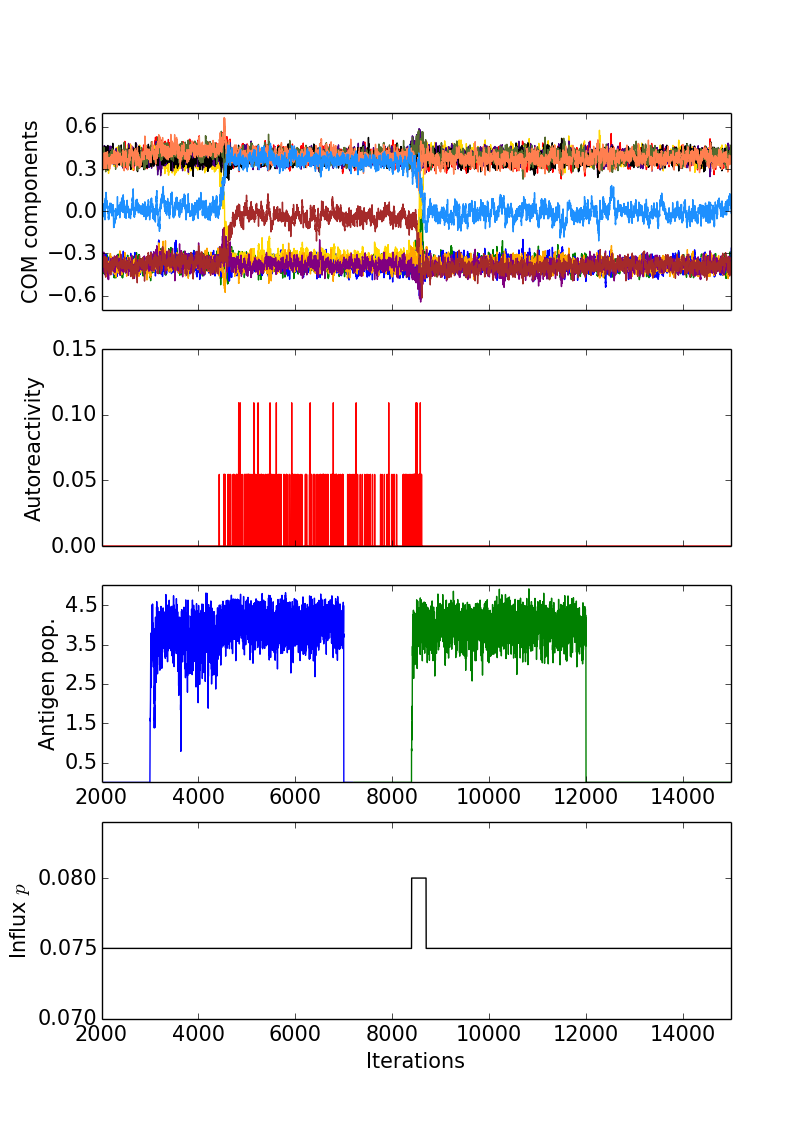}
 	\caption{Protocol for restoring the self-tolerant state by infection with a second antigen. The first antigen population is depicted in blue, the second in green. The increase of $p$ facilitates the transition and the original state is restored.
 	 Same parameters and seed as in Fig.~\ref{Restoring1}.}
 	\label{Restoring2}
 \end{figure}

The entries of the underlined bit positions can not be changed without changing the group membership of $v_1$ or $v_2$ in the original or in the new structure. Since the entries of the bitstrings of $v_1$ and $v_2$ differ in these positions the minimal Hamming distance between $v_1$ and $v_2$ is equal to three. Permuting the entries in the bit positions of $v_1$ and $v_2$ which are not underlined does not change the group membership of the antigens, neither in the original nor in the new structure. Therefore, we find that the maximal number of differing entries between $v_1$ and $v_2$ in this positions is equal to six. Thus, the maximal Hamming distance between $v_1$ and $v_2$ is equal to nine and we get the relation
\begin{linenomath}
\begin{equation}
3 \leq d_H(v_1,v_2) \leq 9.
\label{Hamminddist}
\end{equation}
\end{linenomath}
This shows that the second antigen should not be neighbored to the first one (i.e. it can not be almost complementary), but it is not allowed to be too similar as well. 
\\
Figure \ref{Restoring1} shows the results for a protocol which implements a second infection, trying to restore the original self-tolerant state. 
The first antigen (blue) causes a $T_1$ transition and one observes autoreactive bursts. This population is set to zero, modeling a successful treatment of the infection. Then a second antigen population (green) fulfilling $v_2 \in S_9' \cap S_7$ is inserted. One observes that the center of mass components slightly change but no transition occurs and the autoimmune state remains. 
\\
To facilitate transitions, we implemented the same protocol as in Fig.~\ref{Restoring1} but increased the influx $p$ for a few iterations up to $p=0.08$ which is still far below the critical value $p_c \approx 0.13$. The results of this simulation are presented in Fig.~\ref{Restoring2}.

Here one sees, that the second antigen causes a transition which restores the original state. Therefore, the autoreactivity vanishes and the system is self-tolerant again. 
\section{Conclusion and outlook}
\label{Sec:Conclusion}
In this work we have continued the examination of a minimalistic model of the idiotypic network particularly with regard to the emergence of self-tolerance. We have extended the focus considering failures of self-tolerance which lead to autoimmunity and devised, in the frame of the model, protocols which may lead to a remission of autoimmune conditions.
\\
Simulations have shown that the network evolves towards a self-tolerant state even if the nodes which represent the self are chosen at random. These nodes are permanently occupied from the beginning of the simulation and strongly influence the evolution of the network. If the idiotypes of these nodes do not differ too much, which is a reasonable biological restriction, they are found in groups which  on average only have very few occupied neighbors, thus providing self-tolerance. 
\\
It is of obvious interest to investigate the possible failure of this self-tolerant states by perturbations of the system. This was done using three different approaches. At first it was examined if a slow linear decrease of the influx of new lymphocytes may lead to autoimmune states. This decrease is supposed to model age induced effects, which result in a declining influx of new B-cells to the system. Using this protocol, we could not find transitions from a self-tolerant to an autoimmune state. 
\\
Subsequently we examined the influence of a strong variation of the influx by stopping it completely and setting it back to its original value, which could model a radiation therapy. Here we observed transitions from a self-tolerant state to a state which showed autoreactive bursts.
\\
The second approach is motivated by the observation that certain autoimmune diseases are triggered by infections. We inserted antigen with fixed idiotype into the network, whose population may grow but is suppressed by occupied neighboring nodes. Depending on the group into which the antigen was inserted and its growth parameter $A_\text{max}$ it induced autoimmunity. We classified the observed transitions and calculated the average autoreactivity using a mean-field approach.
\\
The last approach trying to induce an autoimmune state consisted of the repeated random removal of occupied nodes. Here, a certain fraction of all occupied nodes was deleted repeatedly which could mimic the removal of clones due to repeated blood extractions. Simulations revealed that this perturbation may induce an autoimmune state if the frequency of repetition and the fraction of removed clones is high enough. 
\\
We also considered the reverse phenomenon of 'spontaneous' remission, the transition from an autoimmune to a self-tolerant state. Here we again investigated variations of the influx $p$ of new lymphocytes and observed that one can reconstitute the self-tolerant state by applying a certain protocol to $p$ which first increases the influx and then resets it back to its original value.  This approach is in spirit of therapeutic strategies trying to defeat autoimmune diseases which do not make use of immunosuppressive drugs but instead stimulate clones controlling autoreactive lymphocytes \cite{feldmann2005design}.
\\
Furthermore we found that, if the autoimmune state was induced by an infection, a second infection with suitable idiotype can cause a further transition leading back to the original self-tolerant state. The idiotypes of the first and the second antigen are related and not allowed to be too similar or too different.
\\
The findings in this work also emphasize the importance of probabilistic aspects for autoimmune diseases. For some settings, using the same parameters, protocol, and initial conditions repeatedly one observes the emergence of self-tolerant as well as autoimmune states. %This is due to the stochastic nature of the model and reflects that even under the same conditions the individual history of every organism has a strong influence on its evolution.
\\
To get a better understanding of the mechanisms behind autoimmune diseases further theoretical and experimental studies should investigate how the idiotypes of the self are distributed over the base graph. Until now we studied the emergence of self-tolerance by filling a complete group with self or by selecting random nodes which do not have too different idiotypes and occupying them permanently. Although in the frame of our model these approaches lead to the emergence of self-tolerant states  it is not clear how well they represent the real distribution of the self over the network. 
\\
Having a better comprehension of the network structure and the place of self in it would enable us to investigate how autoimmunity develops under the assumption of a dynamical and changing self. This is especially interesting in the context of organ transplantations where rejection can occur due to immunological reactions \cite{KrenskyTransplantRejection1990,SumpterL1129}.
It would also enable us to examine if the idiotypic network comprises a mechanism explaining the above-average number of autoimmune diseases as e.g. systemic lupus erythematosus developed during or after pregnancy \cite{shoenfeld2008mosaic,KhashanPregnancyandAutoimmune2011} or the observation that women who suffer a pre-existing autoimmune disease as e.g. rheumatoid arthritis or multiple sclerosis may experience an amelioration of their condition while they are pregnant \cite{WaldorfAutoimmunePregnancy2008,abramsky1994pregnancy}. 
\\
The proposed model assumes an autonomous functional network of B-lymphocytes. Since B-cells interact with T-cells \cite{lanzavecchia1984antigen,Waldmann01031979} it is of obvious interest to extend the model and include interactions between the network of B- and T-cell clones. This would allow for the investigation of a manifold of new aspects as e.g. the therapy of B-cell mediated autoimmune diseases with engineered T-cells \cite{ellebrecht2016reengineering,leslie2016fighting}.
\\
Experimental studies on mice have revealed that the application of monoclonal antibodies can induce long-term remission of T-cell-mediated autoimmune diseases. This is, for example, the case for type 1 diabetes where CD3-specific antibodies are able to restore self-tolerance \cite{chatenoud2003cd3,belghith2003tgf,chatenoud2002use}. Monoclonal antibodies are also able to promote transplant tolerance in case of organ transplantation providing an alternative for  imunosuppressive therapy and all its side effects \cite{You2012GraftToleranceMonoclonal,starzl1998art}. Considering the growing number of clinically effective monoclonal antibodies \cite{reichert2001monoclonal,reichert2005monoclonal} it is of great importance to get a better comprehension of the consequences of immunomanipulation.
\appendix
\setcounter{figure}{0}
\renewcommand{\thefigure}{A\arabic{figure}}
\section{The critical maximal Hamming distance}   
\label{appendix_max_hamming_distance}                                                          
\begin{figure}[h]
\begin{flushleft}
        \vspace{1em}
	\begin{tikzpicture}
  	\text{$u \in S_1 \hspace{2pt}: \hphantom{}$    }   \tikzoverbrace{\bcircle $\cdots$\bcircle \bcircle$\cdots$\bcircle \bcircle$\cdots$\bcircle \bcircle$\cdots$ \bcircle}{d_M}  \squaree $\, \, \, \,$
  	\end{tikzpicture}
  	 \linebreak
	\begin{tikzpicture}
  	\text{$\hspace{0.3pt}v \in S_i \hspace{3.6pt}: \hphantom{}$    }   \tikzunderbrace{\bfcircle$\cdots$\bfcircle \bfcircle$\cdots$\bfcircle}{i-1} \tikzunderbrace{\bcircle$\cdots$ \bcircle \bcircle $\cdots$\bcircle}{d_M-i+1} \squaree $\, \, \, \,$
  	\end{tikzpicture}
  	\linebreak
  	\linebreak
%   	\linibreak
	\begin{tikzpicture}
  	\text{$\hphantom{\, \, \, \,s }w \hphantom{\, \, \, \, s } \hspace{1.2pt}: \hphantom{}$    }   \tikzunderbrace{\bfcircle$\cdots$\bfcircle}{a} \tikzunderbrace{\bcircle $\cdots$\bcircle}{i-1-a} \tikzunderbrace{\bcircle $\cdots$\bcircle}{\substack{d_M-d_H^\mathrm{max}\\+a+\delta_{z,\blacksquare}}} \tikzunderbrace{\bfcircle$\cdots$ \bfcircle}{\substack{d_H^\mathrm{max}-i\\+1+a\\-\delta_{z,\blacksquare}}}  \hspace{0.15cm}$z$ $\, \, \, \,$
  	\end{tikzpicture}
  	\linebreak
\end{flushleft}
\caption{Group membership of a self node $w$ with Hamming distance $d_H^\text{max}$ to the self-seed $v$. The bitstring of a node $u \in S_1$ is depicted for reference. From this it is possible to infer the maximal $d_H^\text{max}$ for which an arbitrary self node $w$ is always found in the singleton or periphery groups, see text.  Empty and filled circles represent complementary entries of determinant bits, squares represent non-determinant bits. $z$ can be a complementary or a non-complementary entry. $v$ differs in $i-1$ determinant bit positions which are common with $u$; $w$ agrees in $a$ determinant bit positions with complementary entries of $v$. The filled circles in the Kronecker deltas have to be replaced by the corresponding entries, zero or one.
}
\label{Fig:trans_mat}
\end{figure}
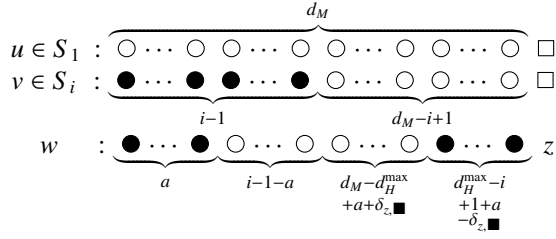

Here we derive the maximal allowed Hamming distance between the self-seed and the self nodes for which the autoreactivity is negligibly small, i.e. all self nodes are found in singleton and periphery groups in the steady state. For the sake of readability, we consider a 12-group pattern only.
The strategy is to pretend to know the critical maximal Hamming distance and use it to calculate the maximal group number of an arbitrary self node with Hamming distance $d_H^\text{max}$ to the self-seed. The group with the largest index not being a core or hole group is $S_5$. Therefore, setting the upper bound of the maximal group number of the self nodes smaller or equal to five we find the upper boundary for the maximal Hamming distance.
\\         
Figure \ref{Fig:trans_mat} shows the bitstrings of a reference node $u \in S_1$, of the self-seed $v \in S_i$ and of a self node $w$ which has the Hamming distance $d_H^{\text{max}}$ to $v$. We can read off the group number $j$ of the self node $w$ by counting the complementary entries in the determinant bit positions, which gives
\begin{linenomath}
\begin{equation}
 j = 2a+d_H^\text{max}-i+2-\delta_{z,\blacksquare}.
\end{equation}
\end{linenomath}
As discussed above, the group index of the self node $w$ should not exceed an upper bound $ j_\text{up}$. Therefore, we determine the maximum of $j$ for fixed  $i$ and $d_H^\text{max}$
\begin{linenomath}
\begin{equation}
 \max_{a,z}\left\{ 2a+d_H^\text{max}+2-i-\delta_{z,\blacksquare} \right\}=i+d_H^\text{max},
\end{equation}
\end{linenomath}
where we used that the maximal value of $a$ is $i-1$ as one concludes from Fig.~\ref{Fig:trans_mat}. This maximum should not exceed the upper bound $j_\text{up}$
\begin{linenomath}
\begin{equation}
i+d_H^\text{max} \leq j_\text{up}.
\end{equation}
\end{linenomath}
The smallest possible choice for $i$ is $i=1$. Furthermore, $j_\text{up}$ is equal to five, such that we arrive at
\begin{linenomath}
\begin{equation}
d_H^\text{max} \leq 5-i=4,
\label{CriticalHammingDistance}
\end{equation}
\end{linenomath}
which agrees with the critical $d_H^\text{max}$ obtained in simulations.
\\
Note that Eq.~(\ref{CriticalHammingDistance}) also predicts in which group the self-seed is in the steady state if we fix $d_H^\text{max}$ and set the upper bound for the group number of the self node to $j_\text{up}=5$. If we choose $d_H^\text{max}=4$ the self-seed is only allowed to be in $S_1$, if we choose $d_H^\text{max}=3$ the self-seed may be in $S_1$ or $S_2$.
\setcounter{figure}{0}
\renewcommand{\thefigure}{B\arabic{figure}}
\section{The transition matrix $\mathbf{T}(a)$}
\label{Derivation of the transitionmatrix}
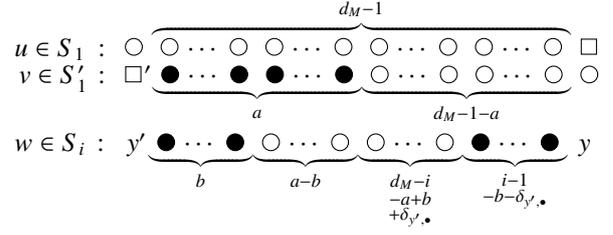
\begin{figure}[h!]
\begin{flushleft}
	\begin{tikzpicture}
  	\text{$u \in S_1 \hspace{2pt}: \hphantom{}$    }    \hspace{0.04cm}\bcircle \tikzoverbrace{\bcircle $\cdots$\bcircle \bcircle$\cdots$\bcircle \bcircle$\cdots$\bcircle \bcircle$\cdots$ \bcircle}{d_M-1}  \squaree $\, \, \, \,$
  	\end{tikzpicture}
  	 \linebreak
	\begin{tikzpicture}
  	\text{$v \in S_1' \hspace{1pt}: \hphantom{}$    }    \hspace{0.03cm}\squaree \hspace{0.15cm}$'$ \tikzunderbrace{\bfcircle$\cdots$\bfcircle \bfcircle$\cdots$\bfcircle}{a} \tikzunderbrace{\bcircle$\cdots$ \bcircle \bcircle $\cdots$\bcircle}{d_M-1-a} \bcircle $\, \, \, \,$
  	\end{tikzpicture}
  	\linebreak
  	\linebreak
%   	\linibreak
	\begin{tikzpicture}
  	\text{$w \in S_i\hspace{1.5pt}: \hphantom{}$    }   \hspace{0.2cm}$y'$\hspace{0.02cm} \tikzunderbrace{\bfcircle$\cdots$\bfcircle}{b} \tikzunderbrace{\bcircle $\cdots$\bcircle}{a-b} \tikzunderbrace{\bcircle $\cdots$\bcircle}{\substack{d_M-i\\-a+b\\+\delta_{y',\bullet}}} \tikzunderbrace{\bfcircle$\cdots$ \bfcircle}{\substack{i-1\\-b-\delta_{y',\bullet}}}\hspace{0.15cm}$y$ $\, \, \, \,$
  	\end{tikzpicture}
  	\linebreak
\end{flushleft}
\caption{Group membership before and after a $T_a$ transition. Depicted are the bitstrings of a node $u \in S_1$ in the initial structure, of a node $v \in S_1'$ in the structure after a $T_a$ transition, and of a node $w$ which belongs to the group $S_i$ in the initial structure. From this it is possible to infer to which group $w$ belongs after the transition, see text. Empty and filled circles represent complementary entries of determinant bits, squares represent non-determinant bits. $v$ differs in $a$ determinant bit positions which are common with $u$; $w$ differs in $i-1$ positions of this kind. From the last line we can read off the number of different entries in the determinant bit positions which determines the group number after transition. The filled circles in the Kronecker deltas have to be replaced by the corresponding entries, zero or one.}
\label{appendix_transmat}
\end{figure}

Here we derive the transition matrix elements $T_{ij}(a)$ in Eq.~(\ref{Transitionmatrix}). Therefore, we take a closer look at the group structure in the systems before and after the transition. Figure~\ref{appendix_transmat} shows the bitstrings of two nodes $u$ and $v$, where $u$ belongs to the group $S_1$ before the transition and $v$ belongs to the group $S_1'$ after the transition. The non-determinant bit has changed its position and the remaining $d_M-1$ determinant bit positions which are common to $u$ and $v$ differ in $a$ entries. Therefore we call this a $T_a$ transition. Furthermore, a bitstring of a node $w$ is depicted which belongs to $S_i$ before the transition because the entries of the determinant bits differ in $i-1$ positions from those of $u$.

Figure \ref{appendix_transmat} allows to read off the group membership of $w$ after the transition. The number of different entries in the determinant bit positions which are common to $v$ and $w$ is
$a-b+i-1-b-\delta_{y',\bullet}+\delta_{y,\bullet}$.
Thus, after the transition, $w$ belongs to the group $S_j'$ where
\begin{linenomath}
\begin{equation}
 j=a-2b+i-\delta_{y',\bullet}+\delta_{y,\bullet}.
 \label{appendix_groupnumber}
\end{equation}
\end{linenomath}
There are $\sum_{y,y'}\binom{a}{b}\binom{d_M-1-a}{i-1-b-\delta_{y',\bullet}}$ nodes in $S_i$ that end up in $S_j'$ after the transition. Using Eq~(\ref{appendix_groupnumber}) to express $b$ in terms of $i$, $j$ and $a$ yields
\begin{linenomath}
\begin{equation}
 T_{ij}(a)=\hspace{-0.7em}\sum_{y,y'=0}^{1}\binom{a}{\frac{a+i-j+y-y'}{2}}\binom{d_M-1-a}{i-1-y'-\frac{a+i-j+y-y'}{2}}.\end{equation}
\end{linenomath} 
Note that the Kronecker deltas are replaced by sums over 0 and 1.
\color{black}
\newline 
\textbf{References}

\bibliographystyle{elsarticle-num} 
\bibliography{bib.bib}

\begin{thebibliography}{10}
\expandafter\ifx\csname url\endcsname\relax
  \def\url#1{\texttt{#1}}\fi
\expandafter\ifx\csname urlprefix\endcsname\relax\def\urlprefix{URL }\fi
\expandafter\ifx\csname href\endcsname\relax
  \def\href#1#2{#2} \def\path#1{#1}\fi

\bibitem{Cruse10Atlas}
J.~M. Cruse, R.~E. Lewis, Atlas of immunology, CRC Press, London/New York,
  2010.

\bibitem{avrameas1983studies}
S.~Avrameas, G.~Dighiero, P.~Lymberi, B.~Guilbert, Studies on natural
  antibodies and autoantibodies, in: Annales de l'Institut Pasteur/Immunologie,
  Vol. 134, Elsevier, 1983, pp. 103--113.

\bibitem{schwartz1986anti}
R.~S. Schwartz, {Anti-DNA antibodies and the problem of autoimmunity}, Cellular
  immunology 99 (1986) 38--43.

\bibitem{tomer1988significance}
Y.~Tomer, Y.~Shoenfeld, The significance of natural autoantibodies,
  Immunological investigations 17 (1988) 389--424.

\bibitem{burnet1959clonal}
F.~Burnet, The clonal selection theory of acquired immunity, Vanderbilt
  University Press, Nashville, 1959.

\bibitem{pereira1986autonomous}
P.~Pereira, L.~Forni, E.-L. Larsson, M.~Cooper, C.~Heusser, A.~Coutinho,
  {Autonomous activation of B and T cells in antigen-free mice}, European
  Journal of Immunology 16~(6) (1986) 685--688.

\bibitem{jerne74network}
N.~Jerne, Towards a network theory of the immune system, Ann. Inst. Pasteur
  Immunol. C 125 (1974) 373.

\bibitem{jerne1984idiotypic}
N.~Jerne, Idiotypic networks and other preconceived ideas, Immunological
  reviews 79~(1) (1984) 5--24.

\bibitem{Jerne85}
N.~Jerne, The generative grammar of the immune system, EMBO Journal 4 (1985)
  847--852.

\bibitem{behn2007idiotypic}
U.~Behn, Idiotypic networks: toward a renaissance?, Immunological reviews
  216~(1) (2007) 142--152.

\bibitem{behn2011idiotype}
U.~Behn, Idiotype network, in: {Encyclopedia of Life Sciences}, John Wiley and
  Sons, Chichester, 2011.

\bibitem{Tauber94}
A.~Tauber, The immune self: theory or metaphor?, Cambridge University Press,
  Cambridge, 1994.

\bibitem{berek1988dynamic}
C.~Berek, C.~Milstein, The dynamic nature of the antibody repertoire,
  Immunological reviews 105~(1) (1988) 5--26.

\bibitem{perelson1997immunology}
A.~S. Perelson, G.~Weisbuch, Immunology for physicists, Reviews of modern
  physics 69~(4) (1997) 1219.

\bibitem{tonegawa1983somatic}
S.~Tonegawa, Somatic generation of antibody diversity, Nature 302~(5909) (1983)
  575--581.

\bibitem{vogelstein1982specific}
B.~Vogelstein, R.~Dintzis, H.~Dintzis, Specific cellular stimulation in the
  primary immune response: a quantized model, Proceedings of the National
  Academy of Sciences 79~(2) (1982) 395--399.

\bibitem{coutinho1989beyond}
A.~Coutinho, Beyond clonal selection and network, Immunological reviews 110~(1)
  (1989) 63--88.

\bibitem{varela1991second}
F.~J. Varela, A.~Coutinho, Second generation immune networks, Immunology today
  12~(5) (1991) 159--166.

\bibitem{coutinho2003walk}
A.~Coutinho, {A walk with Francisco Varela from first-to second-generation
  networks: in search of the structure, dynamics and metadynamics of an
  organism-centered immune system}, Biological research 36~(1) (2003) 17--26.

\bibitem{sim1986t}
G.-K. Sim, I.~Macneil, A.~A. Augustin, T helper cell receptors: idiotypes and
  repertoire, Immunological reviews 90~(1) (1986) 49--72.

\bibitem{UWFC77}
J.~Urbain, M.~Wikler, J.~Franssen, C.~Collignon, Idiotypic regulation of the
  immune system by the induction of antibodies against anti-idiotypic
  antibodies, Proceedings of the National Academy of Sciences of the USA
  74~(11) (1977) 5126--5130.

\bibitem{Hampe12}
C.~S. Hampe, Protective role of anti-idiotypic antibodies in autoimmunity --
  {L}essons for type 1 diabetes, Autoimmunity 45~(4) (2012) 320--331.

\bibitem{Avrameas91}
S.~Avrameas, Natural autoantibodies: from 'horror autotoxicus' to 'gnothi
  seauton', Immunology Today 12~(5) (1991) 154 -- 159.

\bibitem{SG97}
Y.~Shoenfeld, J.~George, Induction of autoimmunity. {A} role for the idiotypic
  network., Annals of the New York Academy of Sciences 815 (1997) 342--9.

\bibitem{Pendergraftetal04}
W.~F. Pendergraft, G.~A. Preston, R.~R. Shah, A.~Tropsha, C.~W. Carter, J.~C.
  Jennette, R.~J. Falk, Autoimmunity is triggered by c{PR}-3(105-201), a
  protein complementary to human autoantigen proteinase-3, Nature Medicine
  10~(1) (2003) 72--79.

\bibitem{McGH05}
K.~L. McGuire, D.~S. Holmes, Role of complementary proteins in autoimmunity: an
  old idea re-emerges with new twists, Trends in Immunology 26~(7) (2005)
  367--372.

\bibitem{TR10}
A.~G. Tzioufas, J.~G. Routsias, Idiotype, anti-idiotype network of
  autoantibodies, Autoimmunity Reviews 9~(9) (2010) 631--633.

\bibitem{RT10}
J.~G. Routsias, A.~G. Tzioufas, B-cell epitopes of the intracellular
  autoantigens {R}o/{SSA} and {L}a/{SSB}: Tools to study the regulation of the
  autoimmune response, Journal of Autoimmunity 35~(3) (2010) 256--264.

\bibitem{Shoenfeld04}
Y.~Shoenfeld, The idiotypic network in autoimmunity: {A}ntibodies that bind
  antibodies that bind antibodies, Nature Medicine 10~(1) (2004) 17--18.

\bibitem{DVK86}
D.~S. Dwyer, M.~Vakil, J.~Kearney, Idiotypic network connectivity and a
  possible cause of myasthenia gravis, Journal of Experimental Medicine 164~(4)
  (1986) 1310--1318.

\bibitem{schmidtchen2012randomly}
H.~Schmidtchen, M.~Th{\"u}ne, U.~Behn, Randomly evolving idiotypic networks:
  Structural properties and architecture, Physical Review E 86~(1) (2012)
  011930.

\bibitem{SV89}
J.~Stewart, F.~J. Varela, Exploring the meaning of connectivity in the immune
  network, Immunological Reviews 110 (1989) 37--61.

\bibitem{kearney1987non}
J.~Kearney, M.~Vakil, N.~Nicholson, {{Non-random VH gene expression and
  idiotype anti-idiotype expression in early B cells}}, in: G.~Kelsoe,
  D.~Schulze (Eds.), Evolution and vertebrate immunity: The antigen receptor
  and MHC gene families, Vol.~1, Texas University Press Austin, 1987, pp.
  175--190.

\bibitem{SVC89}
J.~Stewart, F.~J. Varela, A.~Coutinho, The relationship between connectivity
  and tolerance as revealed by computer simulation of the immune network:
  {S}ome lessons for an understanding of autoimmunity, Journal of Autoimmunity
  2 (1989) 15--23.

\bibitem{sulzer1994central}
B.~Sulzer, J.~L. Van~Hemmen, U.~Behn, Central immune system, the self and
  autoimmunity, Bulletin of mathematical biology 56~(6) (1994) 1009--1040.

\bibitem{leon1998natural}
K.~Le{\'o}n, J.~Carneiro, R.~Per{\'e}z, E.~Montero, A.~Lage, Natural and
  induced tolerance in an immune network model, Journal of theoretical biology
  193~(3) (1998) 519--534.

\bibitem{carneiro1996modela}
J.~Carneiro, A.~Coutinho, J.~Faro, J.~Stewart, {A model of the immune network
  with B-T cell co-operation. I-Prototypical structures and dynamics}, Journal
  of theoretical biology 182~(4) (1996) 513--529.

\bibitem{carneiro1996modelb}
J.~Carneiro, A.~Coutinho, J.~Stewart, {A model of the immune network with B-T
  cell co-operation. II-The simulation of ontogenesis}, Journal of Theoretical
  Biology 182~(4) (1996) 531--547.

\bibitem{Agliari2013}
E.~Agliari, A.~Annibale, A.~Barra, A.~C.~C. Coolen, D.~Tantari, Immune
  networks: multitasking capabilities near saturation, Journal of Physics A:
  Mathematical and Theoretical 46~(41) (2013) 415003.

\bibitem{agliari2015anergy}
E.~Agliari, A.~Barra, G.~Del~Ferraro, F.~Guerra, D.~Tantari, {Anergy in
  self-directed B lymphocytes: a statistical mechanics perspective}, Journal of
  theoretical biology 375 (2015) 21--31.

\bibitem{menshikov2015idiotypic}
I.~Menshikov, L.~Beduleva, M.~Frolov, N.~Abisheva, T.~Khramova, E.~Stolyarova,
  K.~Fomina, The idiotypic network in the regulation of autoimmunity:
  Theoretical and experimental studies, Journal of theoretical biology 375
  (2015) 32--39.

\bibitem{brede2003}
M.~Brede, U.~Behn, Patterns in randomly evolving networks: Idiotypic networks,
  Physical Review E 67~(3) (2003) 031920.

\bibitem{schmidtchen2012Meanfield}
H.~Schmidtchen, U.~Behn, Randomly evolving idiotypic networks: modular mean
  field theory, Physical Review E 86~(1) (2012) 011931.

\bibitem{SWB14}
R.~Schulz, B.~Werner, U.~Behn, Self-tolerance in a minimal model of the
  idiotypic network, Frontiers in immunology 5 (2014) 103.

\bibitem{Saeki20154}
K.~Saeki, H.~M. Doekes, R.~J.~D. Boer, Optimal t cell cross-reactivity and the
  role of regulatory t cells, Journal of Theoretical Biology 375 (2015) 4 --
  12, theories and Modeling of Autoimmunity.
\newblock \href {http://dx.doi.org/10.1016/j.jtbi.2014.11.007}
  {\path{doi:10.1016/j.jtbi.2014.11.007}}.

\bibitem{Blyuss201513}
K.~Blyuss, L.~Nicholson, Understanding the roles of activation threshold and
  infections in the dynamics of autoimmune disease, Journal of Theoretical
  Biology 375 (2015) 13 -- 20, theories and Modeling of Autoimmunity.
\newblock \href {http://dx.doi.org/10.1016/j.jtbi.2014.08.019}
  {\path{doi:10.1016/j.jtbi.2014.08.019}}.

\bibitem{RootBernstein20151}
R.~Root-Bernstein, Towards an integration of mathematical models, theories and
  observations concerning autoimmune diseases, Journal of Theoretical Biology
  375 (2015) 1 -- 3, theories and Modeling of Autoimmunity.
\newblock \href {http://dx.doi.org/10.1016/j.jtbi.2015.04.003}
  {\path{doi:10.1016/j.jtbi.2015.04.003}}.

\bibitem{schmidtchen2006randomly}
H.~Schmidtchen, U.~Behn, Randomly evolving idiotypic networks: analysis of
  building principles, in: Artificial Immune Systems ICARIS, Springer, Berlin
  Heidelberg, 2006, pp. 81--94.

\bibitem{thomas2014complex}
V.~Thomas-Vaslin, A complex immunological idiotypic network for maintenance of
  tolerance, Frontiers in immunology 5 (2014) 369.

\bibitem{shannon2001mathematical}
C.~Shannon, A mathematical theory of communication, The Bell System Technical
  Journal 27~(3) (1948) 379.

\bibitem{linton2004age}
P.~J. Linton, K.~Dorshkind, Age-related changes in lymphocyte development and
  function, Nature immunology 5~(2) (2004) 133--139.

\bibitem{henderson1969treatment}
E.~D. Thomas, R.~Storb, R.~A. Clift, A.~Fefer, F.~L. Johnson, P.~E. Neiman,
  K.~G. Lerner, H.~Glucksberg, C.~D. Buckner, {Bone-Marrow Transplantation},
  New England Journal of Medicine 292~(17) (1975) 895--902.

\bibitem{Bach2002Infections}
J.-F. Bach, {The Effect of Infections on Susceptibility to Autoimmune and
  Allergic Diseases}, New England Journal of Medicine 347~(12) (2002) 911--920.

\bibitem{Rose1998rolfeofinfections}
N.~R. Rose, The role of infection in the pathogenesis of autoimmune disease,
  Seminars in Immunology 10~(1) (1998) 5 -- 13.

\bibitem{Bach2005Infectionsandautoimmdiseases}
J.-F. Bach, Infections and autoimmune diseases, Journal of Autoimmunity 25
  (2005) 74 -- 80.

\bibitem{beyerlein2016infections}
A.~Beyerlein, E.~Donnachie, S.~Jergens, A.-G. Ziegler, {Infections in Early
  Life and Development of Type 1 Diabetes}, JAMA 315~(17) (2016) 1899--1901.

\bibitem{davidson2001autoimmune}
A.~Davidson, B.~Diamond, Autoimmune diseases, New England Journal of Medicine
  345~(5) (2001) 340--50.

\bibitem{Wucherpfennig2001autoimmunitybyinfection}
K.~W. Wucherpfennig, Mechanisms for the induction of autoimmunity by infectious
  agents, The Journal of Clinical Investigation 108~(8) (2001) 1097--1104.

\bibitem{cole1993triggering}
B.~C. Cole, M.~M. Griffiths, {Triggering and exacerbation of autoimmune
  arthritis by the Mycoplasma arthritidis superantigen MAM}, Arthritis \&
  Rheumatism 36~(7) (1993) 994--1002.

\bibitem{Rees1995GuillainBarre}
J.~H. Rees, S.~E. Soudain, N.~A. Gregson, R.~A. Hughes, {Campylobacter jejuni
  Infection and Guillain-Barr\'e Syndrome}, New England Journal of Medicine
  333~(21) (1995) 1374--1379.

\bibitem{mickens1994nonstandard}
R.~E. Mickens, Nonstandard finite difference models of differential equations,
  Vol. 115, World Scientific, Singapore, 1994.

\bibitem{olver2010nist}
F.~W. Olver, {NIST handbook of mathematical functions}, Cambridge University
  Press, 2010.

\bibitem{Kumar2013}
A.~Kumar, Autologous blood transfusion, in: F.~B. Mandell (Ed.), Perioperative
  Management of Patients with Rheumatic Disease, Springer, New York, 2013, pp.
  41--44.

\bibitem{EndelePhD2005}
D.~Endele, {Vor- und Nachteile autologer Transfusionsverfahren}, Ph.D. thesis,
  Eberhard-Karls-Universit{\"a}t zu T{\"u}bingen (2005).

\bibitem{AutologousTransfusionPoposvsky}
M.~Popovsky, B.~Whitaker, N.~Arnold, Severe outcomes of allogeneic and
  autologous blood donation: frequency and characterization, Transfusion 35~(9)
  (1995) 734--737.

\bibitem{blajchman2006clinical}
M.~A. Blajchman, The clinical benefits of the leukoreduction of blood products,
  Journal of Trauma and Acute Care Surgery 60~(6) (2006) 83--90.

\bibitem{feldmann2005design}
M.~Feldmann, L.~Steinman, Design of effective immunotherapy for human
  autoimmunity, Nature 435~(7042) (2005) 612--619.

\bibitem{KrenskyTransplantRejection1990}
A.~M. Krensky, A.~Weiss, G.~Crabtree, M.~M. Davis, P.~Parham,
  {T-Lymphocyte-Antigen Interactions in Transplant Rejection}, New England
  Journal of Medicine 322~(8) (1990) 510--517.

\bibitem{SumpterL1129}
T.~L. Sumpter, D.~S. Wilkes, {Role of autoimmunity in organ allograft
  rejection: a focus on immunity to type V collagen in the pathogenesis of lung
  transplant rejection}, American Journal of Physiology - Lung Cellular and
  Molecular Physiology 286~(6) (2004) 1129--1139.

\bibitem{shoenfeld2008mosaic}
Y.~Shoenfeld, G.~Zandman-Goddard, L.~Stojanovich, M.~Cutolo, H.~Amital,
  Y.~Levy, M.~Abu-Shakra, O.~Barzilai, Y.~Berkun, M.~Blank, et~al., The mosaic
  of autoimmunity: hormonal and environmental factors involved in autoimmune
  diseases--2008, The Israel Medical Association Journal 10~(1) (2008) 8.

\bibitem{KhashanPregnancyandAutoimmune2011}
A.~S. Khashan, L.~C. Kenny, T.~M. Laursen, U.~Mahmood, P.~B. Mortensen, T.~B.
  Henriksen, K.~O'Donoghue, {Pregnancy and the Risk of Autoimmune Disease},
  PLoS ONE 6~(5) (2011) 1--7.

\bibitem{WaldorfAutoimmunePregnancy2008}
K.~M.~A. Waldorf, J.~L. Nelson, {Autoimmune Disease During Pregnancy and the
  Microchimerism Legacy of Pregnancy}, Immunological Investigations 37~(5-6)
  (2008) 631--644.

\bibitem{abramsky1994pregnancy}
O.~Abramsky, Pregnancy and multiple sclerosis, Annals of Neurology 36~(1)
  (1994) 38--41.

\bibitem{lanzavecchia1984antigen}
A.~Lanzavecchia, {Antigen-specific interaction between T and B cells}, Nature
  314~(6011) (1984) 537--539.

\bibitem{Waldmann01031979}
H.~Waldmann, {Interactions between T and B Cells: A Review}, Journal of the
  Royal Society of Medicine 72~(3) (1979) 198--202.

\bibitem{ellebrecht2016reengineering}
C.~T. Ellebrecht, V.~G. Bhoj, A.~Nace, E.~J. Choi, X.~Mao, M.~J. Cho,
  G.~Di~Zenzo, A.~Lanzavecchia, J.~T. Seykora, G.~Cotsarelis, M.~C. Milone,
  A.~S. Payne, {Reengineering chimeric antigen receptor T cells for targeted
  therapy of autoimmune disease}, Science 353~(6295) (2016) 179--184.

\bibitem{leslie2016fighting}
M.~Leslie, Fighting autoimmunity with immune cells, Science 353~(6294) (2016)
  14--14.

\bibitem{chatenoud2003cd3}
L.~Chatenoud, {CD3-specific antibody-induced active tolerance: from bench to
  bedside}, Nature Reviews Immunology 3~(2) (2003) 123--132.

\bibitem{belghith2003tgf}
M.~Belghith, J.~A. Bluestone, S.~Barriot, J.~M{\'e}gret, J.-F. Bach,
  L.~Chatenoud, {TGF-$\beta$-dependent mechanisms mediate restoration of
  self-tolerance induced by antibodies to CD3 in overt autoimmune diabetes},
  Nature medicine 9~(9) (2003) 1202--1208.

\bibitem{chatenoud2002use}
L.~Chatenoud, The use of monoclonal antibodies to restore self-tolerance in
  established autoimmunity, Endocrinology and metabolism clinics of North
  America 31~(2) (2002) 457--475.

\bibitem{You2012GraftToleranceMonoclonal}
S.~You, J.~Zuber, C.~Kuhn, M.~Baas, F.~Valette, V.~Sauvaget, S.~Sarnacki,
  B.~Sawitzki, J.-F. Bach, H.-D. Volk, L.~Chatenoud, {Induction of Allograft
  Tolerance by Monoclonal CD3 Antibodies: A Matter of Timing}, American Journal
  of Transplantation 12~(11) (2012) 2909--2919.

\bibitem{starzl1998art}
T.~Starzl, The art of tolerance, Nature medicine 4~(9) (1998) 1006--1008.

\bibitem{reichert2001monoclonal}
J.~M. Reichert, Monoclonal antibodies in the clinic, Nature biotechnology
  19~(9) (2001) 819--822.

\bibitem{reichert2005monoclonal}
J.~M. Reichert, C.~J. Rosensweig, L.~B. Faden, M.~C. Dewitz, Monoclonal
  antibody successes in the clinic, Nature biotechnology 23~(9) (2005)
  1073--1078.

\end{thebibliography}
\end{document}